\documentclass[a4paper,fleqn,usenatbib]{mnras}

\usepackage{savesym}
\usepackage{txfonts}
\savesymbol{iint}
\savesymbol{iiint}
\savesymbol{iiiint}
\savesymbol{idotsint}

\usepackage[T1]{fontenc}
\usepackage{ae,aecompl}

\usepackage{graphicx}	
\usepackage{subfig}
\usepackage{amsmath}	
\restoresymbol{AMS}{iint}
\restoresymbol{AMS}{iiint}
\restoresymbol{AMS}{iiiint}
\restoresymbol{AMS}{idotsint}
\usepackage{amssymb}	
\usepackage[shortlabels]{enumitem}
\usepackage{xcolor}

\definecolor{darkgreen}{RGB}{0,128,0}

\title[Ejections by SMBH binaries]{Gravitational interactions of stars with supermassive black hole binaries - II. Hypervelocity stars}

\author[S. Darbha et al.]{
Siva Darbha,$^{1}$\thanks{E-mail: siva.darbha@berkeley.edu}
Eric R. Coughlin,$^{2,4}$\thanks{Einstein fellow}
Daniel Kasen$^{1,2,3}$
and Eliot Quataert$^{1,2}$
\\
$^{1}$Department of Physics, University of California, Berkeley, Berkeley, CA 94720, USA\\
$^{2}$Department of Astronomy and Theoretical Astrophysics Center, University of California, Berkeley, Berkeley, CA 94720, USA\\
$^{3}$Nuclear Science Division, Lawrence Berkeley National Laboratory, Berkeley, CA 94720, USA \\
$^4$Columbia Astrophysics Laboratory, Columbia University, New York, NY, 10027, USA
}

\date{Accepted XXX. Received YYY; in original form ZZZ}

\pubyear{2018}

\begin{document}
\label{firstpage}
\pagerange{\pageref{firstpage}--\pageref{lastpage}}
\maketitle

\begin{abstract}

\noindent
Supermassive black holes (SMBHs) in galactic nuclei can eject hypervelocity stars (HVSs). Using restricted three-body integrations, we study the properties of stars ejected by circular, binary SMBHs as a function of their mass ratios $q = M_2 / M_1$ and separations $a$, focusing on the stellar velocity and angular distributions. We find that the ejection probability is an increasing function of $q$ and $a$, and that the mean ejected velocity scales with $q$ and $a$ similar to previous work but with modified scaling constants. Binary SMBHs tend to eject their fastest stars toward the binary orbital plane. We calculate the ejection rates as the binary SMBHs inspiral, and find that they eject stars with velocities $v_\infty > 1000$ km/s at rates of $\sim 4 \times 10^{-2} - 2 \times 10^{-1}$ yr$^{-1}$ for $q = 1$ ($\sim 10^{-4} - 10^{-3}$ yr$^{-1}$ for $q = 0.01$) over their lifetimes, and can emit a burst of HVSs with $v_\infty > 3000$ km/s as they coalesce. We integrate the stellar distributions over the binary SMBH inspiral and compare them to those produced by the ``Hills mechanism'' (in which a single SMBH ejects a star after tidally separating a binary star system), and find that $N \sim 100$ HVS velocity samples with $v_\infty \gtrsim 200$ km/s are needed to confidently distinguish between a binary and single SMBH origin.

\end{abstract}

\begin{keywords}
black hole physics -- galaxies: nuclei -- stars: kinematics and dynamics -- stars: statistics
\end{keywords}



\section{Introduction}
\label{sec:intro}

Supermassive black holes (SMBHs) in the centers of galaxies can eject stars with velocities $\tilde{v} \gtrsim 1000$ km/s \citep{hills88,yu03,brown15}. These hypervelocity stars (HVSs) can be used to probe galactic BHs and their stellar environments. In particular, the properties of HVSs can reveal the presence of a single or binary SMBH in a galaxy's core \citep{yu03}, illuminate a galaxy's star formation history \citep{kollmeier07}, and constrain the shape of a galactic potential generated by both ordinary and dark matter \citep{gnedin05,kenyon14,rossi17}.

Several hypervelocity stars have been observed in the Milky Way \citep{brown05,brown15b}, and the high-precision data from the Global Astrometric Interferometer for Astrophysics (GAIA) \citep{kenyon14,marchetti18,marchetti18b,boubert18,brown18} and the Large Synoptic Survey Telescope (LSST) \citep{lsst09} will likely produce a catalogue of thousands more. Indeed, HVSs are currently detectable only near the Milky Way (a constraint that will continue into the near future), and those observed will likely originate from the Galactic Center (GC) \citep{brown15}. A subset may also arise from satellite galaxies, particularly the Large Magellanic Cloud (LMC) \citep{edelmann05,gualandris07,boubert16,erkal18}, and a small number may come from the Andromeda galaxy (M31) \citep{sherwin08}.

SMBHs can expel HVSs by three general classes of encounters: 1) a stellar binary encounters a single SMBH as a ``slow intruder'' (i.e. the velocity of the binary center-of-mass at infinity relative to the SMBH is lower than the binary orbital velocity; \citealt{hills89}), the black hole replaces one of the stars in an exchange collision, and the dislodged star departs with a high velocity (the ``Hills mechanism,'' \citealt{hills88}); 2) a single star encounters a binary SMBH as a ``slow intruder,'' and is ejected with an enhanced velocity after extracting some of the binary SMBH energy \citep{yu03}; and 3) a star is bound to one SMBH and is ejected after an encounter with the second SMBH \citep{gualandris05,guillochon15}. The rates of ejection and the properties of HVSs differ depending on the means of production.

Although HVSs can be produced by other means, including a supernova explosion by one constituent of a compact binary \citep{blaauw61}, close encounters between single stars \citep{yu03}, and scattering between stars and stellar mass black hole clusters in the potential of a SMBH \citep{miraldaescude00,oleary08}, most are expected to have SMBH origins \citep{yu03,gualandris05,brown15}, and thus the production channels listed above have been investigated in detail. Hypervelocity stars produced by the Hills mechanism have been studied extensively (e.g., \citealt{hills88,yu03,bromley06,kenyon08,rossi14,brown15}). A SMBH binary origin for HVSs (henceforth labeled the ``SMBHB mechanism,'' including both 2 and 3 above) has also received attention \citep{quinlan96,zier01,yu03}, both when the binary ejects incident unbound stars \citep{yu03,sesana06,sesana07b} or bound stars \citep{sesana08}, the latter most often when an IMBH inspirals towards a SMBH \citep{gualandris05,baumgardt06,levin06,sesana07a,sesana08,lockmann08}.

Several features of the velocity distribution of HVSs have been illuminated that can be used to distinguish these two mechanisms. For example, the SMBHB mechanism can eject stars at a rate upwards of $\sim 10$ times higher than the Hills mechanism. The average ejection velocity of the Hills mechanism depends on the mass of the stellar binary, whereas that of the SMBHB mechanism is agnostic to the stellar properties \citep{hills88,yu03}. SMBH binaries can produce velocity distributions with more variable extrema owing to the additional channels and degrees of freedom available in their interactions with stars \citep{gualandris05,baumgardt06,sesana06,sesana07a}. HVSs (all those with $\tilde{v} \gtrsim 1000$ km/s) produced by the Hills mechanism are ejected isotropically, whereas those from the SMBHB mechanism have a more complicated angular behavior: for circular SMBH binaries, HVSs are preferentially ejected in the binary orbital plane for nearly-equal mass ratios and wider binaries, but more isotropically for lower mass ratios and tighter binaries \citep{zier01,baumgardt06,sesana06,levin06}. The Hills mechanism will eject HVSs as long as there is a continuous supply of stellar binaries, whereas the SMBHB mechanism ejects a burst of HVSs in $t \sim 1 - 10$ Myr; this occurs both when a contracting binary SMBH ejects incident unbound stars \citep{sesana06,sesana07b} or inspirals through a stellar cusp \citep{baumgardt06,levin06,sesana08}.

In order to determine the specific mechanism that produces observed HVSs arising from different galaxies, and thus infer the existence or binarity of SMBHs in a given galactic nucleus, one must undertake a systematic study of the properties of HVSs produced by binary SMBHs, and compare them to those produced by the Hills mechanism under realistic conditions. At present, this comparison can help to distinguish between the presence of a single or binary SMBH in the Milky Way, one of its satellite galaxies, or even Andromeda. As astrometry improves, this analysis can be extended to other nearby galaxies.

This paper is the second in a two-part sequence. In the first part, we investigated the statistics of tidal disruption events (TDEs) by binary SMBHs over a range of binary mass ratios and separations \citep{darbha18} \footnote{There were a few minor mistakes in the published version of the first paper. There was a mistake in the error bars in Figures 7, 12, and 14, and in the caption describing them in Figure 12. The notation and terminology were also unclear at times. We have corrected similar work in this paper, which can be used for reference.}, extending the work of \citet{coughlin17} (also see \citealt{coughlin18} for an application to an observed event). In this paper, we study the hypervelocity stars ejected by these SMBH binaries over the same parameter range. The setup and inputs of our simulations partly overlap with those of previous studies, though we focus more exclusively on circular, nearly equal-mass SMBH binaries and have a more densely populated parameter space in mass ratio and separation, so our results can corroborate earlier work and have the potential to reveal new features.

In Section \ref{sec:setup}, we briefly recapitulate our simulation setup; a more detailed overview can be found in \citet{darbha18}. In Section \ref{sec:hvss}, we present the ejection probability and the properties of the ejected stars, namely their velocity and angular distributions. We then present the time-dependent ejection rate as the SMBH binary contracts due to stellar scattering and gravitational wave emission. In Section \ref{sec:comparison}, we compare the integrated distributions for stars ejected by a binary and single SMBH, a more realistic treatment than previously considered in this parameter regime. To conclude, we present the number samples needed to distinguish between the SMBHB and Hills mechanisms, and to identify the progenitor's properties. We summarize our results in Section \ref{sec:conclusion}.

\section{Simulation Setup}
\label{sec:setup}

We study the properties of hypervelocity stars ejected by ``hard'' SMBH binaries when the stars are initially incident from the loss cone in the ``pinhole'' (or ``full loss cone'') regime \citep{frank76,lightman77,cohn78,magorrian99}. 
A binary SMBH becomes ``hard'' at roughly the separation \citep{quinlan96}
\begin{equation}
\tilde{a}_h = \frac{G M_1 M_2}{4 (M_1 + M_2) \sigma^2}
\end{equation}
where $M_1$ and $M_2$ are the masses of the primary and secondary, and $\sigma$ is the one-dimensional velocity dispersion of the stars in the galactic nucleus. We discussed our assumption of ``hard'' binaries and the ``full loss cone'' regime in the first part of this two-paper sequence \citep{darbha18}, and direct the reader there for a more detailed discussion.

Stars incident from the loss cone can be ejected only if they are not first tidally disrupted. A star that approaches a SMBH too closely gets tidally disrupted when the tidal gravity from the SMBH overwhelms the self-gravity of the star. Disruption occurs at the tidal radius, which is roughly equal to $\tilde{r}_t \simeq R_* \left( M_\bullet / M_* \right)^{1/3}$ where $M_\bullet$ is the black hole mass and $M_*$ and $R_*$ are the star's mass and radius. We adopt this definition of the tidal radius and assume that all stars that enter it are completely disrupted, ignoring complications owing to stellar structure \citep{lodato09,guillochon13,mainetti17}.

We outlined our setup and described its domain of validity in our first paper \citep{darbha18}. To briefly summarize, we use Mathematica to simulate stars incident on a binary SMBH in the circular restricted three-body approximation, in the point particle limit, and using Newtonian gravitational potentials. We write our simulation parameters in the units $G = M = \tilde{a} = 1$, where $M = M_1 + M_2$ is the total binary SMBH mass and $\tilde{a}$ is its separation. The SMBH binary is then described by two dimensionless quantities: the mass ratio $q = M_2/M_1$ and the primary's (dimensionless) tidal radius $r_{t1} = \tilde{r}_{t1}/\tilde{a}$. We set the origin of the coordinate system to the binary SMBH center of mass. The stars begin on parabolic (specific energy $\epsilon = 0$) orbits with respect to the origin, and are distributed isotropically over a sphere of radius $r=50$. The specific angular momentum $\ell$ of each star is sampled such that its square is uniformly distributed in the range $\ell^2 \in [0,4]$ (yielding uniformly distributed pericenters $r_p \in [0,2]$), which corresponds to the ``pinhole'' (or ``full loss cone'') regime. An integration terminates if the star crosses the tidal radius of one of the BHs, if it escapes to $r=100$, or if the simulation time reaches $t = 10^4$. A large fraction of the ``escaped'' stars have positive energy and will thus be ``ejected'' from the binary SMBH; we use this terminology throughout to distinguish between these two outcomes. We explore many points in the parameter range $q \in [0.01, 1]$ and $r_{t1} \in [0.001, 0.1]$ ($\tilde{a}/\tilde{r}_{t1} \in [10, 1000]$), and simulate $5 \times 10^6$ encounters for each point. For comparison, we simulated a smaller number of encounters for a few points in our parameter space with the N-body code REBOUND using the IAS15 integrator \citep{rein12,rein15}, and found close agreement with our results.

Though we vary these two binary SMBH quantities in our simulations, we ultimately interpret these in our results as varying $\tilde{a}$ and $M_2$ while holding $\tilde{r}_{t1}$ and $M_1$ fixed. Note that we do not hold the total mass $M$ fixed, in contrast to other studies. In this paper, if the dimensional character of a variable is not clear from the context, then we write dimensioned variables with tildes on top and dimensionless ones without them. Table \ref{tab:scales} presents the different scales we use to define dimensionless variables in different sections of the paper, unless otherwise noted.

\begin{table*}
\centering
\begin{tabular}{|c|c|c|c|c|c|c|}
\hline
Section & Length & Mass & Time & Velocity & Specific energy ($\epsilon$) \\
\hline
Simulation (\ref{sec:setup}) & $\tilde{a}$ & $M = M_1 + M_2$ & $\sqrt{\tilde{a}^3 / GM}$ & $v_\textrm{bin} = \sqrt{GM/\tilde{a}}$ & $GM/\tilde{a}$ \\
\hline
Ejection probability (\ref{subsec:ejprob}) & $\tilde{r}_{t1}$ & & & & \\
 & $\left( a = \tilde{a} / \tilde{r}_{t1} \right)$ & & & & \\
\hline
Ejection properties (\ref{subsec:properties}) & $\tilde{r}_{t1}$ & $M_1$ & & $v_0 = \sqrt{GM_1/\tilde{r}_{t1}}$ & \\
 & $\left( a = \tilde{a} / \tilde{r}_{t1} \right)$ & & & $\left( v = \tilde{v}/v_0 \right)$ & \\
 \hline
Ejection rate (\ref{subsec:inspiral}) & $\tilde{r}_{t1}$ & $M_1$ & $t_0 = \tilde{r}^4_{t1} c^5 / G^3 M^3_1$ & & $GM_1 / \tilde{r}_{t1}$ \\
 & $\left( a = \tilde{a} / \tilde{r}_{t1} \right)$ & & $\left( t = \tilde{t} / t_0 \right)$ & & \\
\hline
Comparison (\ref{sec:comparison}): Binary SMBH & $\tilde{r}_{t1}$ & & & $v_\textrm{bin} = \sqrt{GM/\tilde{a}}$ & \\
 & $\left( a = \tilde{a} / \tilde{r}_{t1} \right)$ & & & $\left( v = \tilde{v}/v_\textrm{bin} \right)$ & \\
Comparison (\ref{sec:comparison}): Binary star & $a_{*0} = 1$ AU & & & $v_* = \sqrt{Gm_\textrm{tot}/\tilde{a}_*}$ & \\
 & $\left( a_* = \tilde{a}_* / a_{*0} \right)$ & & & $\left( v = \tilde{v}/v_* \right)$ & \\
\hline
\end{tabular}
\caption{The quantities used to non-dimensionalize variables (with the dimensions length, mass, time, velocity, and stellar specific energy) in different sections of the paper, unless stated otherwise. Throughout the paper, if the dimensional character of a variable is not clear from the context, then we write dimensioned variables with tildes on top and dimensionless ones without them. The parentheses show the definitions of some variables used in each section.}
\label{tab:scales}
\end{table*}

We take as our main example a primary with mass $M_1 = 10^6 M_\odot$ and stars with solar parameters. The primary's tidal radius is then $\tilde{r}_{t1} = 2.3$ $\mu$pc and the range of separations is $\tilde{a} \in [0.023, 2.3]$ mpc. The characteristic orbital velocity of a binary SMBH is $v_\textrm{bin} = \sqrt{GM/\tilde{a}} \sim 2000$ km/s for $M \sim 10^6 M_\odot$ and $\tilde{a} \sim 10^{-3}$ pc, and the velocities of the stars in the bulge are $\sigma \sim 100$ km$\cdot$s$^{-1}$ for a Milky Way-like galaxy \citep{gultekin09}. Since $\sigma \ll v_\textrm{bin}$, the stars are ``slow intruders'' \citep{hills89} and will be expelled with enhanced velocities in most encounters with the SMBH binary.

For comparison, the distribution of approach distances in our simulations differ from those of \citet{quinlan96} and \citet{sesana06}. We initialize the stars uniformly distributed in pericenter $r_p$, whereas the above authors initialized them uniformly distributed in $b^2$, where $b$ is the impact parameter, which skews the stars towards higher pericenters. We also initialize the stars with zero energy whereas the above authors initialized them on mildly hyperbolic orbits, though this has little impact on our results, as scattering is not sensitive to the initial velocity as long as it is much less than the binary SMBH circular velocity \citep{quinlan96}.

\section{Hypervelocity Stars}
\label{sec:hvss}

In this section, we study in detail the influence of the binary mass ratio and separation on the stellar ejections. We examine the ejection probability (Section \ref{subsec:ejprob}) and the properties of the ejected stars (Section \ref{subsec:properties}), in particular their distributions in velocity and direction.

Throughout this paper, we use the following terminology to distinguish between two possible end states for expelled stars. We label a star as ``escaped'' if it reaches $r_{100} = \tilde{r}_{100} / \tilde{a} \equiv 100$ and terminate the integration, as these stars have effectively left the binary SMBH sphere of influence in most galaxies. We label a star as ``ejected'' if it escapes with positive specific energy $\epsilon = \tilde{\epsilon}/(GM/\tilde{a}) > 0$, and will thus reach $r = \tilde{r}/\tilde{a} \rightarrow \infty$ if the binary SMBH is the only source of gravity. In a multipole expansion of the binary SMBH potential (written in its center-of-mass frame and evaluated at a field point at $r_{100}$), the dipole term vanishes and the ratio of the quadrupole to monopole terms is $\Phi_{\ell=2} / \Phi_{\ell=0} \sim (1 / r_{100})^2 = 10^{-4}$. The monopole term thus dominates, so the specific energy $\epsilon_{100}$ of the ejected stars at $r_{100}$ will be roughly conserved, and their velocities will be $v_\infty = \tilde{v}_\infty / v_\textrm{bin} \simeq \sqrt{2\epsilon_{100}}$ as $r \rightarrow \infty$.

\subsection{Ejection Probability}
\label{subsec:ejprob}

\begin{figure*}
\centering
\subfloat[]{\includegraphics[scale=0.42]{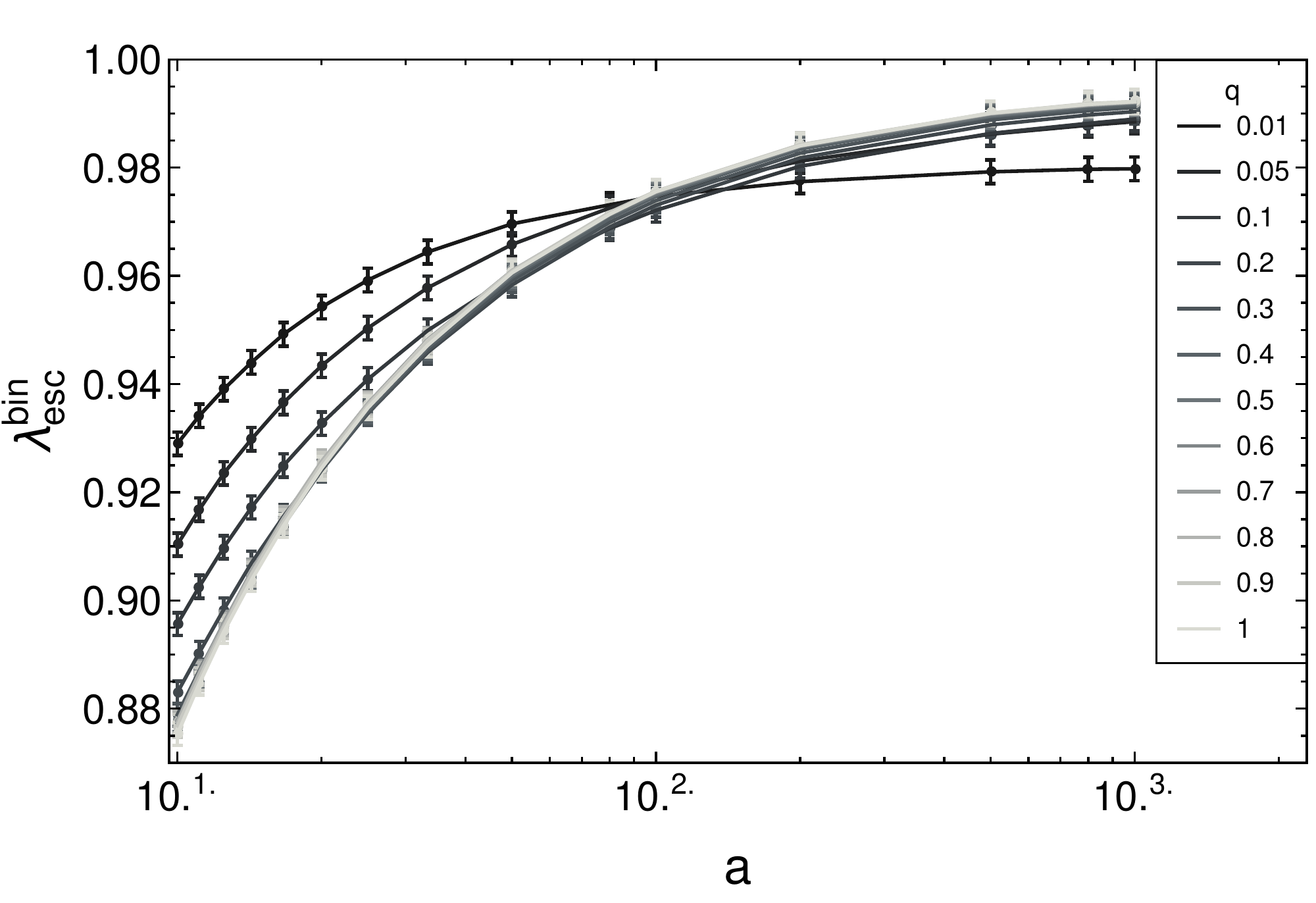}\label{fig:lambdaesca}}\hfill
\subfloat[]{\includegraphics[scale=0.42]{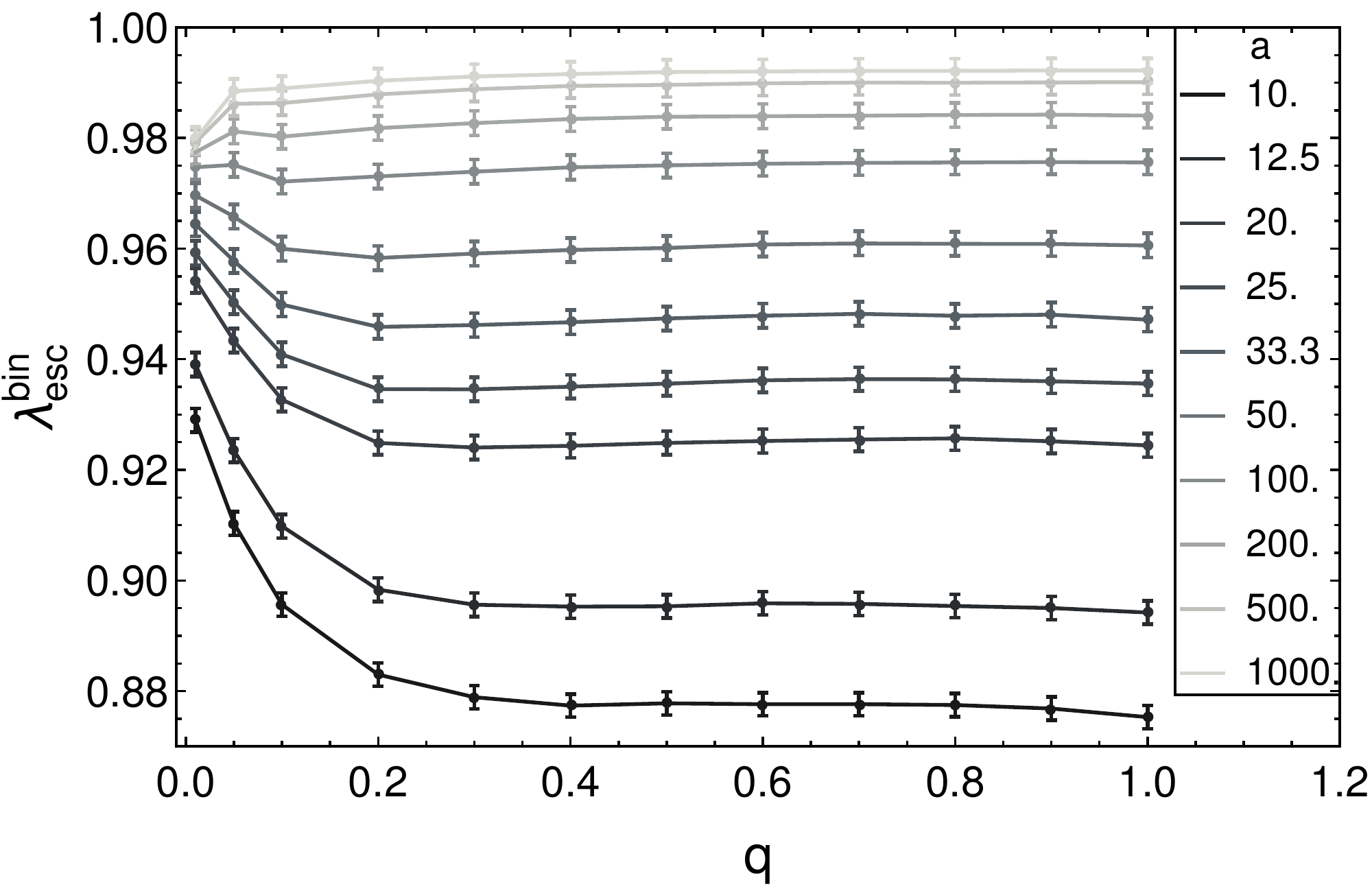}\label{fig:lambdaescb}}
\caption{The escape probability $\lambda_\textrm{esc} = N_\textrm{esc} / N_e$ for the SMBHB mechanism, plotted versus a) $a = \tilde{a}/\tilde{r}_{t1}$ and b) $q = M_2 / M_1$, where $N_\textrm{esc}$ ($N_e = 5 \times 10^6$) is the number of escapes (encounters), and where an ``escape'' occurs when a star crosses the sphere at $\tilde{r}/\tilde{a} = 100$. The error bars have half-widths $5\sigma$, where $\sigma = (\lambda_\textrm{esc} / N_e)^{1/2}$ are the standard errors assuming a Poisson distribution.}
\label{fig:lambdaesc}
\end{figure*}

Figure \ref{fig:lambdaesc} shows the escape probability $\lambda_\textrm{esc} = N_\textrm{esc} / N_e$, where $N_\textrm{esc}$ ($N_e = 5 \times 10^6$) is the number of escapes (encounters). Stars escape the simulation in most encounters ($\geq 86 \%$), and are disrupted ($\leq 12 \%$) or time-out ($\leq 2 \%$) for the rest. The escape probability is largely independent of $q = M_2 / M_1$ for $q \gtrsim 0.2$, and for a given $q$, it decreases as the binary SMBH contracts. At large separations, equal mass SMBH binaries cause more stars to escape, and at small separations the opposite trend holds, with a transition at $a = \tilde{a}/\tilde{r}_{t1} \simeq 100$ where $\lambda_\textrm{esc}$ is roughly independent of $q$ over the entire range we explored.

\begin{figure*}
\centering
\subfloat[]{\includegraphics[scale=0.42]{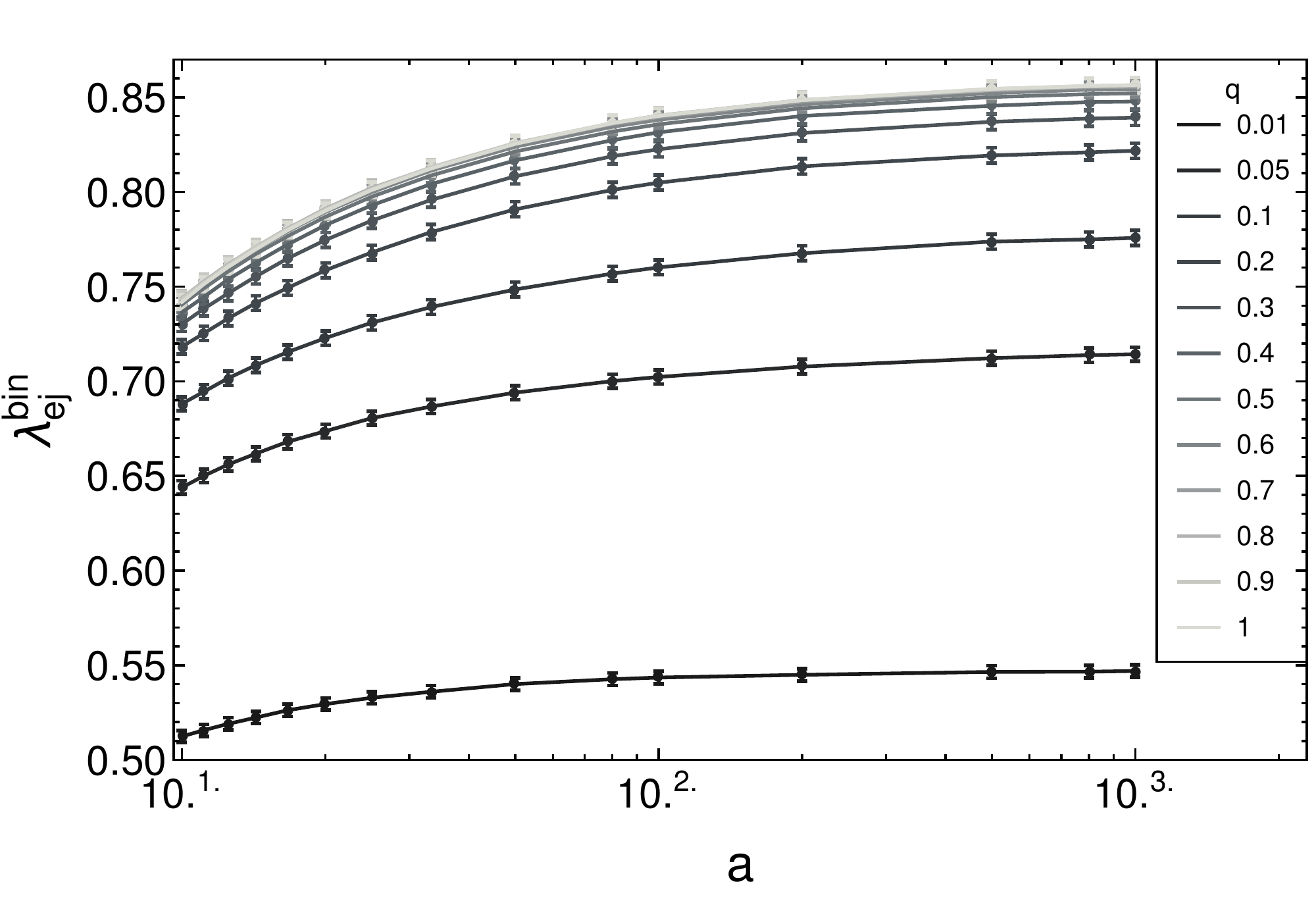}\label{fig:lambdaeja}}\hfill
\subfloat[]{\includegraphics[scale=0.42]{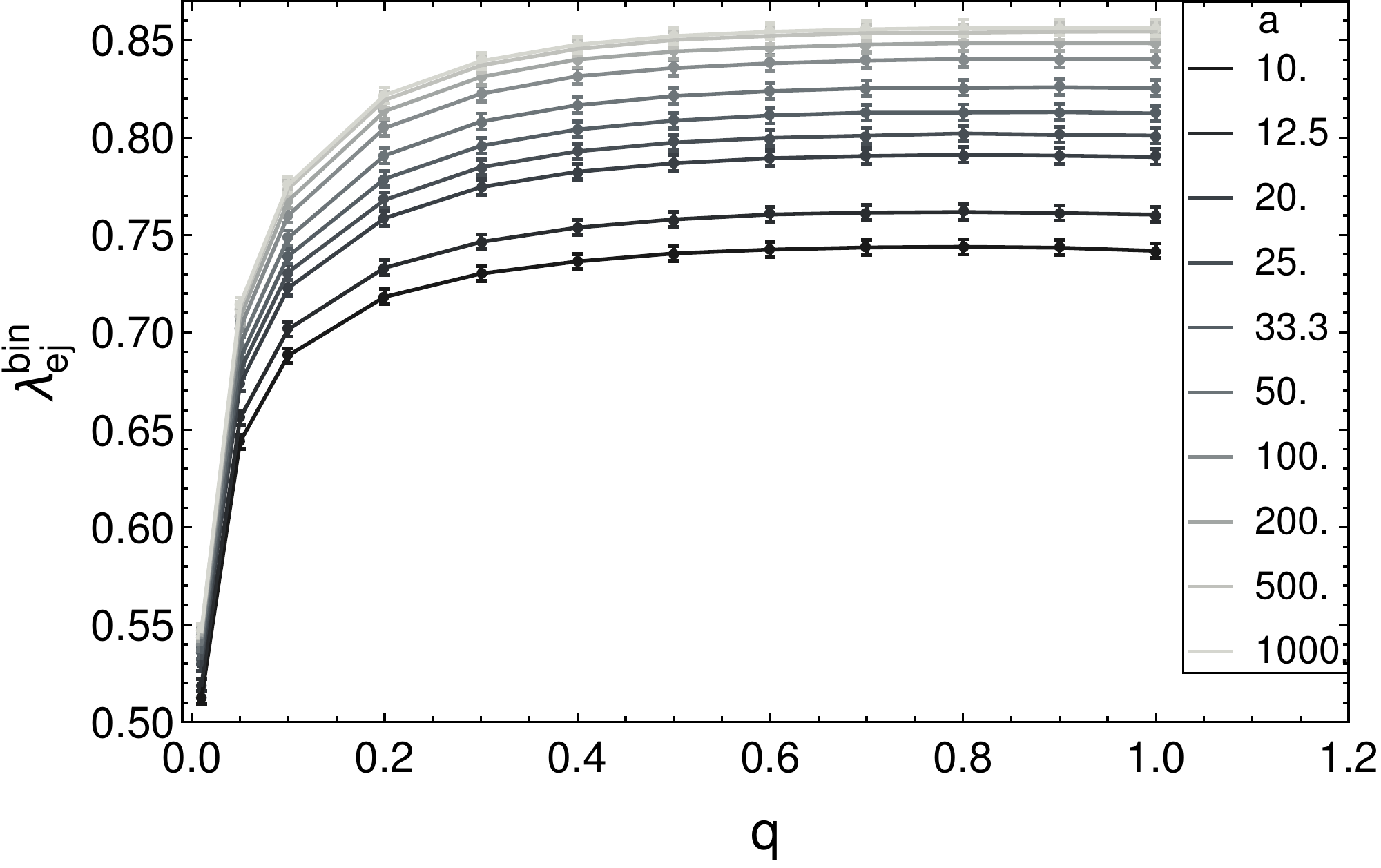}\label{fig:lambdaejb}}
\caption{The ejection probability $\lambda_\textrm{ej} = N_\textrm{ej} / N_e$ for the SMBHB mechanism, plotted versus a) $a = \tilde{a}/\tilde{r}_{t1}$ and b) $q = M_2/M_1$, where $N_\textrm{ej}$ ($N_e = 5 \times 10^6$) is the number of ejections (encounters), and where an ``ejection'' occurs when a star crosses the sphere at $\tilde{r}/\tilde{a} = 100$ with positive energy. The error bars have half-widths $10\sigma$, where $\sigma = (\lambda_\textrm{ej} / N_e)^{1/2}$ are the standard errors assuming a Poisson distribution.}
\label{fig:lambdaej}
\end{figure*}

Figure \ref{fig:lambdaej} shows the ejection probability $\lambda_\textrm{ej} = N_\textrm{ej} / N_e$ for the SMBHB mechanism, where $N_\textrm{ej}$ ($N_e = 5 \times 10^6$) is the number of ejections (encounters). The probability is a monotonically increasing function of both $a$ and $q$, and is roughly independent of $a$ for $a \gtrsim 100$ and of $q$ for $q \gtrsim 0.2$. The probability is lower than the escape probability, since a subset of stars that cross the escape sphere are on very weakly-bound elliptical orbits. This reduction is more pronounced for lower mass ratios, since a lower mass secondary tends to only lightly perturb the incident stars, shifting them from an initial parabolic orbit to a highly elliptical one.

The ejection probabilities we find exhibit a different trend from the mass ejection rates from circular SMBH binaries found by \citet{quinlan96} and \citet{sesana06}, whose rates decrease with increasing $\tilde{a}/\tilde{a}_h$ and whose curves for different $q$ crossover, but this likely arises due to a stricter definition of an ``ejection'' that they adopt while studying the hardening of the binary. We continue to label an ejection in this section as any stars that leave the escape sphere and are unbound to the binary SMBH, whereas the above authors further qualify them as all unbound stars with velocities $\tilde{v} > k\sigma$ for some constant $k$. We fold in information about the velocity of the ejected stars in Section \ref{subsec:properties}, and find a similar dependence as the above authors.

\subsection{Properties of Ejected Stars}
\label{subsec:properties}

We now turn to the properties of the ejected hypervelocity stars, namely their velocity (Section \ref{subsubsec:propertiesvel}) and angular (Section \ref{subsubsec:propertiesmu}) distributions. We rewrite our variables in the units $G = M_1 = \tilde{r}_{t1} = 1$, since these quantities are fixed for different SMBH binary mass ratios and separations. In particular, the velocities are normalized using the scale
\begin{equation}
\begin{split}
v_0 &= \sqrt{\frac{GM_1}{\tilde{r}_{t1}}} \\
&\simeq 4.4 \times 10^4 \textrm{ km} \cdot \textrm{s}^{-1} \left(\frac{M_1}{10^6 M_\odot}\right)^{1/3} \left(\frac{M_*}{1 M_\odot}\right)^{1/6} \left(\frac{R_*}{1 R_\odot}\right)^{-1/2}
\end{split}
\end{equation}
The characteristic circular velocity of a binary SMBH of total mass $M$ and semi-major axis $\tilde{a}$ is then simply $v_\textrm{bh} = \sqrt{GM/\tilde{a}} = v_0 \sqrt{(1+q)/a}$, where $a = \tilde{a}/\tilde{r}_{t1}$, or dimensionally
\begin{equation}
v_\textrm{bh} \simeq 2070 \textrm{ km} \cdot \textrm{s}^{-1} \left(\frac{M}{10^6 M_\odot}\right)^{1/2} \left(\frac{\tilde{a}}{1 \textrm{ mpc}}\right)^{-1/2}
\label{eq:vbh}
\end{equation}
Where necessary, we consider as our central example a primary with mass $M_1 = 10^6 M_\odot$ and stars with solar parameters, yielding $v_0 \simeq 4.4 \times 10^4$ km/s.

\subsubsection{Velocity Distribution}
\label{subsubsec:propertiesvel}

\begin{figure}
\centering
\includegraphics[scale=0.59]{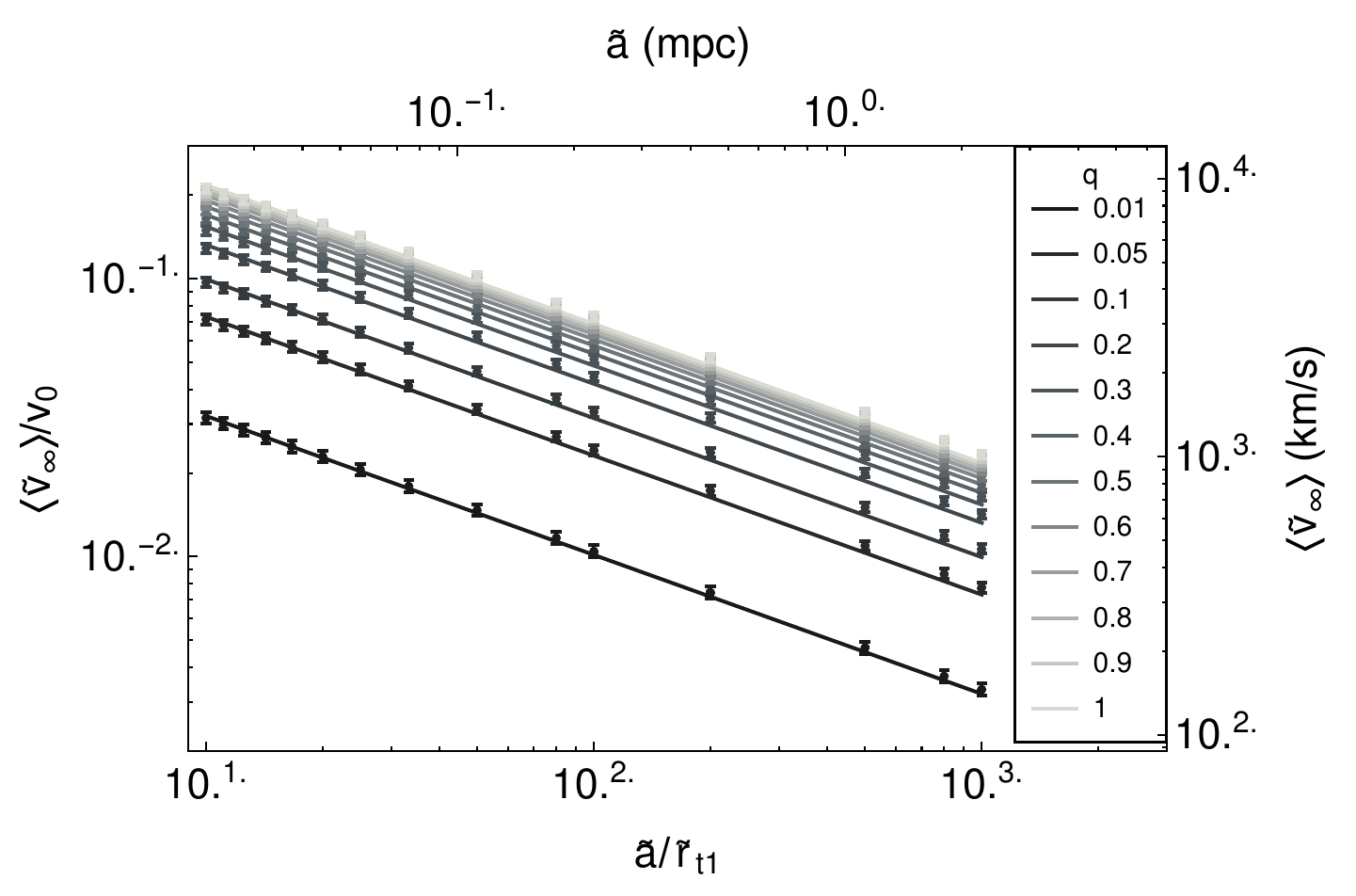}
\caption{
The mean velocity $\langle \tilde{v}_\infty \rangle$ of the ejected stars as a function of the binary SMBH separation $\tilde{a}$ and for several mass ratios $q = M_2 / M_1$. The error bars have half-widths $100\sigma_e$, where $\sigma_e = \sigma/\sqrt{N_\textrm{ej}}$ is the standard error in the mean and $\sigma$ is the standard deviation. The curves that fit the data are given by Eq. \ref{eq:velavg}. The dimensioned axes were calculated for $M_1 = 10^6 M_\odot$ and stars with solar parameters.
}
\label{fig:velvsa}
\end{figure}

Figure \ref{fig:velvsa} shows the mean velocity $\langle \tilde{v}_\infty \rangle$ of the ejected stars. The curves show the data points fit to the function (based on the work of \citealt{yu03}, discussed below)
\begin{equation}
\langle \tilde{v}_\infty \rangle \simeq \kappa v_0 \left(\frac{q}{1+q}\right)^{1/2} \left(\frac{\tilde{a}}{\tilde{r}_{t1}}\right)^{-1/2}
\label{eq:velavg}
\end{equation}
where $\kappa$ is a $q$-dependent dimensionless parameter. The parameter is in the range $0.97 \leq \kappa \leq 1.06$, and we find that it is well described by $\kappa \simeq \alpha e^{-\beta q} + \gamma$ for $q \gtrsim 0.05$, where $\alpha \simeq 0.10$, $\beta \simeq 2.5$, and $\gamma \simeq 0.96$. It decreases for lower $q$, as we find $\kappa \simeq 1.02$ for $q = 0.01$. The mean is thus set by the velocity of the reduced mass $\mu = M_1 M_2 / (M_1 + M_2)$ of the circular SMBH binary, $\langle \tilde{v}_\infty \rangle \sim (G \mu / \tilde{a})^{1/2}$, since an incident star effectively encounters this on average; the parameter $\kappa$ describes the mean deviation from this, which is on the order of a few percent.

\citet{yu03} approximate the mean velocity as $\langle \tilde{v}_\infty \rangle \simeq \sqrt{2 \langle \tilde{\epsilon}_\infty \rangle}$, where they express the mean specific energy of the ejected stars as $\langle \tilde{\epsilon}_\infty \rangle \simeq K G \mu / \tilde{a}$ for a dimensionless constant $K$ that parameterizes the energy the stars extract from the binary SMBH \citep{yu02}. \citet{yu03} find a constant $\kappa \simeq \sqrt{3.2} \simeq 1.79$ (from $K \simeq 1.6$) using the typical hardening rate $H \simeq 16$ found by \citet{quinlan96}. If we use this definition of $\langle \tilde{v}_\infty \rangle$ instead and fit our data to Eq. \ref{eq:velavg}, then we find $1.19 \leq \kappa \leq 1.33$, lower than that used by \citet{yu03}, and the same fitting curve for $q \gtrsim 0.05$ with $\alpha \simeq 0.15$, $\beta \simeq 3.4$, and $\gamma \simeq 1.2$.

\begin{figure}
\centering
\includegraphics[scale=0.58]{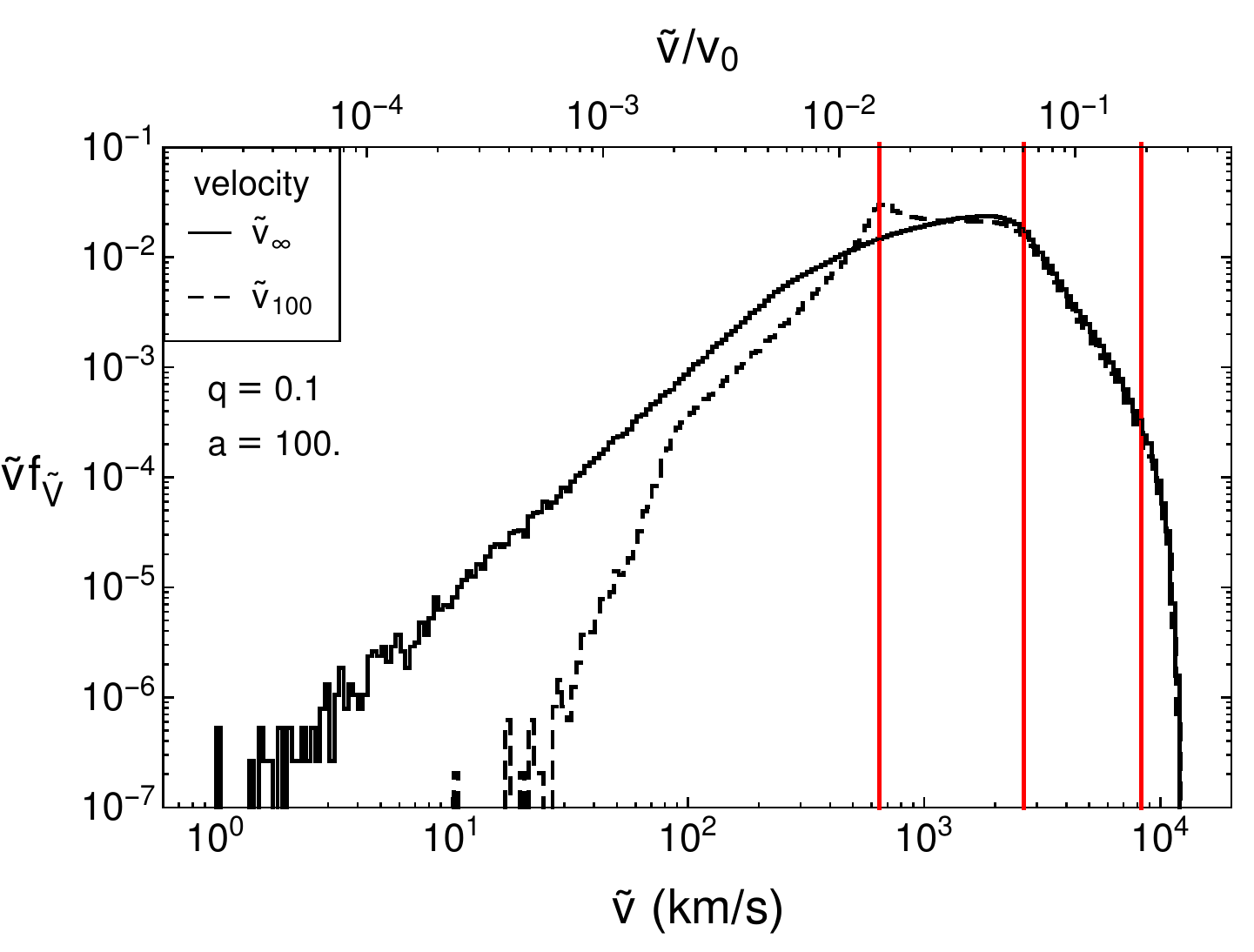}
\caption{The probabilities $\tilde{v}f_{\tilde{V}}$ for $q = 0.1$ and $a = \tilde{a}/\tilde{r}_{t1} = 100$, for the velocity $\tilde{v}_\infty$ of ejected stars at $\tilde{r}/\tilde{a} \rightarrow \infty$ (solid) and the velocity $\tilde{v}_{100}$ of escaped stars at $\tilde{r}/\tilde{a} = 100$ (dashed), where the PDF is $f_{\tilde{V}} \equiv f_{\tilde{V}}(\tilde{v})$. The logarithmic bin widths are $\Delta_{\tilde{v}} = 0.02$. From left to right, the red vertical lines show $\tilde{v}/v_\textrm{bh} = 1/\sqrt{50}$ (the escape velocity at $\tilde{r}/\tilde{a} = 100$); $v_{\textrm{pl},1}/v_0 \simeq k_1 (q/(1+q))^{1/2} (\tilde{a}/\tilde{r}_{t1})^{-1/2}$ where $1 \lesssim k_1 \lesssim 2$ (slightly above the peak, the start of the decreasing power law region); and $v_{\textrm{pl},2}/v_0 \simeq k_2 (1/(1+q))^{1/2} (\tilde{a}/\tilde{r}_{t1})^{-1/2}$ where $1 \lesssim k_2 \lesssim 2$ (the end of the decreasing power law region). Here, $k_1 = k_2 = 2$.}
\label{fig:histvinfescvesc}
\end{figure}

Figure \ref{fig:histvinfescvesc} shows the histograms of the probabilities $\tilde{v}f_{\tilde{V}}$ for the velocity of escaped (at $\tilde{r}/\tilde{a} = 100$) and ejected (at $\tilde{r}/\tilde{a} \rightarrow \infty$) stars, where the probability density function (PDF) is $f_{\tilde{V}} \equiv f_{\tilde{V}}(\tilde{v})$, for $q = 0.1$ and $a = 100$. The distributions incorporate the presence of the binary SMBH potential only. The histogram for $\tilde{v}_{100}$ has a local maximum at $\tilde{v}/v_\textrm{bh} = 1/\sqrt{50}$ (the escape velocity at $\tilde{r}/\tilde{a} = 100$), which is the value marked by the leftmost red vertical line; all stars with velocities to the right of this line are unbound to the binary SMBH and contribute to the distribution for $\tilde{v}_\infty$. For $\tilde{v} \gtrsim 1000$ km/s, the escaped and ejected histograms have identical shapes since the stellar kinetic energies are much larger than the potential energies. The two red vertical lines on the right give the range over which the histograms decrease by a power law $p = \gamma \tilde{v}^\beta$. The middle one shows $v_{\textrm{pl},1}/v_0 \simeq k_1 (q/(1+q))^{1/2} (\tilde{a}/\tilde{r}_{t1})^{-1/2}$, where $k_1$ is a constant in the range $1 \lesssim k_1 \lesssim 2$; this is simply an order-unity multiple of the velocity of the reduced mass, $\tilde{v} = (G \mu / \tilde{a})^{1/2}$, as it occurs slightly above the peak, which roughly coincides with the mean. The right one shows $v_{\textrm{pl},2}/v_\textrm{bh} \simeq k_2 (1/(1+q))$ (or equivalently, $v_{\textrm{pl},2}/v_0 \simeq k_2 (1/(1+q))^{1/2} (\tilde{a}/\tilde{r}_{t1})^{-1/2}$), where $k_2$ is a constant in the range $1 \lesssim k_2 \lesssim 2$; this is simply an order-unity multiple of the velocity of the secondary, $\tilde{v}/v_\textrm{bh} = 1/(1+q)$, which is the largest velocity scale in the system. These features and the locations quoted here apply over our entire parameter range, as presented below.

\begin{figure*}
\centering
\subfloat[]{\includegraphics[scale=0.58]{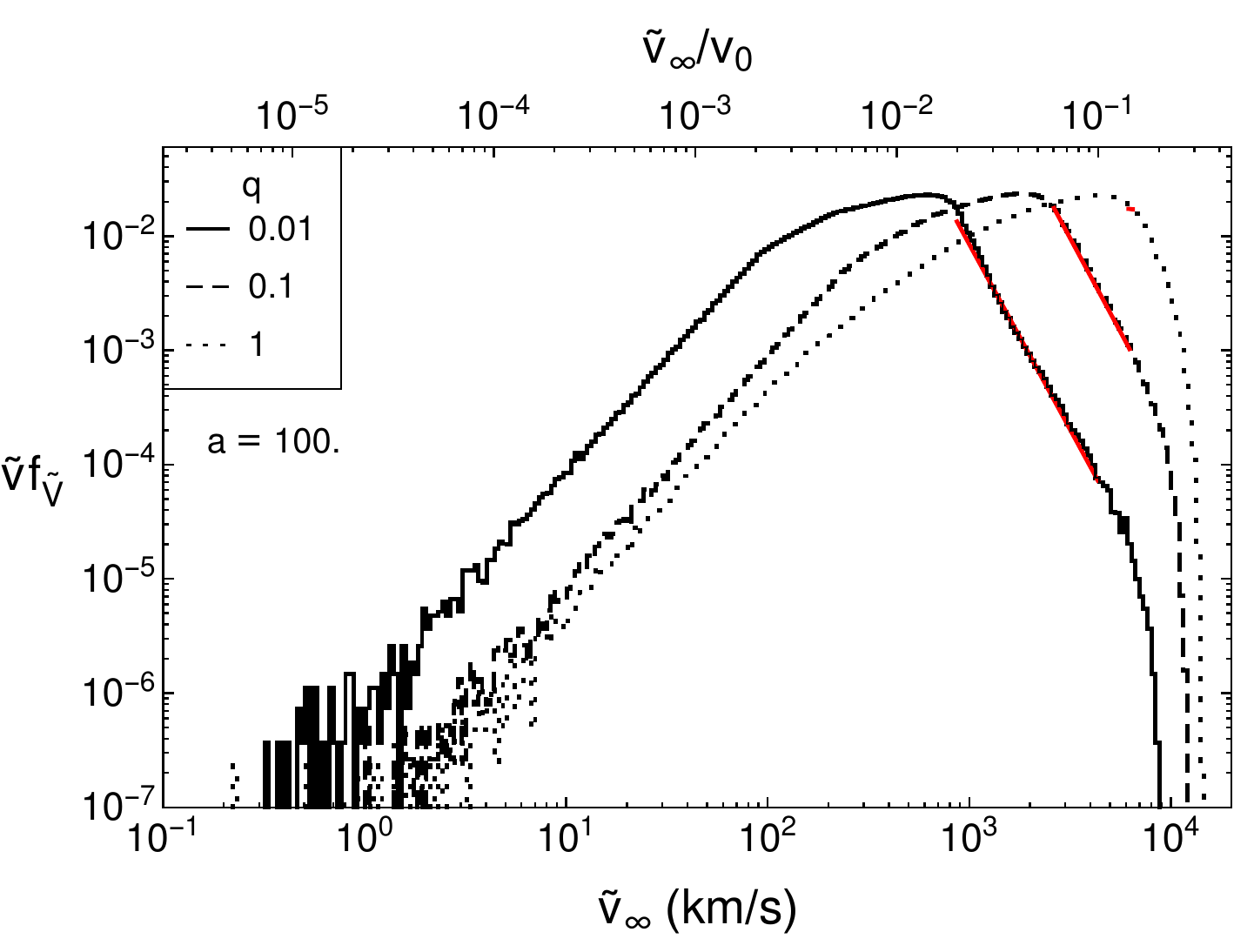}\label{fig:histvinfesca}}\hfill
\subfloat[]{\includegraphics[scale=0.58]{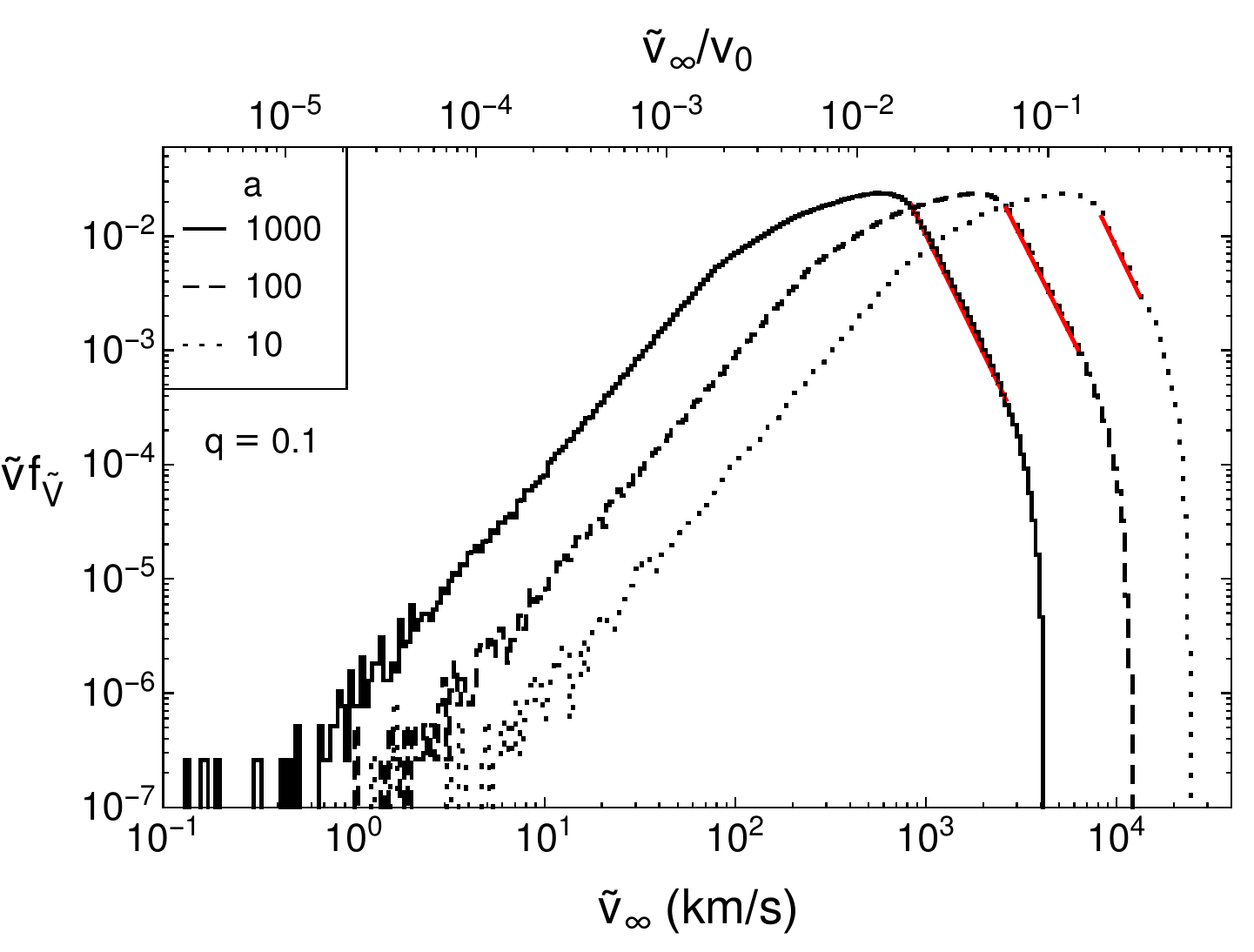}\label{fig:histvinfescb}}
\caption{a) The probabilities $\tilde{v}f_{\tilde{V}}$ for the velocity $\tilde{v}_\infty$ of ejected stars at $\tilde{r}/\tilde{a} \rightarrow \infty$, where the PDF is $f_{\tilde{V}} \equiv f_{\tilde{V}}(\tilde{v})$, for $a = \tilde{a}/\tilde{r}_{t1} = 100$ and $q = 0.01$ (solid), $0.1$ (dashed), and $1$ (dotted). b) The probabilities for $q = 0.1$ and $a = \tilde{a}/\tilde{r}_{t1} = 1000$ (solid), $100$ (dashed), and $10$ (dotted). The logarithmic bin widths are $\Delta_{\tilde{v}} = 0.02$. The red curves show the power law $p = \gamma \tilde{v}_\infty^\alpha$ fit to the regions $k_1 (q/(1+q))^{1/2} (\tilde{a}/\tilde{r}_{t1})^{-1/2} \lesssim \tilde{v}_\infty/v_0 \lesssim k_2 (1/(1+q))^{1/2} (\tilde{a}/\tilde{r}_{t1})^{-1/2}$, where $1 \lesssim k_1, k_2 \lesssim 2$.}
\label{fig:histvinfesc}
\end{figure*}

Figure \ref{fig:histvinfesca} shows the probabilities for the velocity $\tilde{v}_\infty$ of ejected stars (at $\tilde{r}/\tilde{a} \rightarrow \infty$ in the presence of the binary SMBH potential only) for $a = 100$. The peak shifts by about an order of magnitude with a two orders of magnitude change in the mass ratio. The shapes of the distributions depend on the mass ratio. For $q=1$, there is an abrupt drop after the peak of the distribution; the power law region becomes vanishingly small (since $v_{\textrm{pl},1} \simeq v_{\textrm{pl},2}$), and the peak is close to the maximum velocity and considerably higher than the mean in Figure \ref{fig:velvsa}. As $q$ decreases, there is a more gradual decline from the peak to the maximum velocity; the power law region is larger and the peak is closer to the mean. This trend occurs because lower mass ratio SMBH binaries have lower reduced mass velocities and thus impart less energy to the ejected stars on average, but they have higher secondary velocities and can thus still eject stars to high velocities after close encounters. The red curves show $p = \gamma \tilde{v}_\infty^\beta$ fit to the power law region defined above; the specific values of $\gamma$ and $\beta$ are not particularly illuminating, so we do not present them.

Figure \ref{fig:histvinfescb} shows the probability distributions for $q = 0.1$. The peak again shifts by about an order of magnitude with a two order of magnitude change in the separation. The histograms have similar shapes, merely shifted along the $v$-axis for different $a$. Indeed, the histograms coincide when plotted in the units $\tilde{v}_\infty/v_\textrm{bh}$, apart from a slightly lower maximum velocity for $a = 10$. The distributions for the other values of $q$ exhibit this behavior as well.

The similarity in the histograms for different $a$ is a result of the scale invariance with respect to $a$ in our simulation setup and the small tidal disruption probability over our parameter range. The small deviation for $a = 10$ arises because the tidal disruption probability is higher at this separation \citep{darbha18}. Stars incident with low angular momenta are both preferentially disrupted \citep{coughlin17} and preferentially ejected with high velocities since they experience close encounters with the binary SMBH. Since an integration terminates when a star is disrupted, the higher disruption probability implies that fewer low-angular momentum stars are available to undergo high velocity ejections. This deviation is thus physical, and we expect to observe it in realistic velocity spectra.

The histograms in Figure \ref{fig:histvinfesc} are roughly in agreement with the velocity PDFs found by \citet{sesana06} for circular SMBH binaries. In both works, the distributions have the same velocity range and a region of velocities after the peak that are well described by a decreasing power law, and this region is larger for lower mass ratios. However, \citet{sesana06} suggest that the lower bound of the power law region is set by the velocity dispersion of the field stars, which should not be the case, and they use a more complicated fitting function. In addition, the PDFs they find have more kinks above the peak.

\begin{figure}
\centering
\includegraphics[scale=0.58]{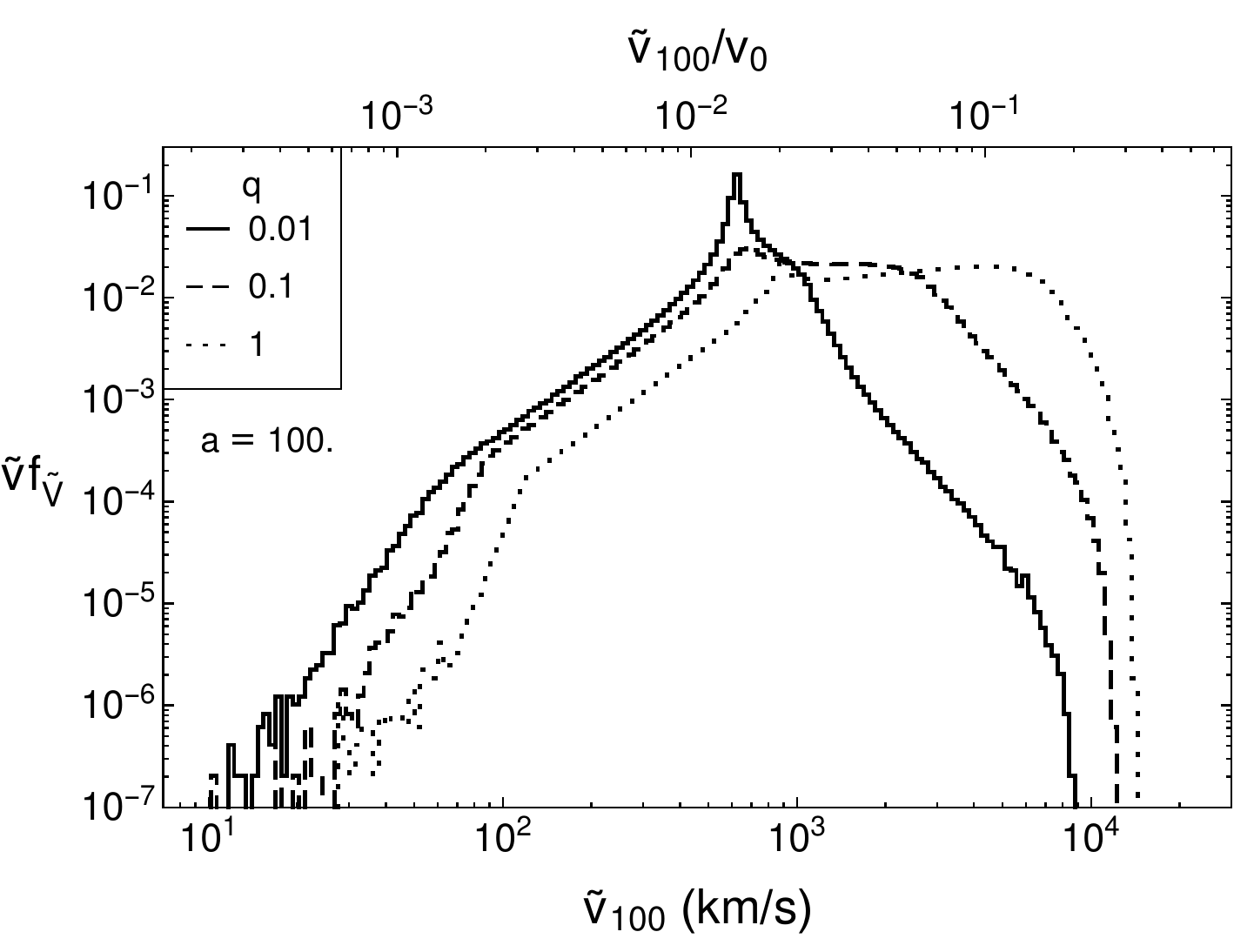}
\caption{The probabilities $\tilde{v}f_{\tilde{V}}$ for the velocity $\tilde{v}_{100}$ of escaped stars at $\tilde{r}/\tilde{a} = 100$, where the PDF is $f_{\tilde{V}} \equiv f_{\tilde{V}}(\tilde{v})$, for $a = \tilde{a}/\tilde{r}_{t1} = 100$ and $q = 0.01$ (solid), $0.1$ (dashed), and $1$ (dotted). The stars with velocities $\tilde{v}_{100} \lesssim v_0 \sqrt{(1+q)/50a}$ are bound to the binary SMBH and will not be ejected to infinity; this velocity corresponds to the position of the lower velocity peak. The logarithmic bin widths are $\Delta_{\tilde{v}} = 0.02$.}
\label{fig:histvesc}
\end{figure}

Figure \ref{fig:histvesc} shows the probability distributions for the velocity $\tilde{v}_{100}$ of escaped stars (at $\tilde{r}/\tilde{a} = 100$ in the presence of the binary SMBH potential only) for $a = 100$. This radius for the escape sphere is less than or roughly equal to the influence radius of the binary SMBH and much less than any galactic length scales, and so these distributions hold regardless of the galactic potential or stellar distribution in the bulge. The shapes of the histograms change dramatically for different $q$, much more than those for $\tilde{v}_\infty$. For $q=1$, the distribution is roughly flat for an order of magnitude change in $\tilde{v}_{100}$, and exhibits a double peaked structure at the two ends of this plateau, which correspond to the two red lines on the left in Figure \ref{fig:histvinfescvesc}, and has a vanishingly small power law region after the higher velocity peak. As $q$ decreases, the lower velocity peak becomes more prominent, the plateau shrinks, and the power law region grows. For $q=0.01$, the lower velocity peak dominates since the low-mass secondary tends to only lightly perturb the incident stars from their $\epsilon=0$ orbits into mildly bound or unbound orbits, which results in the stellar velocities at $\tilde{r}/\tilde{a} = 100$ clustering around the escape velocity. For a given $q$, the histogram has the same shape for different $a$, simply shifted along the $\tilde{v}$-axis as in Figure \ref{fig:histvinfescb}.

Figure \ref{fig:pvelcutoffs} shows the probability that an ejected star has a velocity at $\tilde{r}/\tilde{a} \rightarrow \infty$ of $\tilde{v}_\infty > v_c$, where $v_c$ is a cutoff velocity, for $q = 0.1$. The curves are very similar, simply shifted along the $v_c$-axis, due to the similarity in the underlying distributions (Figure \ref{fig:histvinfesc}). If we consider a cutoff velocity like $v_c \sim 1000$ km/s, then we see similar trends as in the mass ejection rates (defined for $\tilde{v} > k\sigma$ for some $k$) found by \citet{quinlan96} and \citet{sesana06}, albeit with slightly different values. Figure \ref{fig:velcutoffsappendix} shows the probabilities for different values of $q$ in our parameter space.

\subsubsection{Angular Distribution}
\label{subsubsec:propertiesmu}

We parametrize the angle of the ejected stars with the direction cosine $\mu = \cos\theta = v_{z,\infty}/v_\infty$, where $\theta$ is the polar angle measured from the direction normal to the binary SMBH orbital plane. For an isotropic distribution of ejection directions, $\mu$ has a uniform PDF $f_M = 1/2$ for $-1 \leq \mu \leq 1$; note that in this case $\theta$ has a PDF $f_\Theta = (\sin\theta) / 2$ for $0 \leq \theta \leq \pi$. The ejected stars are uniformly distributed in the azimuthal angle $\phi$ when ejected by a circular binary with a randomized phase, so we ignore this distribution. We calculate the angular distributions subject to different velocity cutoffs $v_c$.

\begin{figure*}
\centering
\subfloat[]{\includegraphics[scale=0.43]{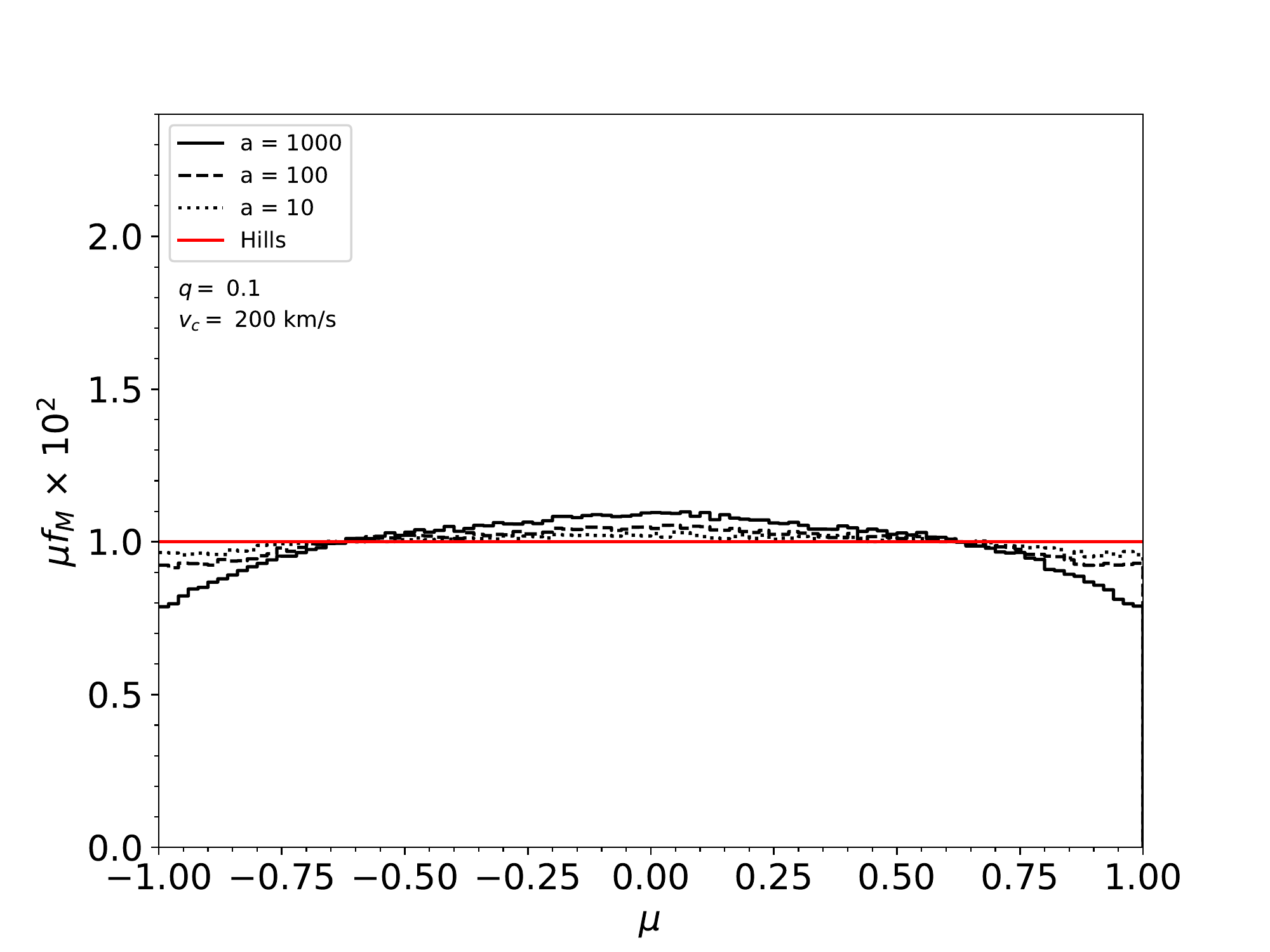}\label{fig:histmua}}\hfill
\subfloat[]{\includegraphics[scale=0.43]{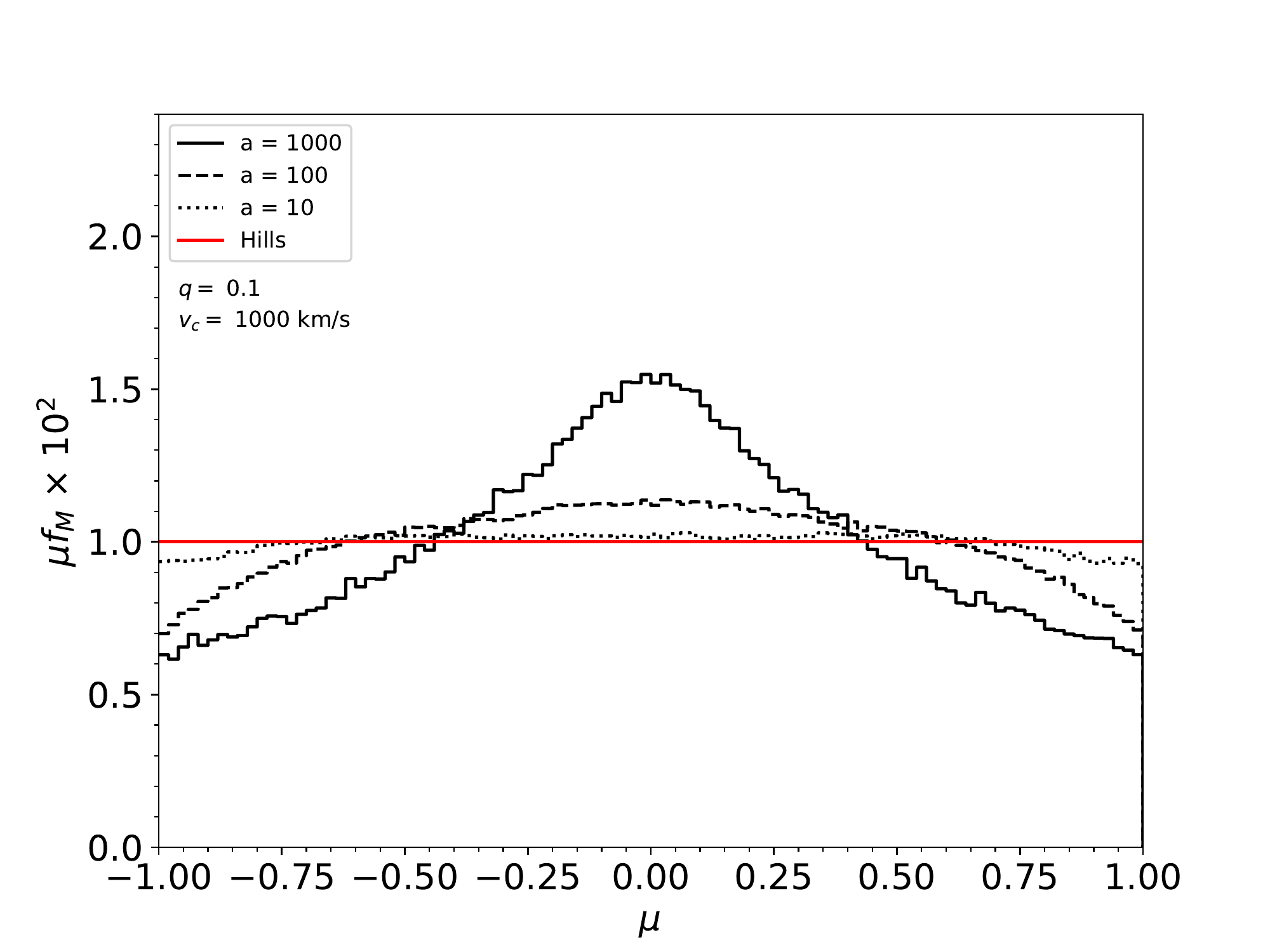}\label{fig:histmub}}
\caption{
The probabilities $\mu f_M$ for the direction cosine $\mu = \cos\theta = v_{z,\infty}/v_\infty$ of ejected stars at $\tilde{r}/\tilde{a} \rightarrow \infty$, for $q = 0.1$ and $a = \tilde{a}/\tilde{r}_{t1} = 1000$ (solid), $100$ (dashed), and $10$ (dotted). The PDF is $f_M \equiv f_M(\mu)$. The variable $\theta$ is the polar angle measured from the direction normal to the binary SMBH orbital plane. The panels show the distributions after applying the velocity cutoffs $\tilde{V} > v_c$, where $v_c$ is a) 200 km/s, and b) 1000 km/s. The linear bin widths are $\Delta_\mu = 0.02$. The red curve shows the uniform distribution for $\mu$ produced by the Hills mechanism.
}
\label{fig:histmu}
\end{figure*}

Figure \ref{fig:histmu} shows the histograms of the probabilities $\mu f_M$, where $f_M \equiv f_M(\mu)$ is the angular PDF, for $q = 0.1$ and two different velocity cutoffs; Figure \ref{fig:histmuappendix} shows this for different values of $q$. If we examine a fixed cutoff $v_c \gtrsim 200$ km/s, then we see a clear trend: for a fixed mass ratio, tighter circular SMBH binaries eject stars more isotropically, whereas wider ones eject them more in the binary SMBH orbital plane; similarly, for a fixed separation, unequal-mass circular binaries (i.e. IMBH/SMBH binaries) eject stars more in the orbital plane, whereas nearly equal-mass ones eject them more isotropically. \citet{sesana06} find the same trend that we observe for fixed $q$ and changing $a$, but the opposite one for fixed $a$ and changing $q$. This difference arises due to the difference in our setups and initial conditions as described previously (Sections \ref{sec:setup} and \ref{subsec:ejprob}).

\begin{figure*}
\centering
\subfloat[]{\includegraphics[scale=0.58]{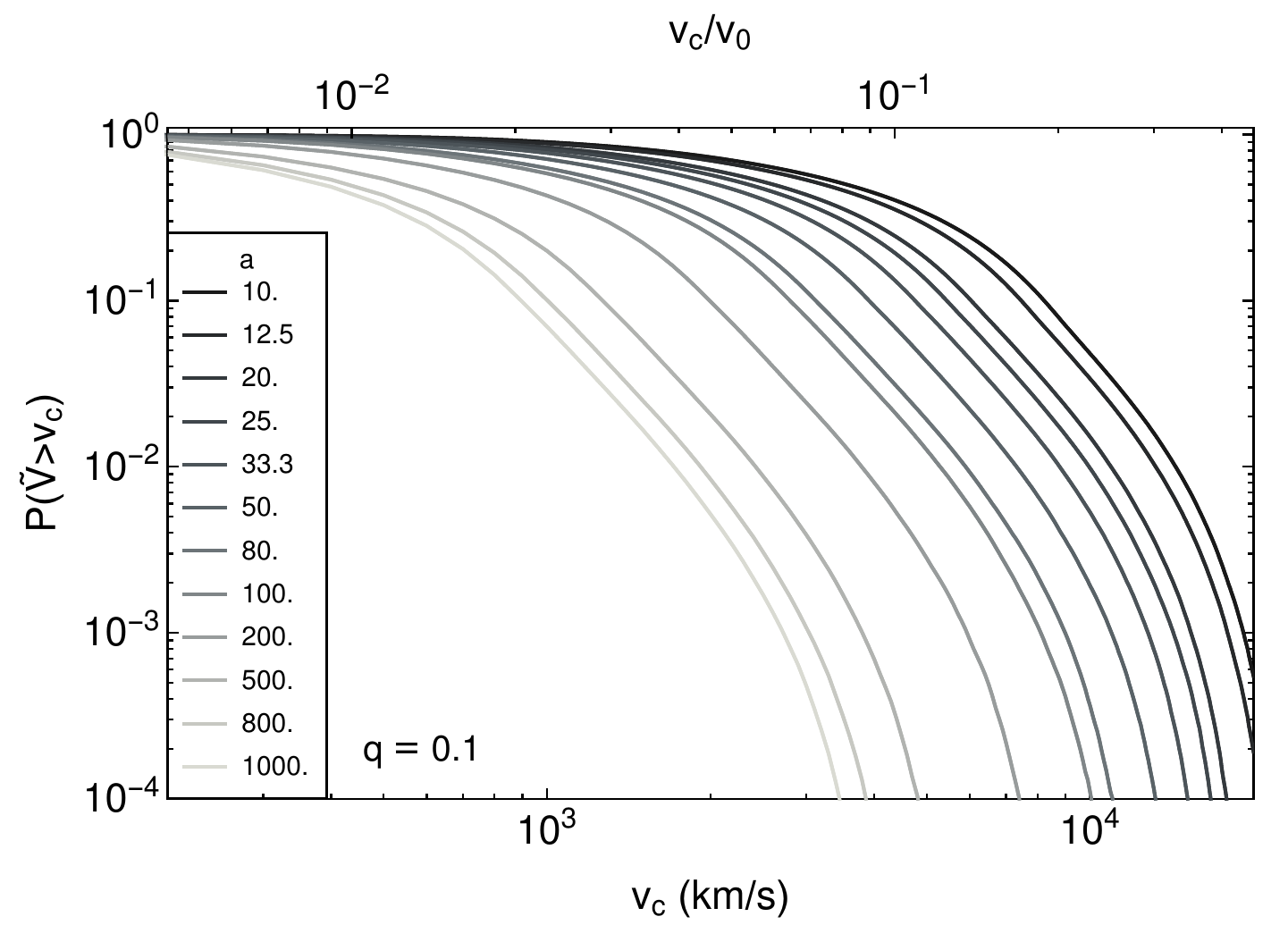}\label{fig:pvelcutoffs}}\hfill
\subfloat[]{\includegraphics[scale=0.58]{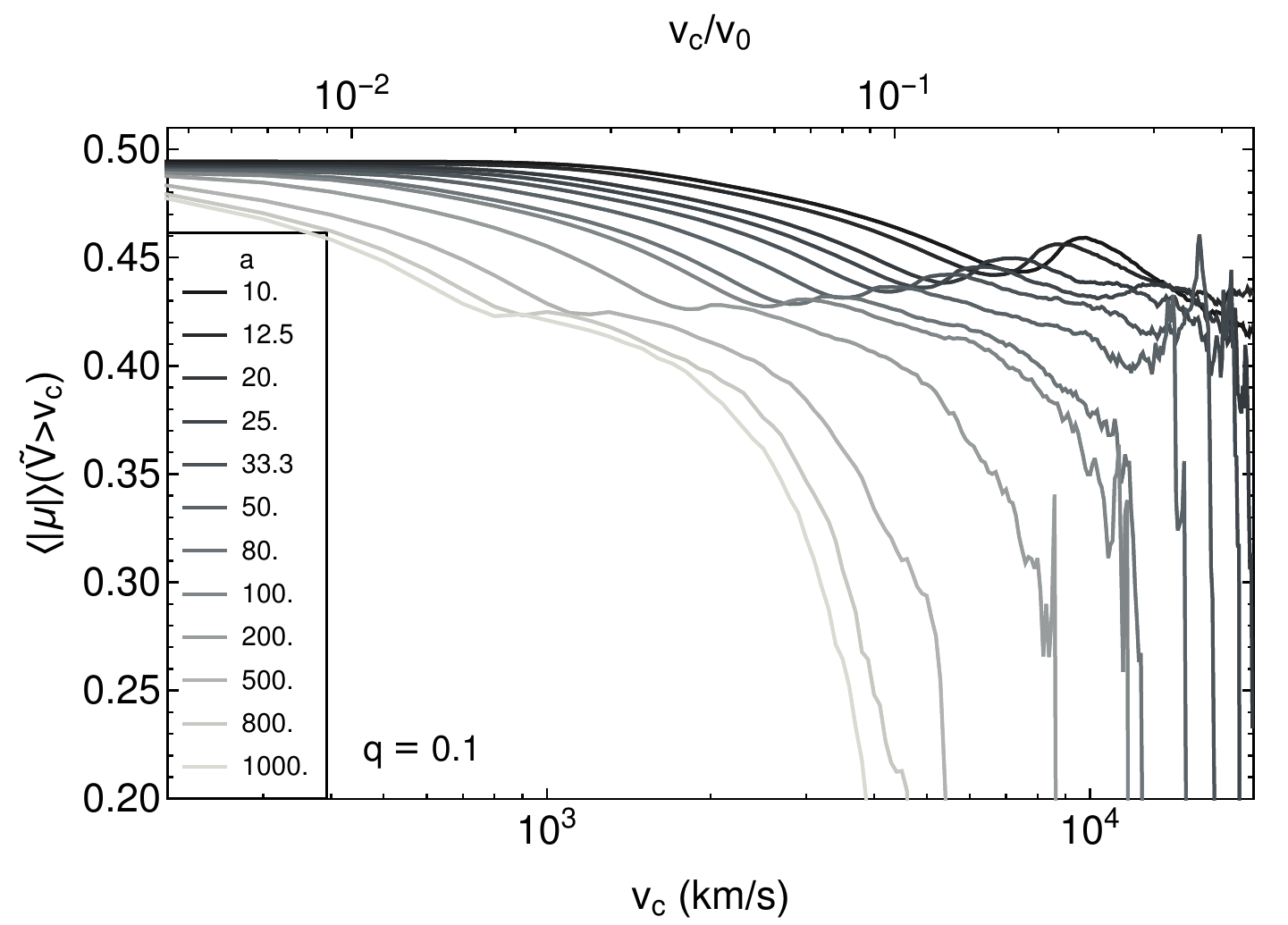}\label{fig:muvelcutoffs}}
\caption{Statistics of ejected stars with a range of velocity cutoffs $v_c$, for $q = 0.1$. a) The probability $P(\tilde{V} > v_c)$ that an ejected star has a velocity at $\tilde{r}/\tilde{a} \rightarrow \infty$ of $\tilde{v}_\infty > v_c$. b) The orientation of ejected stars with velocities $\tilde{v}_\infty > v_c$. The orientation is parameterized by $\mu = \cos\theta = v_{z,\infty} / v_\infty$, where $\theta$ is the polar angle measured from the direction normal to the binary SMBH orbital plane. 
For an isotropic distribution, $\langle \left| \mu \right| \rangle = 0.5$.}
\label{fig:velcutoffs}
\end{figure*}

Figure \ref{fig:muvelcutoffs} shows the normed mean direction of the ejected stars with velocity $\tilde{v}_\infty > v_c$ for $q = 0.1$; Figure \ref{fig:velcutoffsappendix} expands this to the other $q$ in our parameter space. For a sufficiently high cutoff, a binary SMBH preferentially emits stars near its orbital plane, with higher cutoffs needed for tighter binaries and higher mass ratios. However, this trend is not quite monotonic with velocity for a fixed $q$ and $a$. The curves exhibit a small bump towards more isotropic emission in a region around $v_{\textrm{pl},1}/v_0 \simeq k_1 (q/(1+q))^{1/2} (\tilde{a}/\tilde{r}_{t1})^{-1/2}$, shown by the middle red vertical line in Figure \ref{fig:histvinfescvesc}, where the size of the region correlates with the size of the power law region of the underlying velocity distribution. This suggests that the stars with velocities just after the peaks are preferentially emitted near the orbital plane. There is also a spike at high velocities, though this is likely due to noise from the low statistics there. These two features can in principle be used to constrain the binary SMBH mass ratio and separation.

\subsection{Time-Dependent Ejection Rate}
\label{subsec:inspiral}

In this subsection, we present the time-dependent ejection rate of hypervelocity stars by a binary SMBH contracting due to stellar scattering and gravitational wave emission. We examined the time-dependent tidal disruption rate in \citet{darbha18}; for a more detailed discussion of the dynamics of the inspiral, we refer the reader there (or to earlier work by \citealt{zier01}). Here, we briefly summarize our setup and modify it to describe HVSs.

We use the units $G = M_1 = \tilde{r}_{t1} = 1$ to write the dimensionless binary SMBH separation $a$ and the stellar specific energy $\epsilon$ and angular momentum $\ell$. We define the timescale
\begin{equation}
t_0 = \frac{\tilde{r}^4_{t1} c^5}{G^3 M^3_1} \simeq 7.7 \times 10^{-1} \textrm{ y} \left(\frac{M_1}{10^6 M_\odot}\right)^{-5/3} \left(\frac{M_*}{M_\odot}\right)^{-4/3} \left(\frac{R_*}{R_\odot}\right)^4
\end{equation}
to define the dimensionless time $t = \tilde{t}/t_0$. We consider a primary mass $M_1 = 10^6 M_\odot$ and stars with solar parameters in what follows.

In these units, the total inspiral rate is
\begin{equation}
\frac{da}{dt} = \left(\frac{da}{dt}\right)_\textrm{ss} + \left(\frac{da}{dt}\right)_\textrm{gw}
\label{eq:dadt}
\end{equation}
The rate of coalescence due to gravitational wave (gw) emission by two point particles on a circular orbit is \citep{peters64}
\begin{equation}
\left(\frac{da}{dt}\right)_\textrm{gw} = -\frac{64}{5} \frac{q(1+q)}{a^3}
\label{eq:dadtgw}
\end{equation}
The inspiral rate due to stellar scattering (ss) is
\begin{equation}
\left(\frac{da}{dt}\right)_\textrm{ss} = -\frac{2M_*}{M_1} \phi t_0 \frac{a^2}{q} \langle \Delta \epsilon_* \rangle (q,a) \lambda_\textrm{ej}(q,a)
\label{eq:dadtss}
\end{equation}
where $\langle \Delta \epsilon_* \rangle$ is the average change in the specific energy of the ejected stars and $\lambda_\textrm{ej}$ is the stellar ejection probability, both found from our simulations; $M_* / M_1$ is the ratio of the stellar mass to the primary mass; and $\phi$ is the stellar injection rate from the binary SMBH loss cone (in units of y$^{-1}$). This is simply $\phi(q,a) = 2a (1+q) \phi_0$, assuming the stars are incident from a full loss cone repopulated through two-body relaxation and are thus uniformly distributed in $\ell^2$ \citep{frank76,lightman77,cohn78,magorrian99}, where $\phi_0 \sim 10^{-4}$ yr$^{-1}$ is the fiducial rate for a SMBH of mass $M_\bullet = 10^6 M_\odot$ \citep{magorrian99,wang04,stone16}. The time-dependent ejection rate for stars with $\tilde{v}_\infty > v_c$ is then $\dot{n}_\textrm{ej}(\tilde{v}_\infty > v_c) = \phi \lambda_\textrm{ej}(\tilde{v}_\infty > v_c)$, where $\lambda_\textrm{ej}(\tilde{v}_\infty > v_c)$ depends on $q$ and $a$. We calculate $a(t)$ and $\dot{n}_\textrm{ej}$ as the binary SMBH contracts from $a = 1000$ ($\tilde{a} = 2.3$ mpc) to $10$ ($\tilde{a} = 0.023$ mpc).

\begin{figure}
\centering
\subfloat[]{\includegraphics[scale=0.58]{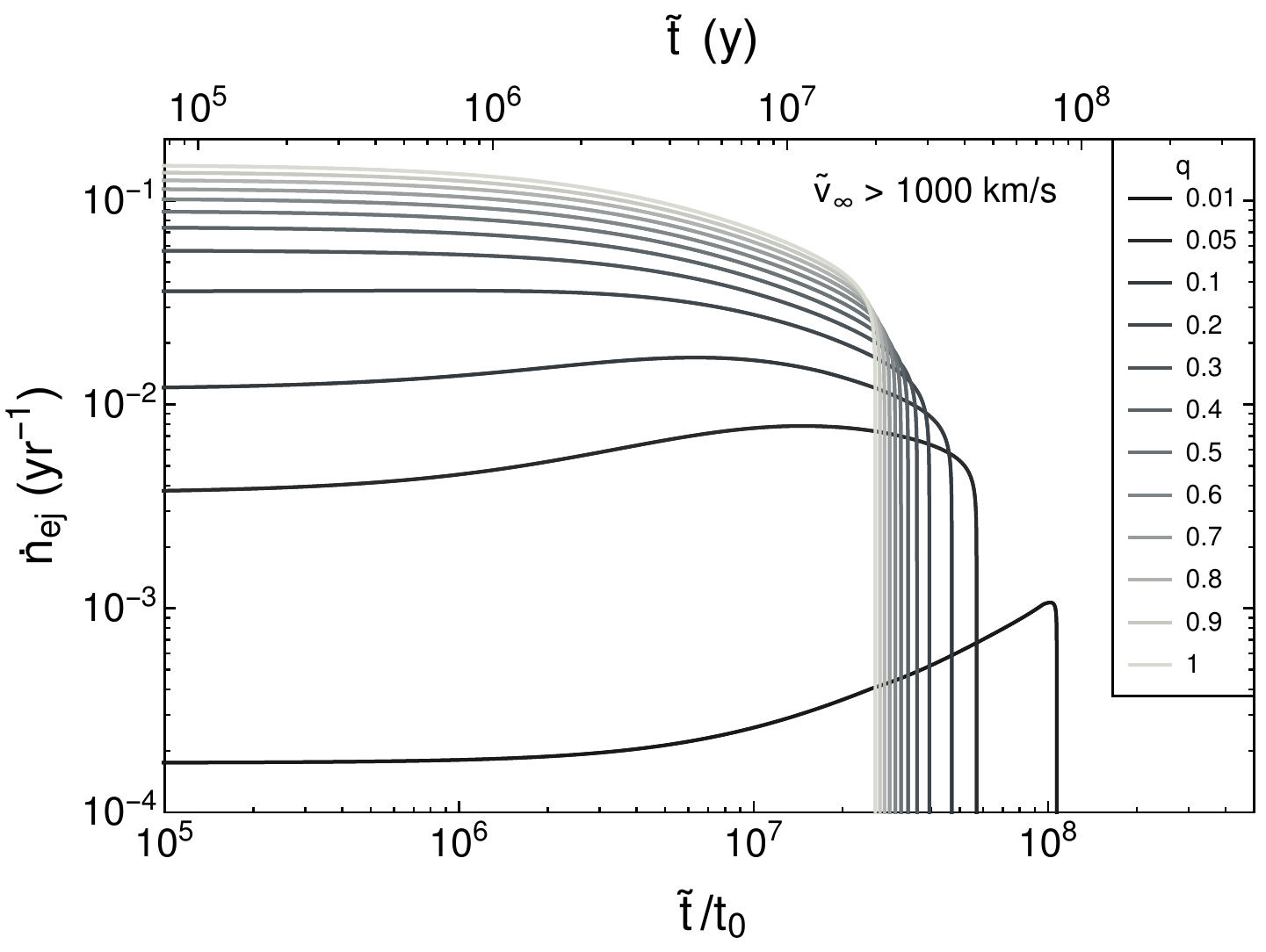}\label{fig:dndta}}\\
\subfloat[]{\includegraphics[scale=0.58]{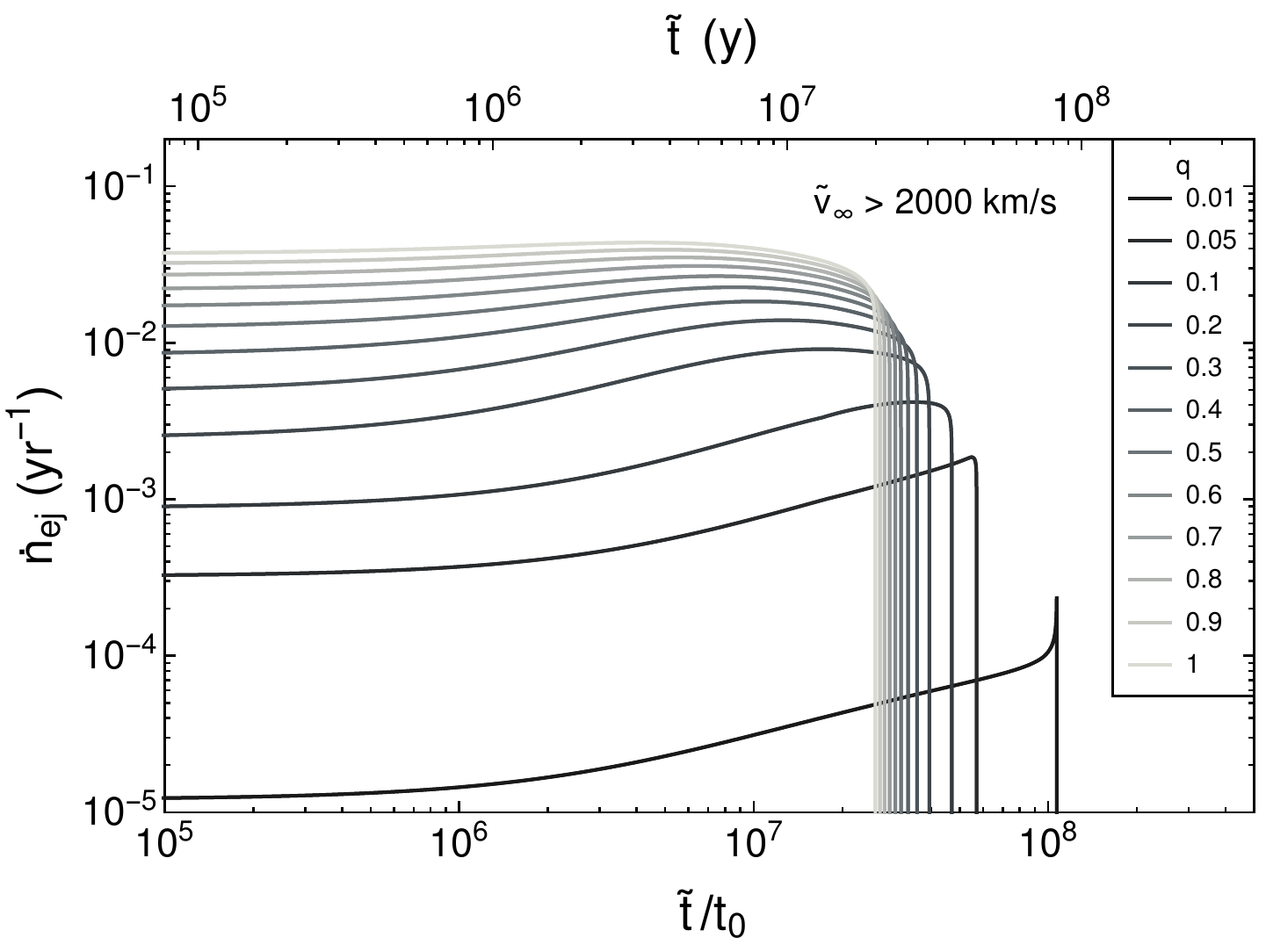}\label{fig:dndtb}}\\
\subfloat[]{\includegraphics[scale=0.58]{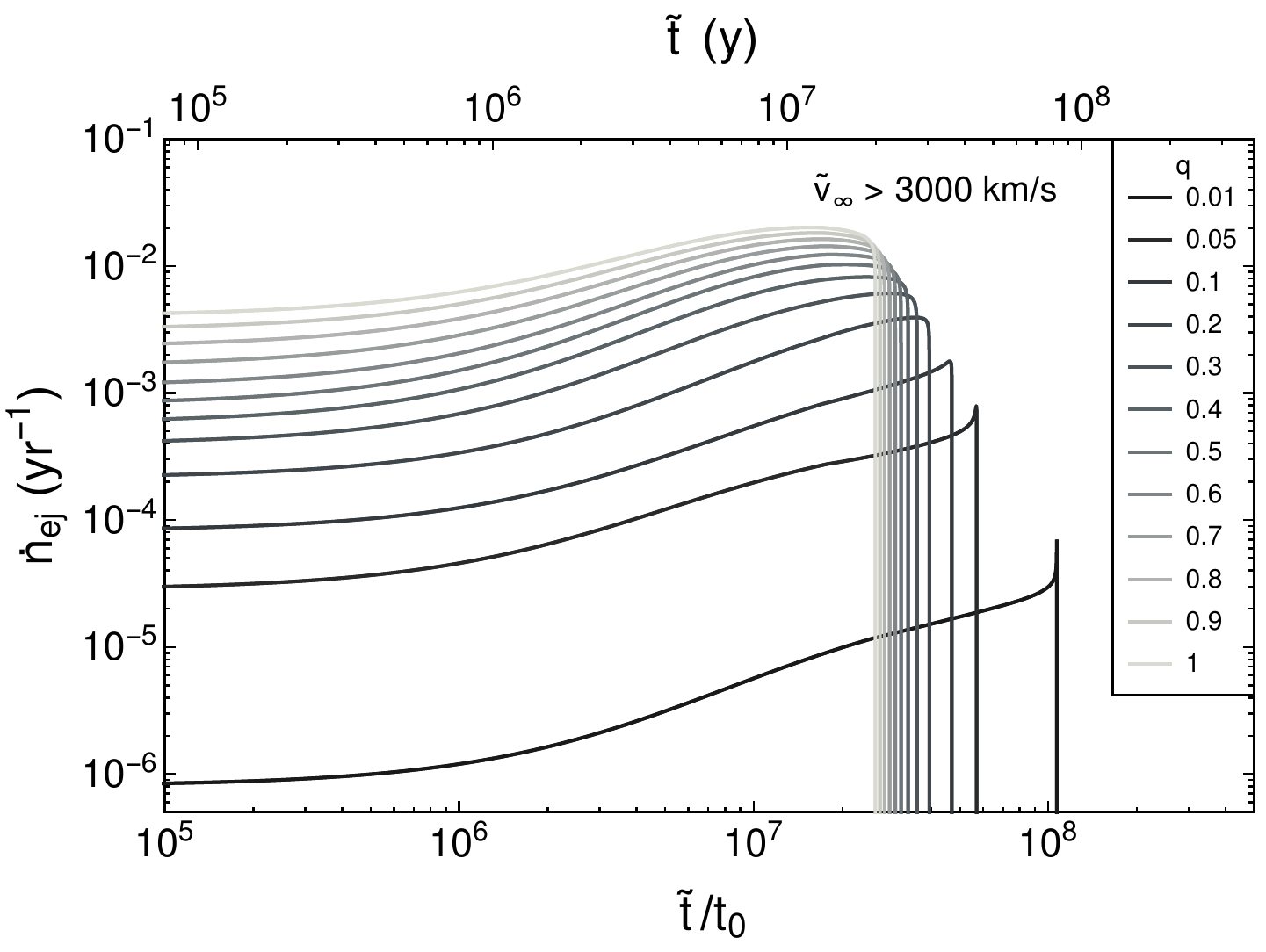}\label{fig:dndtc}}\\
\caption{The time-dependent ejection rate $\dot{n}_\textrm{ej}$ of stars with velocities $\tilde{v}_\infty > v_c$, where $v_c$ is a) $1000$ km/s, b) $2000$ km/s, and c) $3000$ km/s, for $M_1 = 10^6 M_\odot$ and Sun-like stars, starting at an initial separation of $a = \tilde{a}/\tilde{r}_{t1} = 1000$ ($\tilde{a} = 2.3$ mpc).}
\label{fig:dndt}
\end{figure}

Figure \ref{fig:dndt} shows $\dot{n}_\textrm{ej}$ for stars with different velocity cutoffs $v_c$. The rates are in the range $\sim 10^{-3} - 10^{-1}$ yr$^{-1}$ for $q = 1$, and drop to $\sim 10^{-6} - 10^{-3}$ yr$^{-1}$ for $q = 0.01$. The SMBH binaries eject a burst of stars with $\tilde{v}_\infty \gtrsim 3000$ km/s as they are about to coalesce, for our full range of mass ratios. For $v_c \simeq 1000$ km/s, the ejection rate actually declines monotonically for binaries with high mass ratios ($q \gtrsim 0.2$), and only those with low mass ratios ($q \lesssim 0.05$) exhibit a burst at late times. The ejection rate transitions between these two phases at $v_c \simeq 2000$ km/s, where the rate remains roughly constant over the binary SMBH lifetime for $q \gtrsim 0.5$.

This behavior arises due to a trade-off: as the SMBH binary contracts, it ejects higher velocity stars (Figures \ref{fig:histvinfesc}, \ref{fig:velcutoffsappendix}), but when gravitational radiation begins to dominate, the binary ejects fewer stars overall since it contracts more rapidly and spends less time at a given separation (Eq. \ref{eq:dadtgw}). For low mass ratios, the first effect is dominant; a $q=0.01$ SMBH binary at $a = 1000$ ejects few stars with velocities $\tilde{v} > v_c \sim 1000$ km/s (Figure \ref{fig:velcutoffsappendix}), but as it contracts, it ejects more stars above this cutoff, larger than the number suppressed by the gravitational wave coalescence. For nearly-equal mass ratios, the second effect is dominant; a $q=1$ SMBH binary at $a = 1000$ already ejects most stars with velocities $\tilde{v} > v_c \sim 1000$ km/s (Figure \ref{fig:velcutoffsappendix}), and though it ejects slightly more in this range as the binary SMBH contracts, the rapid coalescence suppresses the possibility of ejections, and the ejection rate decreases.

A burst thus arises as a result of the decreasing binary SMBH separation alone, in a time $\tilde{t} \sim 10 - 100$ Myr, given that stars are continually injected from a full loss cone. Earlier work hinted at this result \citep{sesana06}, though here we explicitly fold in the SMBH binary coalescence rate, which washes out the effect for nearly equal-mass binaries when $v_c = 1000$ km/s. Previous studies have calculated a burst of HVSs over the lifetime of a binary SMBH, but under different physical conditions. A stable binary SMBH that ejects incident unbound stars will produce a surge of HVSs when the binary first becomes hard and expels the stars present in the loss cone \citep{sesana07b}. An IMBH, inspiraling through a stellar cusp due to dynamical friction, can also produce a burst when it reaches the center where the cusp is most dense, promptly ejecting the high concentration of stars and leaving behind a depleted region \citep{baumgardt06,levin06,sesana08}.

\section{Comparison with the Hills Mechanism}
\label{sec:comparison}

Hypervelocity stars exhibit different properties depending on the mechanism by which they are produced, and can thus reveal information about possible massive BH binarity and the distribution of stars in a galactic nucleus. Here, we compare the distributions of HVSs ejected by a single and binary SMBH. We first calculate the integrated distributions for the two production mechanisms with different parameters (Section \ref{subsec:integrateddists}), and then use parameter estimation to study the number of samples required to distinguish between them (Section \ref{subsec:modelfitting}).

To study the properties of HVSs ejected by the Hills mechanism \citep{hills88}, we simulated $2 \times 10^5$ encounters between a binary star and an isolated SMBH using the N-body code REBOUND with the IAS15 integrator \citep{rein12,rein15}. Due to scale invariance, we ran our simulations with the simulation parameters in the units $G = m_\textrm{tot} = \tilde{a}_* = 1$, where $m_\textrm{tot} = m_1 + m_2$ is the total mass of the binary star system with primary (secondary) mass $m_1$ ($m_2$) and $\tilde{a}_*$ is the incident binary star separation. We ran our simulations for Sun-like stars, $m_1 = m_2 = 1 M_\odot$, and a BH of mass $M_\bullet = 10^6 M_\odot$. The origin of the coordinate system was set to the three-body center of mass. Each binary star was initialized with its center of mass on a parabolic orbit at a distance of $r = 2000$ relative to the SMBH, and with an isotropically distributed orientation. The pericenter of the parabolic orbit was sampled uniformly in the interval $r_p \in [0,175]$; we used a uniform distribution since we treated the binary loss cone in the ``pinhole'' regime, and we set this range since the probability of a Hills ejection drops to zero for $r_p \geq 175$ \citep{hills88}. We stopped the integration if a star escaped to $r = 4000$ or if the simulation time reached $t = 10^4$; we ignored the possibility of a tidal disruption as the probability of one is small \citep{mandel15}. We classify an encounter as a Hills ejection if at least one of the stars crosses the escape sphere at $r = 4000$, one star is unbound and the other is bound to the SMBH, and the stellar binary energy is positive (i.e. the two stars are on a relative unbound orbit). We only recorded ejected stars, so we do not need to distinguish between ``ejected'' and ``escaped'' stars here as we did for the SMBHB mechanism in Section \ref{subsec:ejprob}. We find the same ejection probabilities versus pericenter distance as in previous studies, and the same mean ejected velocity of $\langle \tilde{v}_\infty \rangle \sim \tilde{a}^{-1/2}_* m_\textrm{tot}^{1/3} M^{1/6}_\bullet$, where $M_\bullet$ is the black hole mass \citep{hills88,bromley06}.

\subsection{Integrated Distributions}
\label{subsec:integrateddists}

Although the average properties of ejected HVSs can constrain the nature of their progenitors (Section \ref{sec:hvss}, and \citealt{yu03,gualandris05}), the distributions can reveal more complete information. In addition, HVSs observed at a given epoch were produced over an interval of time, and were thus likely sampled from an integrated PDF for each ejection mechanism. If the stars were ejected by the Hills mechanism, they were sampled from a PDF integrated over a range of incident binary star separations. If they were ejected by a binary SMBH, they may have been sampled from one integrated over the binary SMBH lifetime (Section \ref{subsec:inspiral}). In particular, for the Milky Way, GC-origin HVSs observed with velocities $\tilde{v} \sim 1000$ $\textrm{km} \cdot \textrm{s}^{-1}$ at Galactocentric distances $\tilde{r} \sim 100$ kpc, roughly the current limit of HVS distance measurements \citep{boubert18,brown18}, would have been produced $t \sim 100$ Myr ago (ignoring the Galactic potential). If there was a binary SMBH with separation $\tilde{a} \sim 2$ mpc in the GC in the recent past, then it would have coalesced in a time $T \sim 30 - 100$ Myr (Figure \ref{fig:dndt}), and the HVSs observed at large distances would have arisen from a PDF integrated over the binary SMBH lifetime. In contrast, if the binary SMBH had a larger separation, then it would contract more slowly and would not have coalesced in the past $t \sim 100$ Myr, and the HVSs observed would have arisen roughly from the PDF of a binary SMBH with a single set of parameters. The other local sources of HVSs have similar behavior.

In previous work, authors have studied the integrated distributions in the context of an IMBH inspiraling towards a SMBH through a stellar cusp, both for the velocity \citep{baumgardt06,sesana07a} and the ejection direction \citep{zier01,baumgardt06,levin06}, and have compared these with the Hills mechanism. Here, we compare the integrated distributions produced by nearly-equal mass binary SMBHs that eject incident unbound stars with those produced by the Hills mechanism.

\subsubsection{Velocity Distribution}
\label{subsubsection:integrateddistsvel}

We first consider the velocity distribution of HVSs ejected by the Hills mechanism. To simplify our calculations, we define the dimensionless binary star separation $a_* = \tilde{a}_*/a_{*0}$ where the length scale $a_{*0} = 1$ AU, and the dimensionless velocity $v = \tilde{v}/v_*$ where $v_*(a_*) = \sqrt{Gm_\textrm{tot} / \tilde{a}_*} = \sqrt{Gm_\textrm{tot} / a_* a_{*0}}$ is the characteristic circular velocity of a binary star with total mass $m_\textrm{tot}$ and semi-major axis $\tilde{a}_*$, or dimensionally
\begin{equation}
v_* \simeq 133 \textrm{ km} \cdot \textrm{s}^{-1} \left(\frac{m_\textrm{tot}}{2M_\odot}\right)^{1/2} \left(\frac{\tilde{a}_*}{0.1 \textrm{ AU}}\right)^{-1/2}
\end{equation}

\begin{figure}
\centering
\includegraphics[scale=0.43]{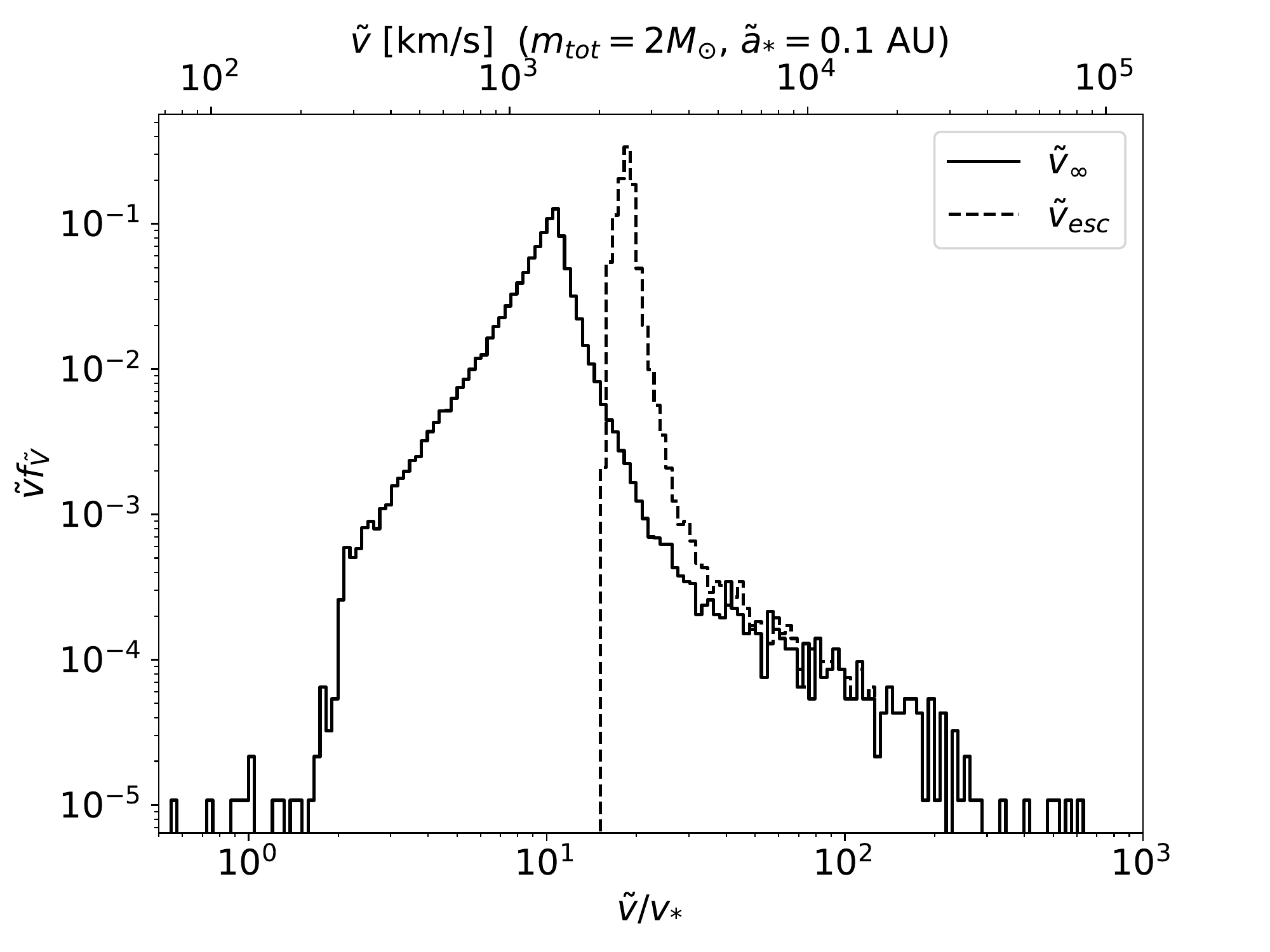}
\caption{The probability distribution $\tilde{v}f_{\tilde{V}}$ for the velocity of stars ejected by the Hills mechanism at the escape sphere $r = 4000$ (dashed) and at $r \rightarrow \infty$ (solid), where the PDF is $f_{\tilde{V}} \equiv f_{\tilde{V}}(\tilde{v}|\textrm{ej.},a_*)$. The ratio of the black hole mass to the total stellar binary mass is $M_\bullet / m_\textrm{tot} = 5 \times 10^5$. The upper $\tilde{v}$-axis gives the velocity in km/s for a binary star with mass $m_\textrm{tot} = 2M_\odot$ and semi-major axis $\tilde{a}_* = 0.1$ AU.}
\label{fig:histvhillsmech}
\end{figure}

The velocity PDF for ejected stars given an initial binary star separation $a_*$ is $f_{\tilde{V}}(\tilde{v}|\textrm{ej.},a_*)$, where the random variable describing the end state is $E = \textrm{ej.}$ for an ejected star. This can be rewritten as $f_{\tilde{V}}(\tilde{v}|\textrm{ej.},a_*) = (1 / v_*(a_*)) f_V( \tilde{v}/v_*(a_*) | \textrm{ej.},a_* )$. Due to scale invariance, the latter distribution simplifies to $f_V( \tilde{v}/v_*(a_*) | \textrm{ej.},a_* ) = f_V( \tilde{v}/v_*(a_*) | \textrm{ej.} )$, i.e. $v = \tilde{v}/v_*$ is conditionally independent of $a_*$ given an ejection. Figure \ref{fig:histvhillsmech} shows the velocity PDFs for Hills ejections at the escape radius and at $r \rightarrow \infty$, for a black hole mass $M_\bullet / m_\textrm{tot} = 5 \times 10^5$. When the stars are infinitely far from the SMBH, the distribution is peaked at $v_\infty = \tilde{v}_\infty / v_* \simeq 10$ and decreases asymmetrically away from this point. This conforms to the data presented by \citet{bromley06}, who approximate this central region in the $v_\infty$ distribution with a Gaussian. The mean velocity decreases by a factor of $\sim 2$ as the stars escape the gravitational potential of the black hole. The distribution is wider for $v_\infty$ since the $v$-axis is log-scaled, and thus the velocities in the central bin for $v_\textrm{esc}$ get redistributed into a range of bins for $v_\infty$.

The PDF for the ejected velocities integrated over a range of stellar binary separations is $f_{\tilde{V}}(\tilde{v}|\textrm{ej.}) = \int f_{\tilde{V}}(\tilde{v}|\textrm{ej.},a_*) f_{A_*}(a_*) da_*$. The ejection probability is independent of $a_*$, $P(\textrm{ej.}|a_*) = P(\textrm{ej.})$, due to scale invariance. In our model, the stellar binaries in the galactic bulge have semi-major axes distributed as $f_{A_*}(a_*) = K/a_*$ \citep{heacox98,kobulnicky07}, i.e. a uniform distribution in $\log a_*$, where $K$ is a normalization constant given by $K = [\ln\left(a_{*,\textrm{max}} / a_{*,\textrm{min}}\right)]^{-1}$ for a minimum (maximum) separation of $a_{*,\textrm{min}}$ ($a_{*,\textrm{max}}$). The survivability of a binary star in three-body encounters in the bulge depends on the ratio of its circular velocity to the stellar velocity dispersion \citep{hills89}, and the encounter rate (and thus the local stellar density). If $v_* \gtrsim \sigma$, where $\sigma$ is the 1D stellar velocity dispersion, then the binary star's binding energy will likely increase in an encounter; if $v_* \lesssim \sigma$, then its binding energy will likely decrease in an encounter, either through the binary star widening or an exchange collision, ultimately leading to a dissociation after many such encounters. For a Milky Way-like galaxy, $\sigma \sim 100$ km/s \citep{gultekin09}, and the velocity condition roughly becomes $a_{*,\textrm{max}} \lesssim 0.1$ for binary stars with total mass $m_\textrm{tot} = 2M_\odot$. Many binary stars in the Galaxy have velocities less than $\sigma$ due to the local density condition, so we take our upper bound to be slightly higher at $a_{*,\textrm{max}} \simeq 1$, though the exact upper bound is unimportant since it simply translates into a different lower bound in the ejection velocities. If its separation becomes small, a stellar binary will undergo mass transfer or experience a merger, which imposes a rough lower bound $a_{*,\textrm{min}} \gtrsim 0.001$. The integrated PDF is then
\begin{equation}
f_{\tilde{V}}(\tilde{v}|\textrm{ej.}) = K \int_{\ln a_{*,\textrm{min}}}^{\ln a_{*,\textrm{max}}} \frac{1}{v_*(a_*)} f_V\left( \left. \frac{\tilde{v}}{v_*(a_*)} \right| \textrm{ej.} \right) d(\ln a_*)
\label{eq:fvintegratedhillsmech}
\end{equation}

\begin{figure*}
\centering
\subfloat[]{\includegraphics[scale=0.43]{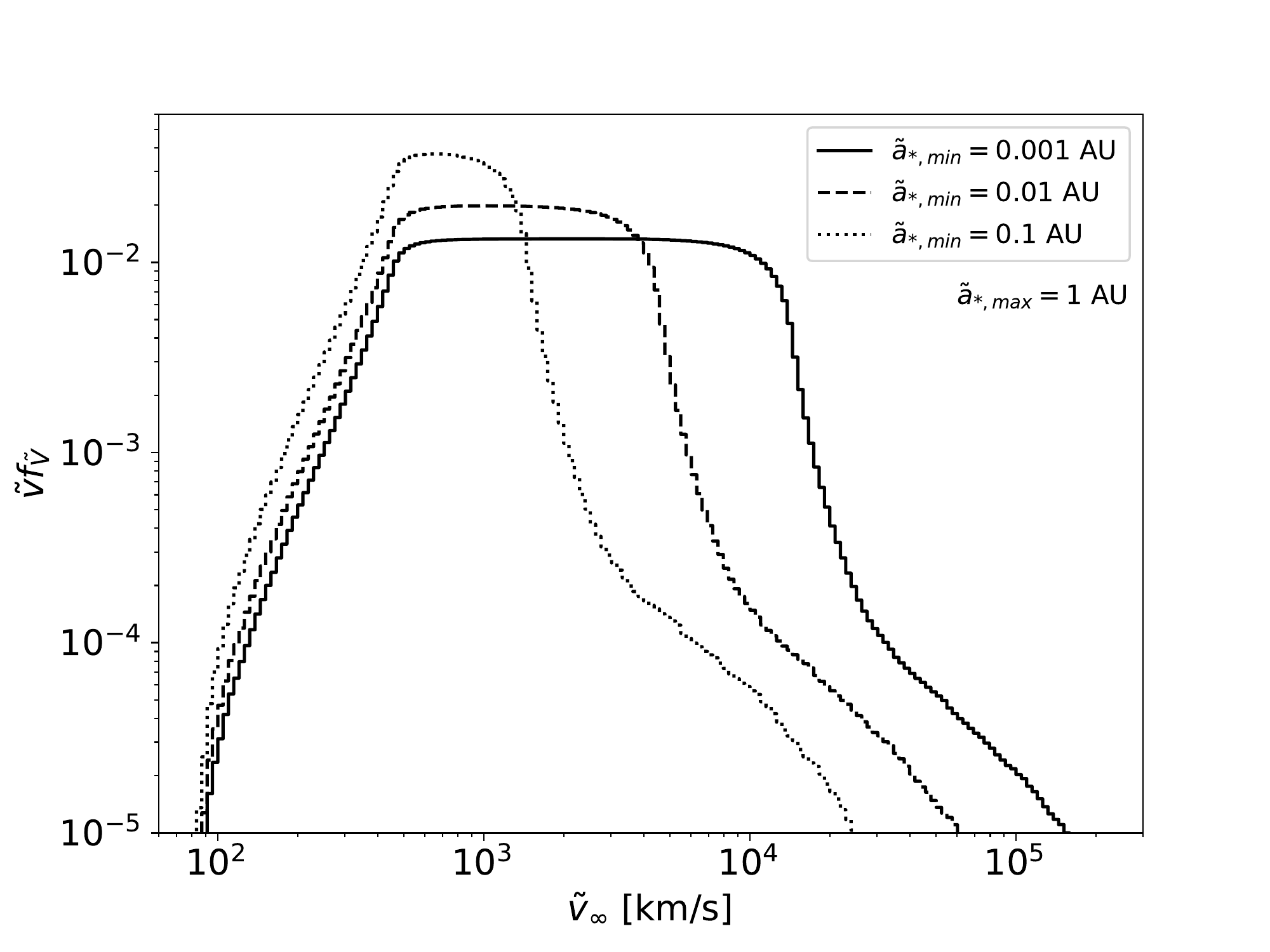}\label{fig:histvinfhillsmechintegrated}}\hfill
\subfloat[]{\includegraphics[scale=0.43]{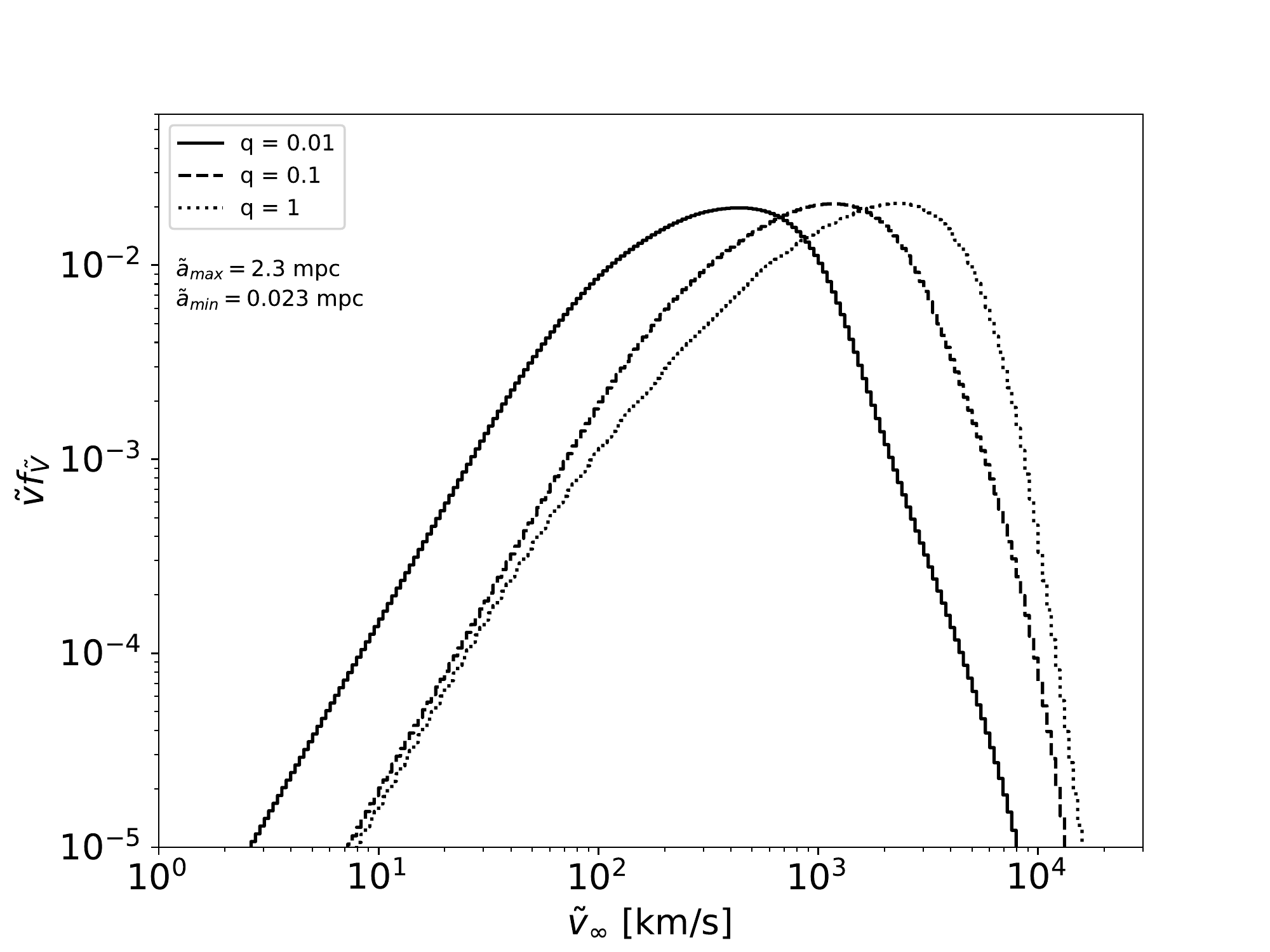}\label{fig:histvinfbinbhintegrated}}
\caption{The integrated probability distributions $\tilde{v}f_{\tilde{v}}$ for the velocity $\tilde{v}_\infty$ of ejected stars. a) The distributions for stars ejected by the Hills mechanism, using the BH mass $M = 10^6 M_\odot$ and stellar masses $m_1 = m_2 = M_\odot$. The incident binary stars are distributed in semi-major axis as $f_{\tilde{A}_*}(\tilde{a}_*) \sim 1/\tilde{a}_*$ between $\tilde{a}_{*,\textrm{max}} = 1$ AU and $\tilde{a}_{*,\textrm{min}} = 0.001$ AU (solid), $0.01$ AU (dashed), and $0.1$ AU (dotted). The integrated PDFs $f_{\tilde{V}} \equiv f_{\tilde{V}}(\tilde{v}|\textrm{ej.})$ are given in Eq. \ref{eq:fvintegratedhillsmech}. b) The distributions for stars ejected by a binary SMBH contracting from $\tilde{a}_\textrm{max} = 2.3$ mpc ($a_\textrm{max} = 1000$) to $\tilde{a}_\textrm{min} = 0.023$ mpc ($a_\textrm{min} = 10$) due to stellar scattering and gravitational wave emission, using the primary mass $M_1 = 10^6 M_\odot$ and Sun-like stars, for $q = 0.01$ (solid), $0.1$ (dashed), and $1$ (dotted). The integrated PDFs $f_{\tilde{V}} \equiv f_{\tilde{V}}(\tilde{v}|\textrm{ej.},q)$ are given in Eq. \ref{eq:fvintegratedbinbh}. }
\label{fig:histvinfintegrated}
\end{figure*}

Figure \ref{fig:histvinfhillsmechintegrated} shows the integrated velocity distribution for stars ejected by the Hills mechanism for three different ranges of binary star separations. The lower that the range of binary star separations can extend, the higher the ejected velocities can reach, as expected from energy conservation in the exchange collision. If the full range $a_* \sim 0.001 - 1$ of separations are possible, then the distribution becomes flat over $\tilde{v}_\infty \sim 10^3 - 10^4$ km/s. Otherwise, it is flat over a smaller range of lower velocities.

The potential of the host galaxy can, of course, modify this distribution. \citet{rossi14} thoroughly studied the velocity distributions in the Galactic halo of HVSs ejected (presumably) by the Hills mechanism, including the contribution of the Galactic potential. The authors considered the same distribution of binary star separations as in this work, $f_{\tilde{A}_*}(\tilde{a}_*) \sim 1/\tilde{a}_*$, though with $\tilde{a}_{*,\textrm{max}} \simeq 20$ mpc ($\sim 4000$ AU) and $\tilde{a}_{*,\textrm{min}} \simeq (1 - 10) R_*$ ($\sim 0.005 - 0.05$ AU for $R_* = R_\odot$), and two different distributions for the component masses. In the full loss cone regime, they find that the velocity peak occurs at $\tilde{v} \sim 800$ km/s for equal-mass stellar binaries with component masses $m_* = 3M_\odot$, which appears consistent with our result for $a_{*,\textrm{min}} = 0.01$ if we were to include deceleration from a galactic potential (most stars with $\tilde{a}_{*,\textrm{max}} \gtrsim 1$ AU would not escape the BH potential, so a different upper limit would not influence the peak location). In addition, they found that the velocity distribution before the peak depends solely on the Galactic potential, and after the peak on the properties of the stellar binaries. This suggests that we examine and compare the high velocity regions of the velocity distributions, as they may encode information about the BH sources if the stellar populations are known.

We next turn to the velocity distribution of HVSs ejected by a binary SMBH. We write the dimensionless binary SMBH separation $a = \tilde{a}/\tilde{r}_{t1}$, and define the dimensionless velocity $v = \tilde{v}/v_\textrm{bh}$ where $v_\textrm{bh}(q,a) = v_0 \sqrt{(1+q)/a}$ is the characteristic binary SMBH circular velocity given in Eq. \ref{eq:vbh}. The velocity distribution for a given binary SMBH mass ratio $q$ and separation $a$ can be expressed as $f_{\tilde{V}}(\tilde{v}|\textrm{ej.},q,a)$, where the random variable describing the end state is $E = \textrm{ej.}$ for an ejected star; Figure \ref{fig:histvinfesc} shows this distribution for several binary SMBH parameters.

To obtain an integrated PDF for the ejected velocities, we must integrate over the lifetime of the binary SMBH as it contracts. The details of the inspiral are given in Section \ref{subsec:inspiral}. The integrated PDF for a given $q$ is $f_{\tilde{V}}(\tilde{v}|\textrm{ej.},q) = f_{\tilde{V}EQ}(\tilde{v},\textrm{ej.},q) / f_{EQ}(\textrm{ej.},q)$, where $f_{\tilde{V}EQ}(\tilde{v},\textrm{ej.},q) = \int f_{\tilde{V}}(\tilde{v}|\textrm{ej.},q,a) P(\textrm{ej.}|q,a) f_A(a|q) f_Q(q) da$ and $f(\textrm{ej.},q) = \int P(\textrm{ej.}|q,a) f_A(a|q) f_Q(q) da$. The velocity distribution for a given $q$ and $a$ can be rewritten as $f_{\tilde{V}}(\tilde{v}|\textrm{ej.},q,a) = f_V(\tilde{v}/v_\textrm{bh}(q,a)|\textrm{ej.},q,a) / v_\textrm{bh}(q,a)$. Though we have the values of this distribution for the discrete values of $a$ in our parameter space, we can obtain a continuous version in this case by approximating $f_V(\tilde{v}/v_\textrm{bh}(q,a)|\textrm{ej.},q,a) \simeq f_V(\tilde{v}/v_\textrm{bh}(q,a)|\textrm{ej.},q,a'=100)$, since the shape of $f_V(v|\textrm{ej.},q,a')$ is largely independent of $a'$ (see Figure \ref{fig:histvinfesc} and the related discussion in the text). The ejection probability was previously labeled as $\lambda_\textrm{ej}(q,a) \equiv P(\textrm{ej.}|q,a)$ (Figure \ref{fig:lambdaej}). The PDF for the semi-major axis as the (circular) binary SMBH contracts is simply $f_A(a|q) = \left\lvert (dt/da)_q \right\rvert / T(q)$, where the binary SMBH lifetime is $T(q) = \int \left\lvert (dt/da)_q \right\rvert da$, and $(dt/da)_q$ is the inverse of the total contraction rate in Eq. \ref{eq:dadt}. We integrate the distribution from $a_\textrm{max} = 1000$ to $a_\textrm{min} = 10$. The integrated PDF is then
\begin{equation}
\begin{split}
& f_{\tilde{V}}(\tilde{v}|\textrm{ej.},q) =\\
& \frac{\int_{\ln a_\textrm{min}}^{\ln a_\textrm{max}} \left[ \frac{f_V\left(\left.\frac{\tilde{v}}{v_\textrm{bh}(q,a)}\right|\textrm{ej.},q,a'=100\right)}{v_\textrm{bh}(q,a)} \right]  g(\textrm{ej.},q,a) a d(\ln a)}{\int_{\ln a_\textrm{min}}^{\ln a_\textrm{max}} g(\textrm{ej.},q,a) a d(\ln a)}
\end{split}
\label{eq:fvintegratedbinbh}
\end{equation}
where $g(\textrm{ej.},q,a) \equiv \lambda_\textrm{ej}(q,a) \left\lvert \left(\frac{dt}{da}\right)_q \right\rvert$. We note that the $p(q)$ terms and $T(q)$ terms each canceled out.

Figure \ref{fig:histvinfbinbhintegrated} shows the integrated velocity distribution of the ejected stars for three different values of $q$. The histograms are similar in shape, with the differences between them arising for the same reasons as in the underlying histograms for fixed separations (Figure \ref{fig:histvinfesca}). The distributions have a precise peak and a power-law decay away from the peak. Importantly, the binary SMBH coalesces rapidly by gravitational wave emission beginning around $a \sim 100$ \citep{darbha18}, so the high velocities found in the distributions for $a \lesssim 100$ are suppressed (Figure \ref{fig:histvinfescb}).

The distributions produced by a binary SMBH are distinct in shape from those produced by the Hills mechanism, although this may not fully persist when including the influence of the galactic potential \citep{rossi14}. In addition, for a binary SMBH with $q = 0.1 - 1$ and a binary star distribution with $a_{*,\textrm{min}} = 0.01$, the two overlap and are difficult to distinguish. If stellar binaries in the bulge can shrink to $a_{*,\textrm{min}} \sim 0.001$, then they can reach much higher velocities than those produced by a binary SMBH; this high velocity signature can thus reveal the presence of an isolated SMBH.

We can compare our results with earlier studies, which calculated the velocity distributions as an IMBH inspirals through a stellar cusp towards an SMBH (Sgr A$^*$, specifically). In \citet{baumgardt06}, the IMBH ($M_2 = 10^3 M_\odot - 10^4 M_\odot$) begins on a circular orbit about the SMBH ($M_1 = 3 \times 10^6 M_\odot$) at $\tilde{a} = 100$ mpc and inspirals to $\tilde{a} \lesssim 1$ mpc. They find that the distributions drop off around $\tilde{v} \sim 2000$ km/s and are largely insensitive to the IMBH mass, and that the binary evolves to high eccentricities by the time it stalls if the IMBH mass is $M_2 \sim 10^4 M_\odot$. In \citet{sesana07a}, the IMBH ($q = 1/729$) begins on an eccentric orbit ($e = 0.9$) at $\tilde{a} \sim 30$ mpc and inspirals until it stalls at $\tilde{a} \sim 4$ mpc. The authors additionally fold in the influence of a model galactic potential; they find the distribution peaks at $\tilde{v}_\infty \sim 700$ km/s. In contrast, our work examines the domain in which the central stellar density has been depleted and further contraction arises from stars incident from a full loss cone; for our primary mass $M_1 = 10^6 M_\odot$, the binary SMBH begins at $\tilde{a} = 2.3$ mpc. For $q = 0.01$, though, the distributions we find are similar to those in both of the works above, with slightly lower ejection velocities due to our lower primary mass. This suggests that it is difficult to use a HVS velocity spectrum to distinguish binary SMBHs in the early slingshot stage from those in the late slingshot stage with a full loss cone.

In the two studies above, the authors also compared their results with integrated distributions from the Hills mechanism. \citet{baumgardt06} analyze fixed stellar binary separations, and find that (for a SMBH with mass $M_\bullet = 3.5 \times 10^6 M_\odot$) the distributions can be peaked at higher velocities than those from binary SMBHs if the stellar binaries are compact ($\tilde{a}_* \lesssim 0.05$ AU). This is clear from Figure \ref{fig:histvinfintegrated} and conforms to our result, though they achieve slightly higher velocities with a higher minimum separation since, using the data of \citet{gualandris05}, they consider more massive stars with $M_* = 3M_\odot$ (the average ejection velocity scales as $\langle \tilde{v}_\infty \rangle \sim m_\textrm{tot}^{1/3} \tilde{a}^{-1/2}$; \citealt{hills88}). \citet{sesana07a} study both flat and lognormal distributions for the stellar binaries, and find that the Hills mechanism tends to eject stars with lower velocities than a binary SMBH, the opposite of our result. The difference may arise because the authors limit their binary stars to a high minimum separation, say $\tilde{a}_{*,\textrm{min}} \sim 0.1$ AU, in which case our results would be in agreement (Figure \ref{fig:histvinfhillsmechintegrated}), though this is unclear.

\subsubsection{Angular Distribution}
\label{subsubsection:integrateddistsmu}

We now calculate the integrated angular distribution from the two HVS ejection mechanisms. We calculate the distributions subject to different velocity cutoffs $v_c$; we do this since an observed sample of ejected stars may exhibit such a cutoff due to the difficulty in observing low velocity stars, particularly in the galaxy core where most are likely located, and since applying such cutoffs often yields distinguishing features.

For the Hills mechanism, isotropic incident stars produce isotropic ejected stars, and thus the angular PDF is $f_M(\mu | \textrm{ej.}, \tilde{V}>v_c) = 1/2$ for all velocity cutoffs $v_c$ and minimum binary star separations $a_{*,\textrm{min}}$. For the SMBHB mechanism, we calculate the integrated angular PDF $f_M(\mu | \textrm{ej.},q,\tilde{V}>v_c)$ using Bayesian inference, as we did with the velocity distributions in Section \ref{subsubsection:integrateddistsvel}. We obtain
\begin{equation}
\begin{split}
& f_M(\mu | \textrm{ej.},q,\tilde{V}>v_c) =\\
& \frac{\int_{\ln a_\textrm{min}}^{\ln a_\textrm{max}} f_M(\mu | \textrm{ej.},q,a,\tilde{V}>v_c) h(\textrm{ej.},q,a,v_c) a d(\ln a)}{\int_{\ln a_\textrm{min}}^{\ln a_\textrm{max}} h(\textrm{ej.},q,a,v_c) a d(\ln a)}
\end{split}
\label{eq:fmuintegrated}
\end{equation}
where $f_M(\mu | \textrm{ej.},q,a,\tilde{V}>v_c)$ is the angular PDF given that an ejection has occurred for a given $q$, $a$, and velocity cutoff $\tilde{V} > v_c$ (Figure \ref{fig:histmu}); $h(\textrm{ej.},q,a,v_c) \equiv P(\tilde{V}>v_c | \textrm{ej.},q,a) P(\textrm{ej.}|q,a) f_A(a|q)$; $P(\tilde{V}>v_c | \textrm{ej.},q,a)$ is the probability that a star has a velocity $\tilde{V} > v_c$ given that an ejection has occurred for a given $q$ and $a$ (Figures \ref{fig:pvelcutoffs} and \ref{fig:velcutoffsappendix}); and $P(\textrm{ej.}|q,a)$ and $f_A(a,q)$ are the same as in Section \ref{subsubsection:integrateddistsvel} above. We evaluate this integral by discretizing it at the values of $a$ in our parameter space.

\begin{figure*}
\centering
\subfloat[]{\includegraphics[scale=0.43]{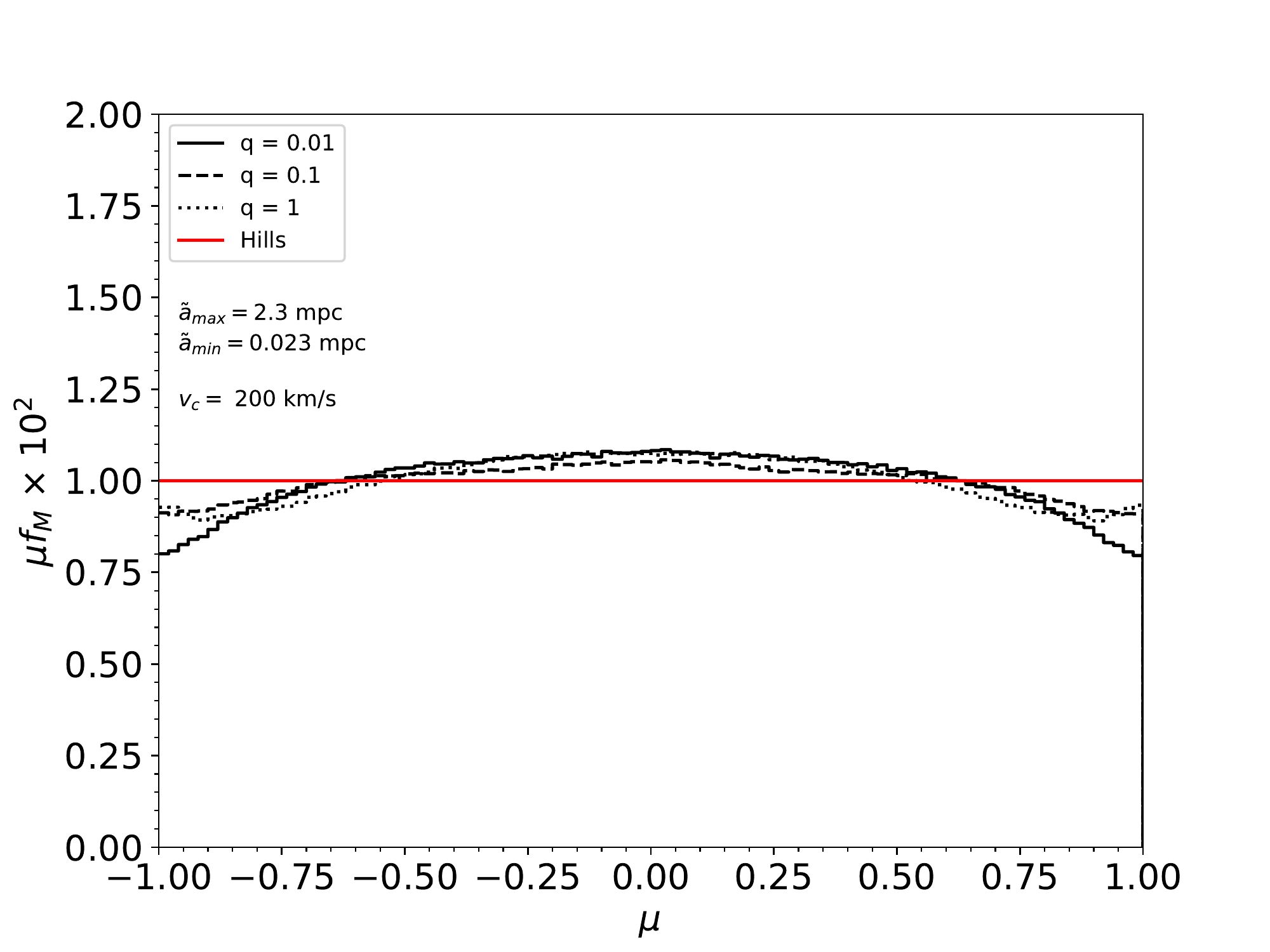}\label{fig:histmuvc200integrated}}\hfill
\subfloat[]{\includegraphics[scale=0.43]{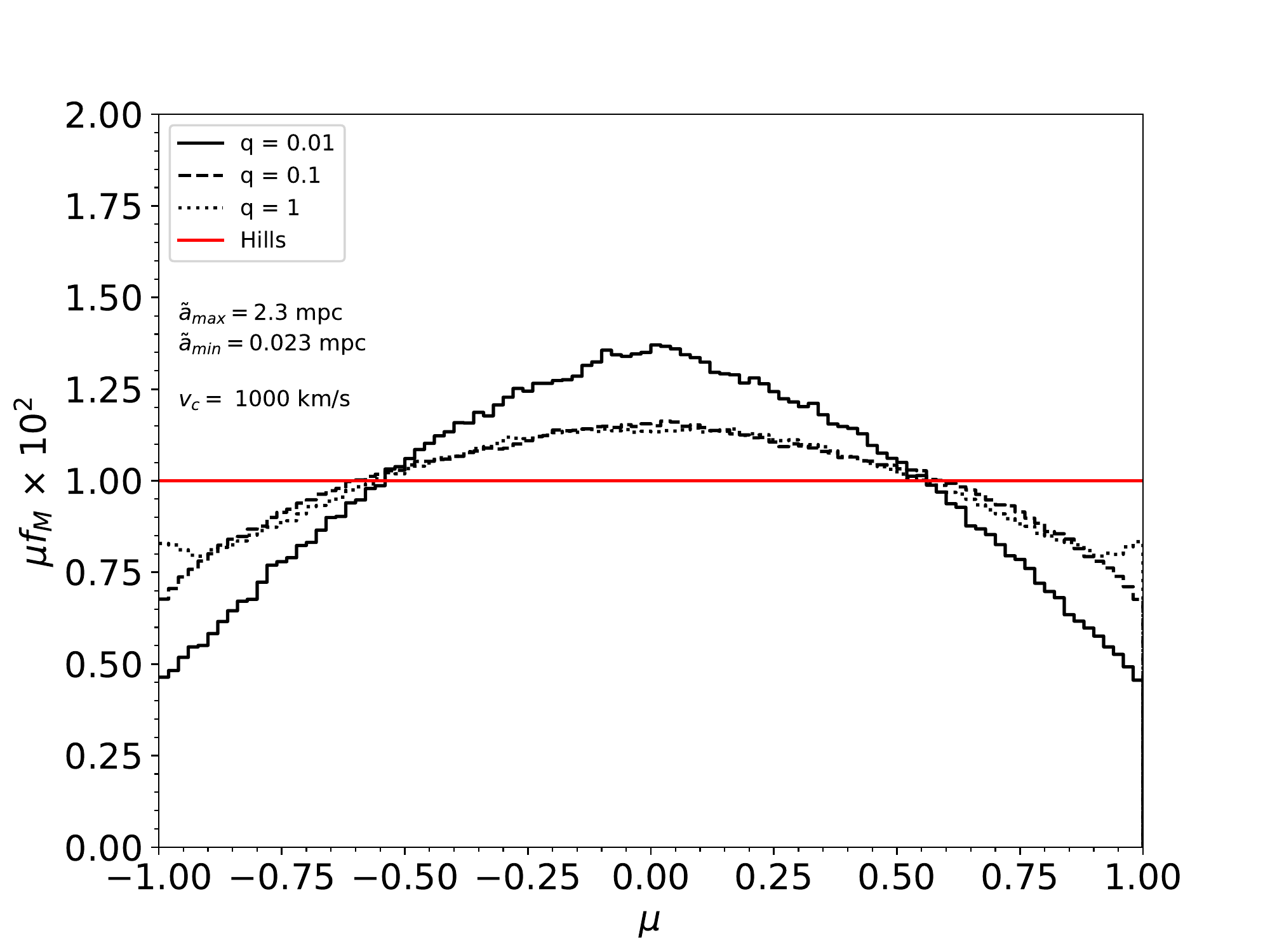}\label{fig:histmuvc1000integrated}}
\caption{
The integrated angular distributions $\mu f_M$ for the direction cosine $\mu$ of stars ejected by the SMBHB mechanism with velocities $\tilde{V} > v_c$, where the binary SMBH contracts from $\tilde{a}_\textrm{max} = 2.3$ mpc ($a_\textrm{max} = 1000$) to $\tilde{a}_\textrm{min} = 0.023$ mpc ($a_\textrm{min} = 10$) due to stellar scattering and gravitational wave emission. We consider a primary with mass $M_1 = 10^6 M_\odot$ and Sun-like stars, and the mass ratios $q = 0.01$ (solid), $0.1$ (dashed), and $1$ (dotted). The velocity cutoffs $v_c$ are a) $200$ km/s, and b) $1000$ km/s. The integrated PDFs $f_M \equiv f_M(\mu|\textrm{ej.},q,\tilde{V}>v_c)$ are given in Eq. \ref{eq:fmuintegrated}. The red curve shows the uniform distribution for $\mu$ produced by the Hills mechanism.
}
\label{fig:histmuintegrated}
\end{figure*}

Figure \ref{fig:histmuintegrated} shows the integrated angular distribution for two different velocity cutoffs. 
For a given $q$, ejected stars are increasingly concentrated in the binary SMBH orbital plane as one applies higher cutoffs, where lower $q$ require a lower cutoff to begin to see this effect, as we found for the distributions for fixed $q$ and $a$ (Section \ref{subsubsec:propertiesmu}). Indeed, for a given $q$ and $v_c$, the integrated distributions are very similar to the corresponding ones for $a=100$ (Figures \ref{fig:histmu} and \ref{fig:histmuappendix}). This is because $a \sim 100$ is the closest separation at which the binary spends considerable time ejecting stars before rapidly coalescing by gravitational radiation \citep{darbha18}, and thus the separation that produces the largest deviation from isotropic emission.

Previous studies have examined some features of the ejection angles during the binary SMBH inspiral, but focused on a binary SMBH embedded in a stellar cusp \citep{zier01,baumgardt06,levin06}, a different setup from this work. The above authors find different trends as compared to our work. \citet{baumgardt06} found that the ejected stars, with no velocity cutoff, are more isotropically distributed for a lower mass secondary SMBH, and exhibit an overdensity in the binary SMBH orbital plane for a higher mass secondary. \citet{levin06} found that the ejected stars exhibit a small anisotropy as the binary shrinks.

\subsection{Model Fitting}
\label{subsec:modelfitting}

In this subsection, we use Bayesian parameter estimation to calculate the number of HVS samples required to distinguish between the different integrated distributions, examining first the velocity distributions (Section \ref{subsubsec:modelfittingvel}) and second the angular distributions (Section \ref{subsubsec:modelfittingmu}). In what follows, we suppose that we have $N$ samples of HVSs. Let the random variable $D$ describe the class of distributions corresponding to a given SMBH progenitor, $D = \{ d_\textrm{smbhb}, d_\textrm{hills} \}$. Let $\Theta$ label the parameter that parametrizes each class, so that $\Theta = Q$ for $D = d_\textrm{smbhb}$, and $\Theta = A_{*,\textrm{min}}$ for $D = d_\textrm{hills}$. Note that $f_\Theta(q|d_\textrm{hills}) = f_\Theta(a_{*,\textrm{min}}|d_\textrm{smbhb}) = 0$. We drop the random variable $E = \textrm{ej.}$. We quantify the accuracy of a fit using the Bayesian odds ratio $OR$ that our set of samples arises from one model $(d,\theta)$ as opposed to another $(d',\theta')$.

\subsubsection{Estimation with Velocity Samples}
\label{subsubsec:modelfittingvel}

We first consider the velocities $\tilde{V}_1, \hdots, \tilde{V}_N$ of the samples. The odds ratio $OR$ given the observed velocities is
\begin{align}
OR & = \frac{f_{D,\Theta}(d,\theta | \tilde{v}_1,\hdots,\tilde{v}_N)}{f_{D,\Theta}(d',\theta' | \tilde{v}_1,\hdots,\tilde{v}_N)}\\
& = \frac{f_{\tilde{V}_1,\hdots,\tilde{V}_N}(\tilde{v}_1,\hdots,\tilde{v}_N | d,\theta) f_\Theta(\theta|d) P(D=d) }{f_{\tilde{V}_1,\hdots,\tilde{V}_N}(\tilde{v}_1,\hdots,\tilde{v}_N | d',\theta') f_\Theta(\theta'|d') P(D=d')}
\label{eq:oddsratiovel}
\end{align}
where $f_{\tilde{V}_1,\hdots,\tilde{V}_N}(\tilde{v}_1,\hdots,\tilde{v}_N | d,\theta)$ is the joint velocity PDF. The samples are independent, so $f_{\tilde{V}_1,\hdots,\tilde{V}_N}(\tilde{v}_1,\hdots,\tilde{v}_N | d,\theta) = f_{\tilde{V}_1}(\tilde{v}_1 | d,\theta) \hdots f_{\tilde{V}_N}(\tilde{v}_N | d,\theta)$, where $f_{\tilde{V}_i}(\tilde{v}_i | d,\theta)$ is the integrated PDF for a single sample (Eq. \ref{eq:fvintegratedhillsmech} and Figure \ref{fig:histvinfhillsmechintegrated} for the Hills mechanism; Eq. \ref{eq:fvintegratedbinbh} and Figure \ref{fig:histvinfbinbhintegrated} for the SMBHB mechanism). We have no prior knowledge about the SMBH progenitor, so $P(D=d_\textrm{smbhb}) = P(D=d_\textrm{hills}) = 1/2$, and $f_\Theta(q|d_\textrm{smbhb}) = f_\Theta(a_{*,\textrm{min}}|d_\textrm{hills}) = 1/3$ since we consider three discrete parameter values for each class. The odds ratio thus simplifies to
\begin{equation}
OR = \frac{f_{\tilde{V}_1}(\tilde{v}_1 | d,\theta) \hdots f_{\tilde{V}_N}(\tilde{v}_N | d,\theta)}{f_{\tilde{V}_1}(\tilde{v}_1 | d',\theta') \hdots f_{\tilde{V}_N}(\tilde{v}_N | d',\theta')}
\label{eq:oddsratiovelfinal}
\end{equation}

As discussed in Section \ref{subsec:integrateddists}, a sample of ejected stars may exhibit a velocity cutoff $v_c$. If we analyze HVSs with velocities $\tilde{V} > v_c$, then we can calculate the odds ratio as above by simply replacing the PDF for the full distribution $d$ with that for the truncated distribution $d(\tilde{V} > v_c)$, namely $f_{\tilde{V}_i}(\tilde{v}_i | d, \theta) \rightarrow f_{\tilde{V}_i}(\tilde{v}_i | d(\tilde{V}_i > v_c), \theta) = f_{\tilde{V}_i}(\tilde{v}_i | d, \theta, \tilde{V}_i > v_c)$. The truncated PDF is
\begin{equation}
f_{\tilde{V}_i}(\tilde{v}_i | d, \theta, \tilde{V}_i > v_c) =
\begin{cases}
\frac{f_{\tilde{V}_i}(\tilde{v}_i | d, \theta)}{1 - F_{\tilde{V}_i}(v_c | d, \theta)}  &,\, \tilde{v}_i > v_c\\
0 &,\, \textrm{else}
\end{cases}
\end{equation}
where $F_{\tilde{V}_i}(v | d, \theta) \equiv \int_{-\infty}^{v_c} f_{\tilde{V}_i}(\tilde{v}_i | d, \theta) d\tilde{v}_i$ is the cumulative distribution function (CDF), which is evaluated at $v_c$ in the denominator. To simplify our calculation, we discretize our PDFs with the same logarithmic bin widths used in Figure \ref{fig:histvinfintegrated}.

\begin{figure*}
\centering
\subfloat[]{\includegraphics[scale=0.43]{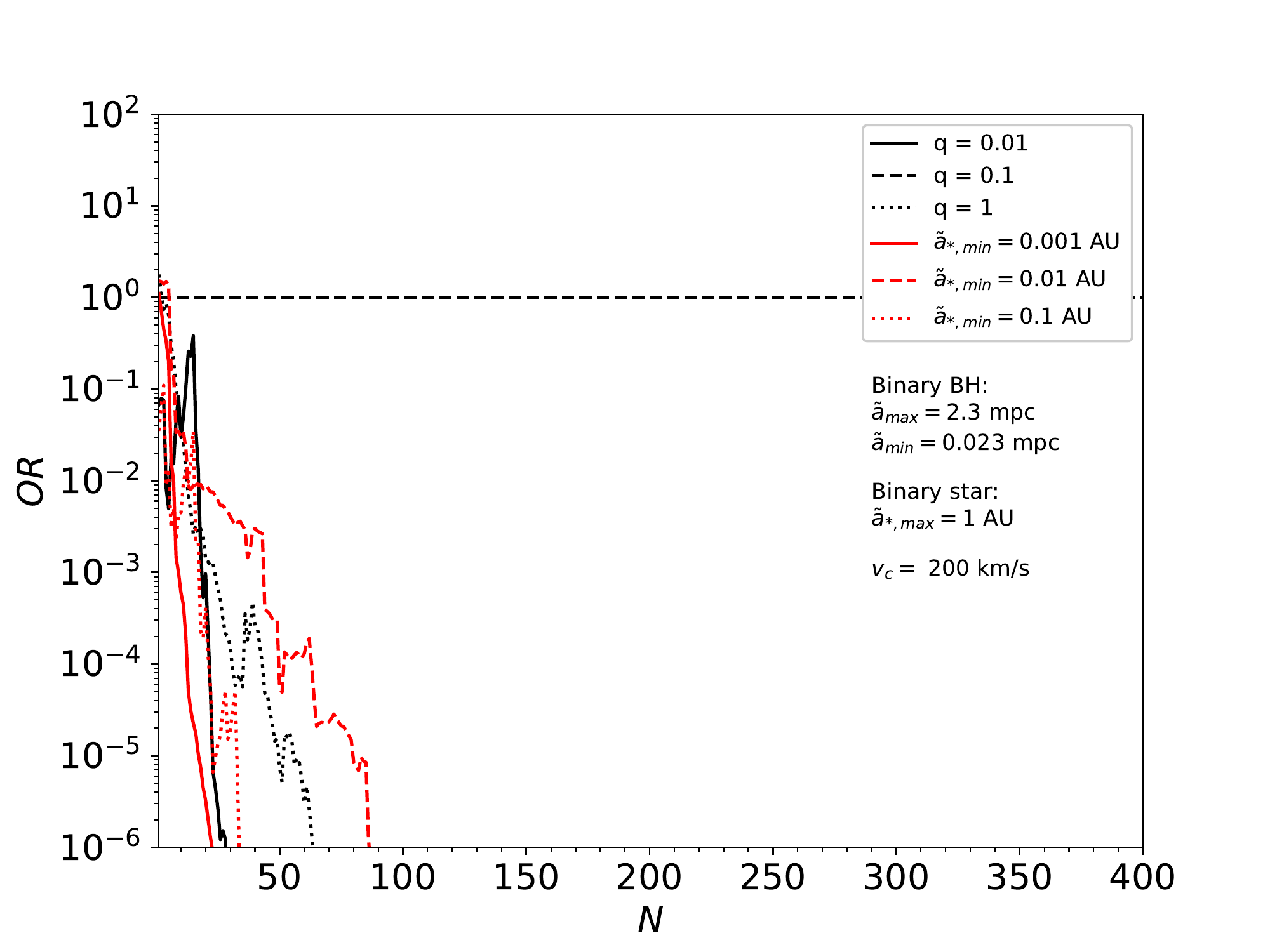}\label{fig:oddsratiovelvc200}}\hfill
\subfloat[]{\includegraphics[scale=0.43]{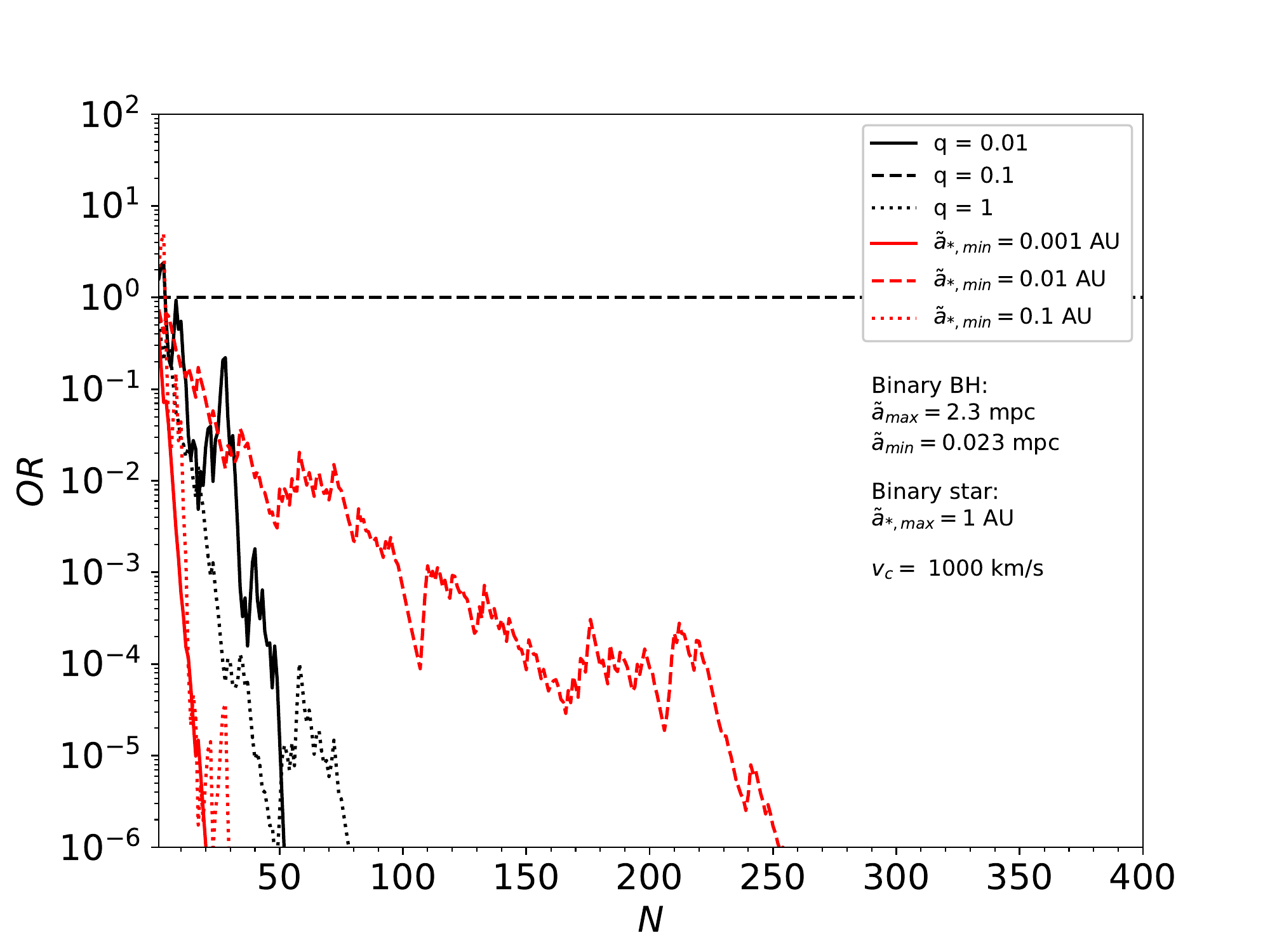}\label{fig:oddsratiovelvc1000}}
\caption{The velocity-sampled Bayesian odds ratio $OR$ (Eq. \ref{eq:oddsratiovelfinal}) as a function of $N$ randomly drawn samples. The samples were drawn from the distribution ($D = d_\textrm{smbhb}$, $\Theta = q = 0.1$), and the odds ratio for each distribution is calculated relative to it. The panels correspond to samples drawn from the truncated PDF with $\tilde{V}_i > v_c$, where $v_c$ is a) $200$ km/s and b) $1000$ km/s.}
\label{fig:oddsratiovel}
\end{figure*}

Figure \ref{fig:oddsratiovel} shows the velocity-sampled odds ratio as a function of $N$ for two different velocity cutoffs $v_c$, where the samples are drawn from the distribution ($D = d_\textrm{smbhb}$, $\Theta = Q = 0.1$) and the odds ratio is calculated relative to this distribution. The true velocity distribution can be distinguished from the others most easily if the full distributions are known; in this case, only $N \sim 100$ samples are needed to obtain $OR \lesssim 10^{-6}$. In the intermediate range of cutoffs $200$ km/s $\lesssim v_c \lesssim$ $1000$ km/s, roughly $N \sim 100 - 300$ samples are needed to distinguish $D = d_\textrm{smbhb}$ with $q = 0.1$ from $D = d_\textrm{hills}$ with $a_{*,\textrm{min}} = 10^{-2}$, since these two distributions overlap heavily here (Figure \ref{fig:histvinfintegrated}), but only $N \sim 50 - 100$ samples are needed to distinguish it from the other mechanisms. \citet{sesana07a} find that $N \gtrsim 100$ samples are needed to identify the ejection mechanism, and our numbers are in agreement for the above range of cutoffs. For $v_c \sim 2000$ km/s, $N \sim 400$ samples are required to discriminate the SMBHB mechanism with different values of $q$, though only $N \sim 300$ samples are needed to rule out the Hills mechanism. For $v_c \gtrsim 4000$ km/s, the velocity distributions for the SMBHB mechanism with different $q$ become very similar \ref{fig:histvinfbinbhintegrated}, and thousands of samples are needed to distinguish them.

If the samples are drawn from $D = d_\textrm{smbhb}$ with $Q = 0.01$ or $1$, then we find similar results of $N \sim$ hundreds of samples for $v_c \lesssim 2000$ km/s, and $N \sim$ thousands for $v_c \gtrsim 4000$ km/s. If the samples are drawn from $D = d_\textrm{hills}$ with $A_{*,\textrm{min}} = 10^{-2}$, then we find analogous results to $D = d_\textrm{smbhb}$ with $\Theta = Q = 0.1$ for $v_c \lesssim 1000$ km/s, though for $v_c \gtrsim 2000$ km/s only $N \sim 50 - 100$ samples are needed to distinguish this ejection mechanism from the others, since the high velocity behavior in Hills ejections is fairly distinct for a given minimum binary star separation.

\subsubsection{Estimation with Angular Samples}
\label{subsubsec:modelfittingmu}

We next consider the direction cosines $M_1, \hdots, M_N$ of the samples. We proceed as above, but set $f_\Theta(a_{*,\textrm{min}}|d_\textrm{hills}) = 1$ here since the Hills mechanism ejects stars isotropically regardless of the minimum binary star separation. The odds ratio $OR$ given the observed directions is then
\begin{equation}
OR = \frac{f_{M_1}(\mu_1 | d,\theta) \hdots f_{M_N}(\mu_N | d,\theta) f_\Theta(\theta|d)}{f_{M_1}(\mu_1 | d',\theta') \hdots f_{M_N}(\mu_N | d',\theta') f_\Theta(\theta'|d')}
\label{eq:oddsratiomu}
\end{equation}
where $f_{M_i}(\mu_i | d,\theta)$ is the angular PDF for a single sample. If we generalize this by applying any velocity cutoff $v_c$, then the PDF is $f_{M_1}(\mu_1 | d(\tilde{V}>v_c),\theta) = f_{M_1}(\mu_1 | d,\theta,\tilde{V}>v_c)$ (Eq. \ref{eq:fmuintegrated} and the paragraph preceding it, and Figure \ref{fig:histmuintegrated}).

\begin{figure*}
\centering
\subfloat[]{\includegraphics[scale=0.43]{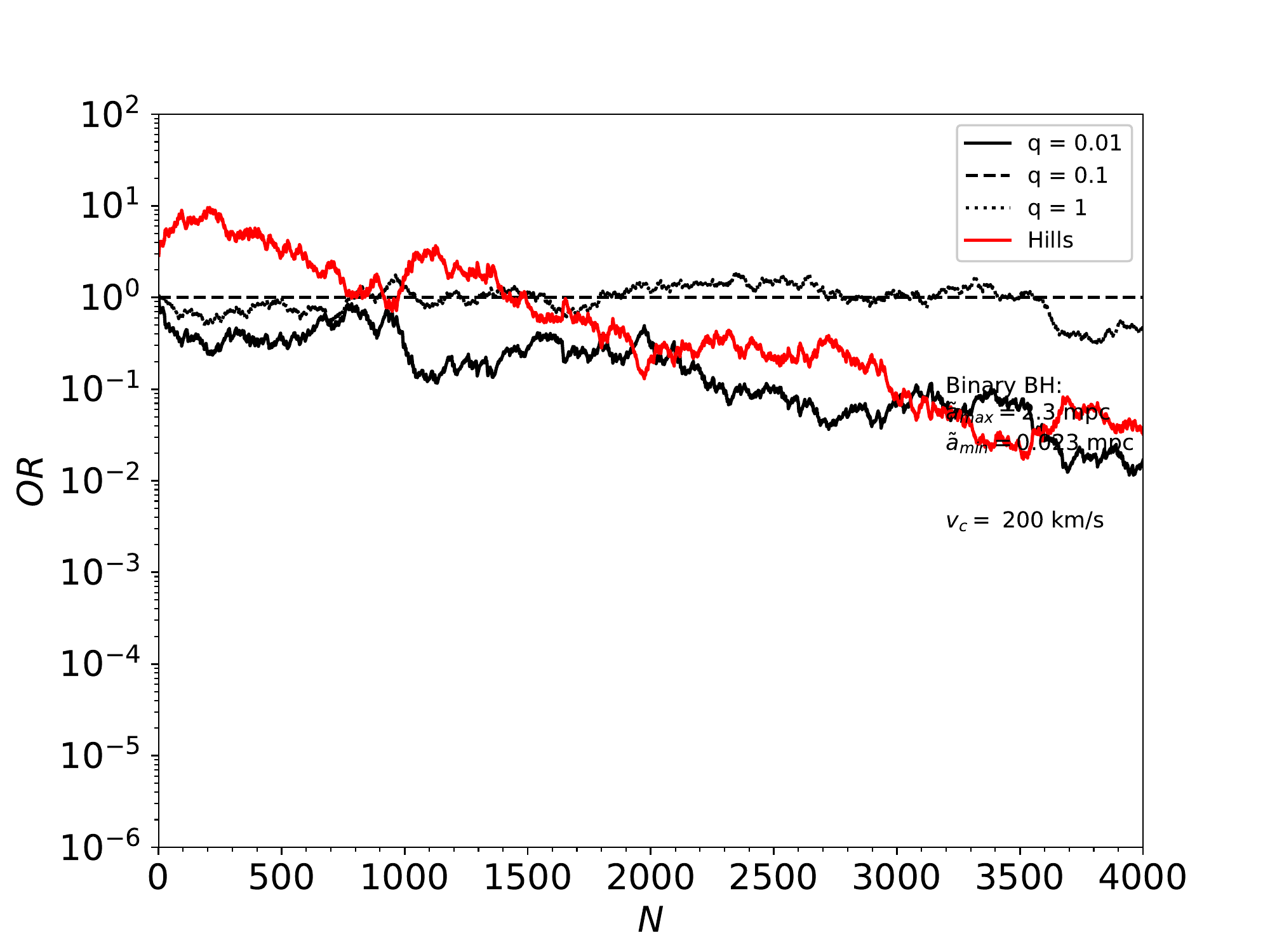}\label{fig:oddsratiomuvc200}}\hfill
\subfloat[]{\includegraphics[scale=0.43]{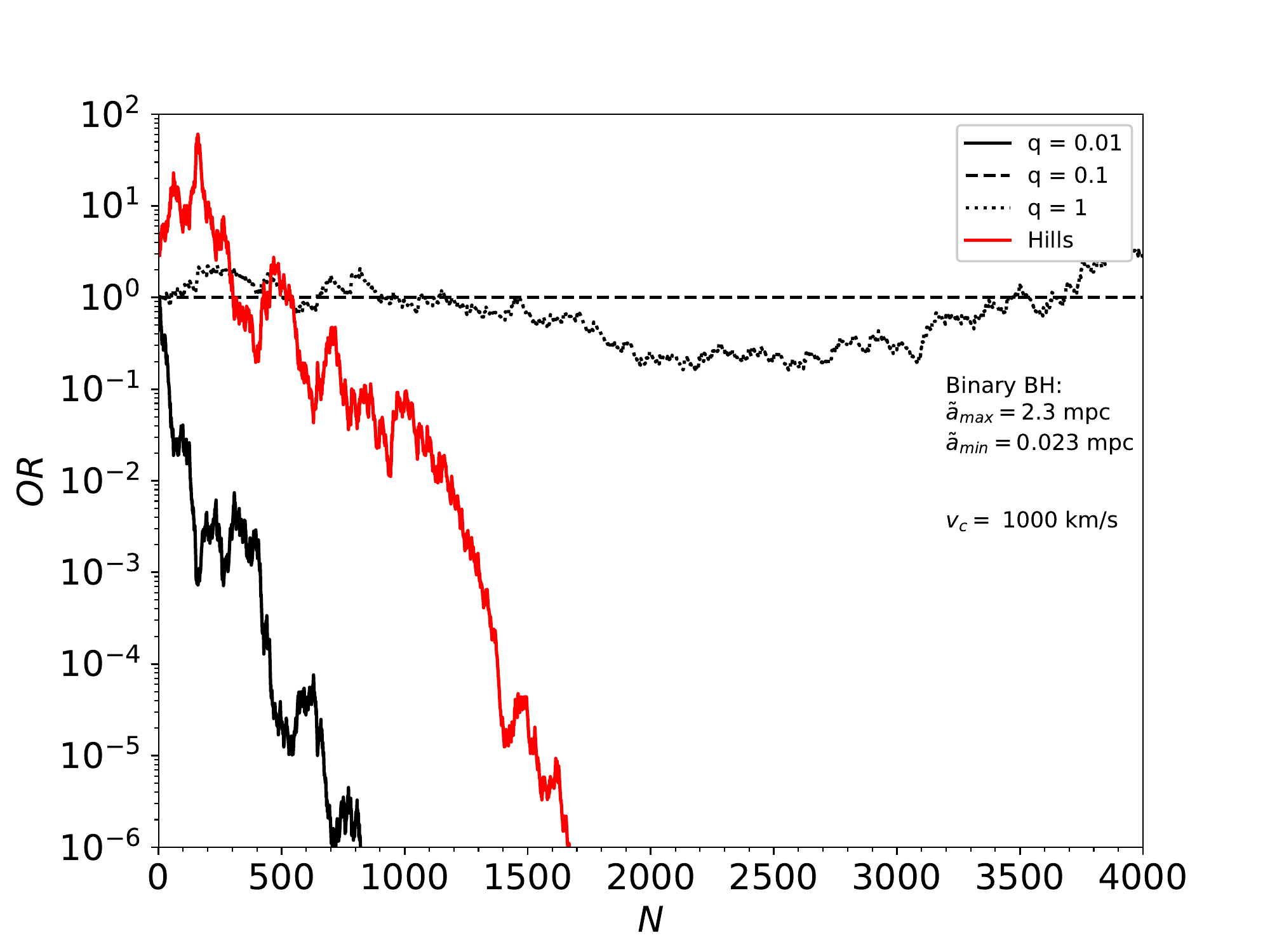}\label{fig:oddsratiomuvc1000}}
\caption{The direction-sampled Bayesian odds ratio $OR$ (Eq. \ref{eq:oddsratiomu}) as a function of $N$ randomly drawn samples. The samples were drawn from the distribution ($D = d_\textrm{smbhb}$, $\Theta = q = 0.1$), and the odds ratio for each distribution is calculated relative to it. The panels correspond to samples drawn from the truncated PDF with $\tilde{V}_i > v_c$, where $v_c$ is a) $200$ km/s, and b) $1000$ km/s.}
\label{fig:oddsratiomu}
\end{figure*}

Figure \ref{fig:oddsratiomu} shows the direction-sampled odds ratio as a function of $N$ for two different velocity cutoffs, where the samples are drawn from the distribution ($D = d_\textrm{smbhb}$, $\Theta = Q = 0.1$) and the odds ratio is calculated relative to this distribution. The odds ratio calculated from the direction cosines requires many more samples to discriminate between the ejection mechanisms than that calculated from the velocity distribution. In general, thousands of samples are required regardless of $v_c$. For $v_c \lesssim 200$ km/s, the distributions are all exactly or nearly isotropic, and require many thousands of samples to separate. As one examines increasingly higher velocity cutoffs, fewer samples are needed to distinguish between the true distribution and some of the other distributions, but at a sufficiently high $v_c$, the distributions for the different $q$ all become peaked in the orbital plane and require many thousands of samples to separate. If the samples are drawn from $D = d_\textrm{smbhb}$ with $Q = 0.01$ or $1$, then we find analogous results, albeit with different velocities at which the behavior transitions. If the samples are drawn from $D = d_\textrm{hills}$, then the behavior is as expected: many thousands of samples are required for $v_c \lesssim 200$ km/s, and the number of required samples decreases with increasing $v_c$, reaching roughly $N \sim 1000$ samples for $v_c \sim 2000$ km/s.

\section{Summary and Conclusions}
\label{sec:conclusion}

Galactic supermassive black holes can eject hypervelocity stars by several mechanisms, and the properties of the ejected stars can reveal information about their black hole progenitors and the galaxy in which they reside. In this paper, we studied encounters between unbound (parabolic, zero-energy) stars incident from a full loss cone and (hard, circular) binary SMBHs, and examined the properties of the ejected HVSs as a function of the binary SMBH mass ratio and separation. Where necessary, we considered a binary SMBH with primary mass $M_1 = 10^6 M_\odot$ and an isolated SMBH with mass $M_\bullet = 10^6 M_\odot$, and stars with solar parameters. We found several features of HVSs from binary SMBHs that both corroborate earlier work and reveal detailed behavior:

\begin{enumerate}[1., leftmargin=*]

\item The ejection probabilities are in the range $\lambda_\textrm{ej}^\textrm{bin} \simeq 0.5 - 0.86$, and are monotonically increasing functions of $a$ and $q$ (where $a = \tilde{a}/\tilde{r}_{t1}$ is the dimensionless binary SMBH separation and $q = M_2/M_1$ is its mass ratio). The probabilities are largely independent of $a$ for $a \gtrsim 100$, and of $q$ for $q \gtrsim 0.2$. These are lower than the probabilities at which stars reach the escape sphere of our simulations, $\tilde{r}/\tilde{a} = 100$, which roughly corresponds to the influence radius of the SMBHs.

\item The mean velocity of the stars ejected by a binary SMBH is well described by $\langle v_\infty \rangle \simeq \kappa v_0 (q/(1+q))^{1/2} (\tilde{a}/\tilde{r}_{t1})^{-1/2}$, where the constant of proportionality $\kappa$ is $q$-dependent. We find it to be in the range $0.97 \leq \kappa \leq 1.06$ for our parameter space, which is less than the estimate $\kappa \simeq 1.8$ found by \citet{yu03} (using $\langle v_\infty \rangle \simeq \sqrt{2 \epsilon_\infty}$ and the numerical results of \citealt{quinlan96}).

\item  A binary SMBH preferentially emits stars near its orbital plane, where binaries with lower separations and higher mass ratios require higher velocity cutoffs to observe this effect (Figure \ref{fig:histmu} and \ref{fig:histmuappendix}); this trend is not monotonic, as the mean polar angle of ejected stars shows some bumps towards isotropy as a function of ejection velocity (Figure \ref{fig:muvelcutoffs}). The locations of these bumps suggest that stars are preferentially emitted near the orbital plane if they have velocities just after the peak of the velocity distribution (Figure \ref{fig:histvinfesc}).

\item As the binary SMBHs in our parameter range contract, they eject stars with velocities $\tilde{v}_\infty \gtrsim 1000$ km/s at a rate $\sim 4 \times 10^{-2} - 2 \times 10^{-1}$ yr$^{-1}$ for $q = 1$ ($\sim 10^{-4} - 10^{-3}$ yr$^{-1}$ for $q = 0.01$) (Figure \ref{fig:dndta}). For our entire range of $q$, the binary SMBHs emit a burst of HVSs with $\tilde{v}_\infty > 3000$ km/s as they are about to coalesce; for lower velocity cutoffs, only those with low mass ratios exhibit a burst at late times (Figure \ref{fig:dndt}).

\item The ejected star velocity distribution integrated over the lifetime of the ejecting binary SMBH has a well-defined peak and power-law decay away from the peak, in contrast to the generally flatter distribution for stars ejected by the Hills mechanism (integrated over typical binary star separations in the surrounding bulge) (Section \ref{sec:comparison}). The former will yield lower velocities than the latter if the stellar binaries can reach a minimum separation of $\tilde{a}_{*,\textrm{min}} \sim 0.001$ AU; will be more difficult to distinguish if $\tilde{a}_{*,\textrm{min}} \sim 0.01$ AU; and will yield higher velocities if $\tilde{a}_{*,\textrm{min}} \sim 0.1$ AU (Figure \ref{fig:histvinfintegrated}).

\item From a set of HVS samples, one can more efficiently identify the SMBH progenitor by performing parameter estimation using the sample velocities as opposed to the directions, and probing as much of the velocity distribution as possible (i.e. applying a velocity cutoff $\tilde{V} > v_c$ with $v_c \lesssim 200$ km/s). Roughly $N \sim 100$ samples are required when examining the velocities with $v_c \lesssim 200$ km/s (Figure \ref{fig:oddsratiovel}), whereas thousands of samples are required when examining the directions with any velocity cutoff (Figure \ref{fig:oddsratiomu}).

\end{enumerate}

We focused on stars ejected directly by a SMBH progenitor, though HVSs (and more general ``runaway stars'') can be produced through several other channels (Section \ref{sec:intro}). Of the many alternatives, only the star-BH cluster scenario can produce HVSs with velocities and at rates even approaching those of the Hills mechanism or a binary SMBH \citep{yu03,oleary08}. Indeed, the ejection rates by binary SMBHs that we find (Figure \ref{fig:dndt}) are several orders of magnitude larger than those by star-BH cluster scattering found by \citet{oleary08}.

We analyzed the ejection of incident unbound stars from a full loss cone. The details of the merger process may depopulate the loss cone in the far field, suppressing the rate of incident low angular momentum stars. The binary SMBH components will likely have bound stars at the earlier stage of the inspiral \citep{baumgardt06,sesana08}, so the absence of a full loss cone after the binary SMBH depletes the inner part of the cusp would make these bound stars the last source of HVSs.

We presented the velocities of ejected stars in the potential of the BH progenitors only. The shape of the galaxy can, of course, affect the properties of the ejected stars \citep{rossi14,sesana07a,sesana07b}. In particular, a galactic potential may homogenize some of the HVS properties, making it difficult to distinguish between different SMBH origins.

In Section \ref{sec:comparison}, we compared HVSs arising from either single stars incident on a binary SMBH or binary stars incident on an isolated SMBH. In addition to these, if stellar binaries are incident on a SMBH binary, then one component of the SMBH binary can eject hypervelocity stars via the Hills mechanism \citep{coughlin18b}. This additional process should be considered when examining the properties of HVSs to determine their SMBH progenitors; we leave this investigation to another paper.

Binary SMBHs can also eject hypervelocity binary stars \citep{lu07,sesana09,coughlin18b,wang18}. The stellar binaries typically depart with modified semi-major axes and increased eccentricities, and thus have a shorter merger timescale than before the encounter \citep{coughlin18b}. Isolated SMBHs can eject hypervelocity binary stars as well by disrupting stellar triple systems \citep{perets09}. As with single HVSs, the properties of hypervelocity binary stars may depend on their origins and also serve as a probe of central black holes.

\section*{Acknowledgements}

We thank the anonymous reviewer for helpful comments. This research used resources of the National Energy Research Scientific Computing Center, a Department of Energy Office of Science User Facility supported by the Office of Science of the U.S. Department of Energy under Contract No. DE-AC02-05CH11231. ERC was supported by the National Aeronautics and Space Administration through the Einstein Fellowship Program, Grant PF6-170150. This work was supported by the National Science Foundation under Grant No. 1616754. This work was supported in part by a Simons Investigator award from the Simons Foundation (EQ) and the Gordon and Betty Moore Foundation through Grant GBMF5076. Simulations in this paper made use of the REBOUND code which can be downloaded freely at http://github.com/hannorein/rebound.




\bibliographystyle{mnras}
\bibliography{references} 




\appendix

\section{Additional Figures}
\label{sec:appendixa}


In this appendix, we expand some of the figures presented in the paper to our full parameter space.

\begin{figure*}
\centering
\subfloat{\includegraphics[scale=0.43]{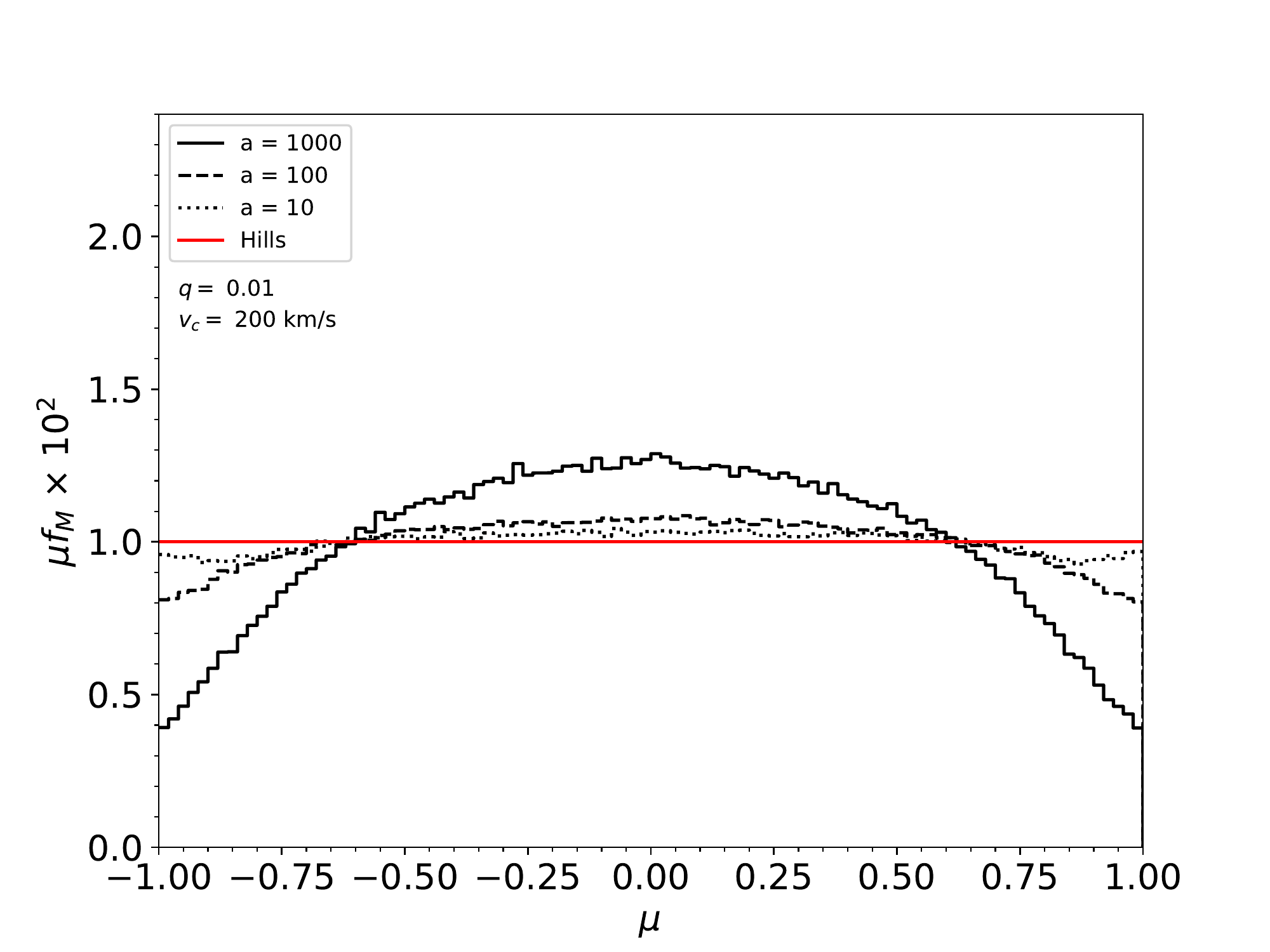}}\hfill
\subfloat{\includegraphics[scale=0.43]{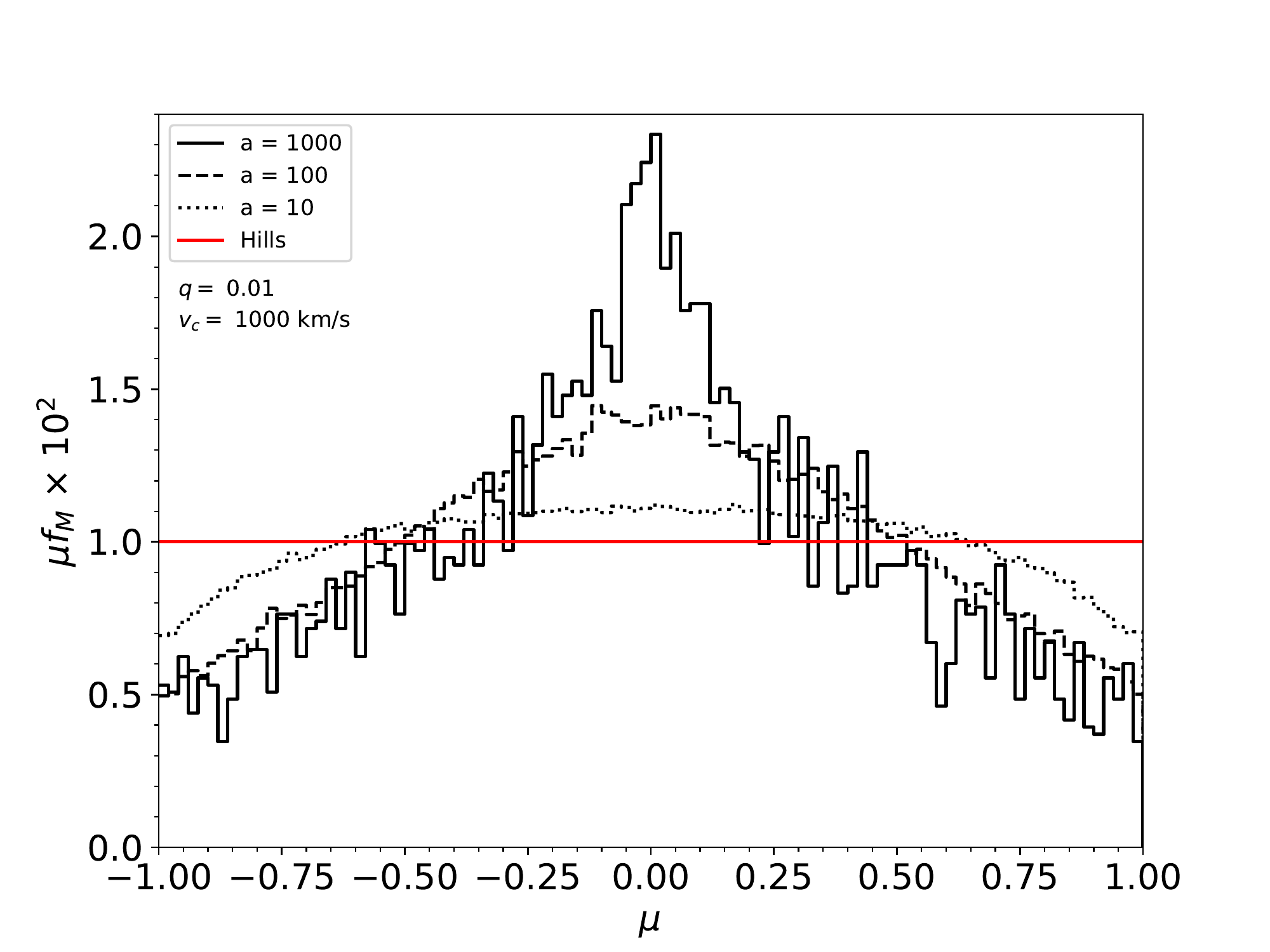}}\\
\subfloat{\includegraphics[scale=0.43]{figures/mu-binbh-q_0_1-vc_200.pdf}}\hfill
\subfloat{\includegraphics[scale=0.43]{figures/mu-binbh-q_0_1-vc_1000.pdf}}\\
\subfloat{\includegraphics[scale=0.43]{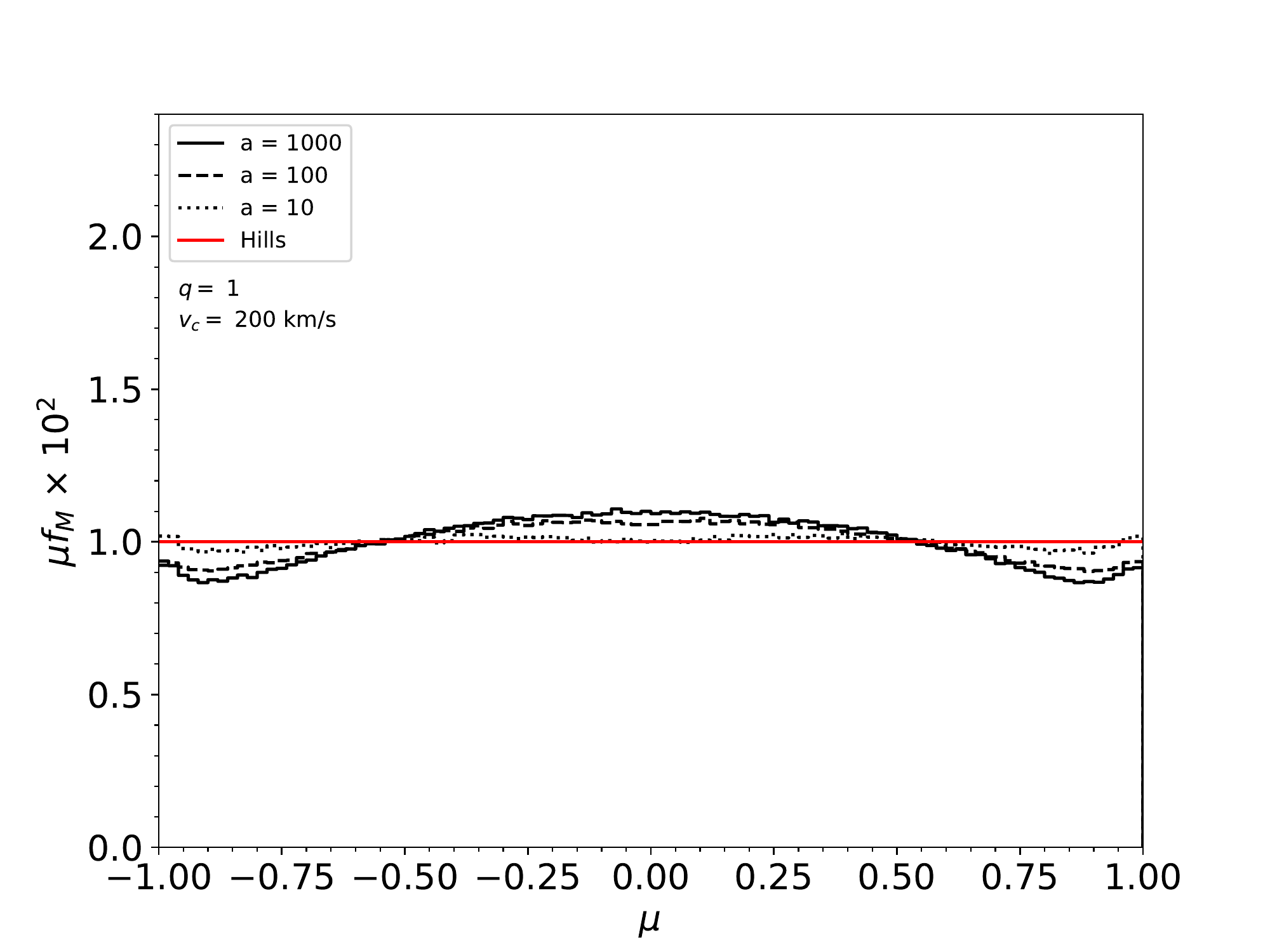}}\hfill
\subfloat{\includegraphics[scale=0.43]{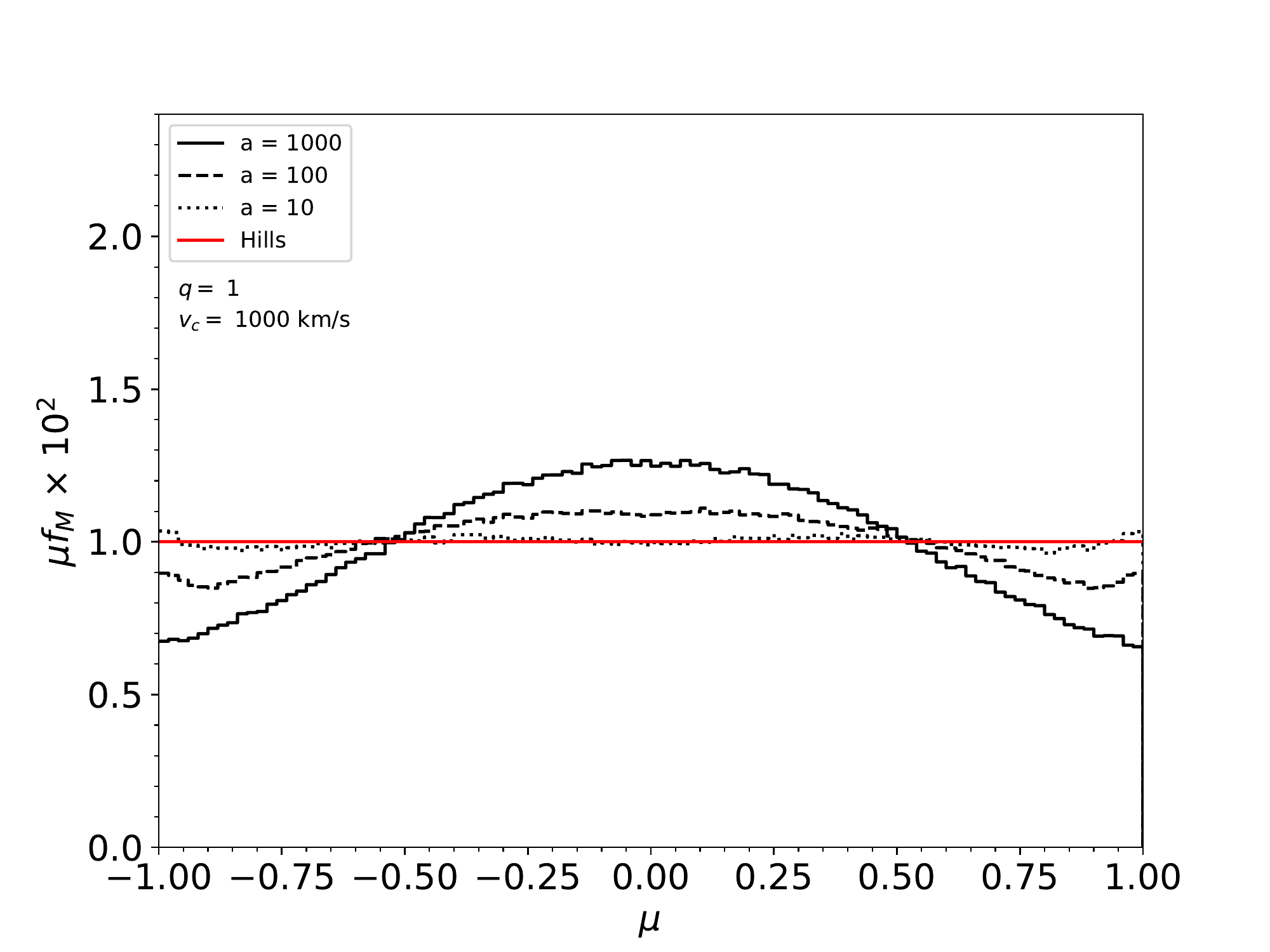}}
\caption{
The probabilities $\mu f_M$ for the direction cosine $\mu = \cos\theta = v_{z,\infty}/v_\infty$ of ejected stars at $\tilde{r}/\tilde{a} \rightarrow \infty$ for different binary SMBH mass ratios $q$ and velocity cutoffs $\tilde{V} > v_c$. The PDF is $f_M \equiv f_M(\mu)$. The variable $\theta$ is the polar angle measured from the direction normal to the binary SMBH orbital plane. The top panels show the results for $q = 0.01$, the center ones for $q = 0.1$, and the bottom ones for $q = 1$. The left panels show ejections with $v_c = 200$ km/s and the right ones show those with $v_c = 1000$ km/s. The linear bin widths are $\Delta_\mu = 0.02$. The red curve shows the uniform distribution for $\mu$ produced by the Hills mechanism.
}
\label{fig:histmuappendix}
\end{figure*}

\begin{figure*}
\centering
\subfloat{\includegraphics[scale=0.58]{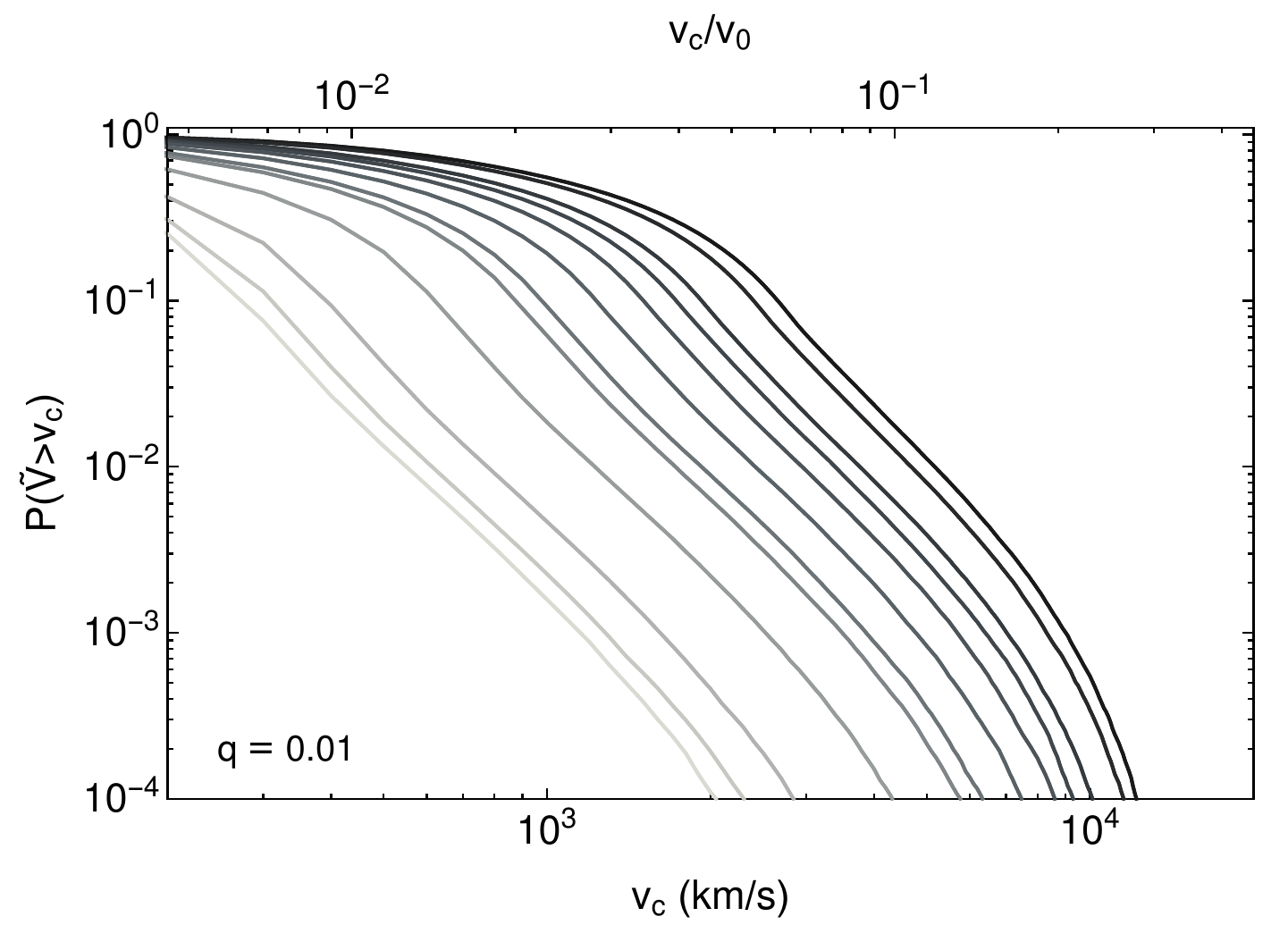}}\hfill
\subfloat{\includegraphics[scale=0.58]{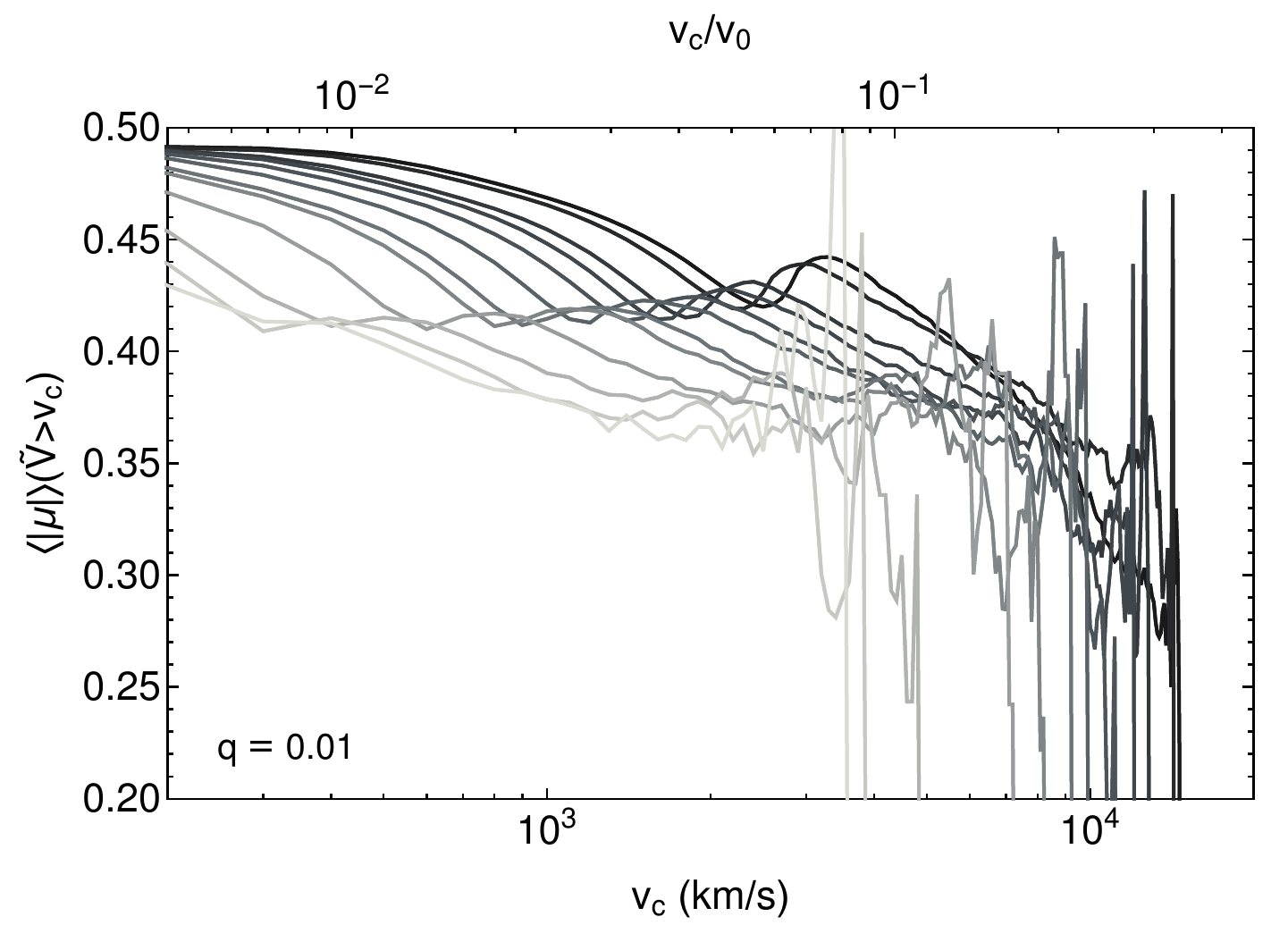}}\\
\subfloat{\includegraphics[scale=0.58]{figures/pvelcutoffs-q_0_1.pdf}}\hfill
\subfloat{\includegraphics[scale=0.58]{figures/muabsmeanvelcutoffs-q_0_1.pdf}}\\
\subfloat{\includegraphics[scale=0.58]{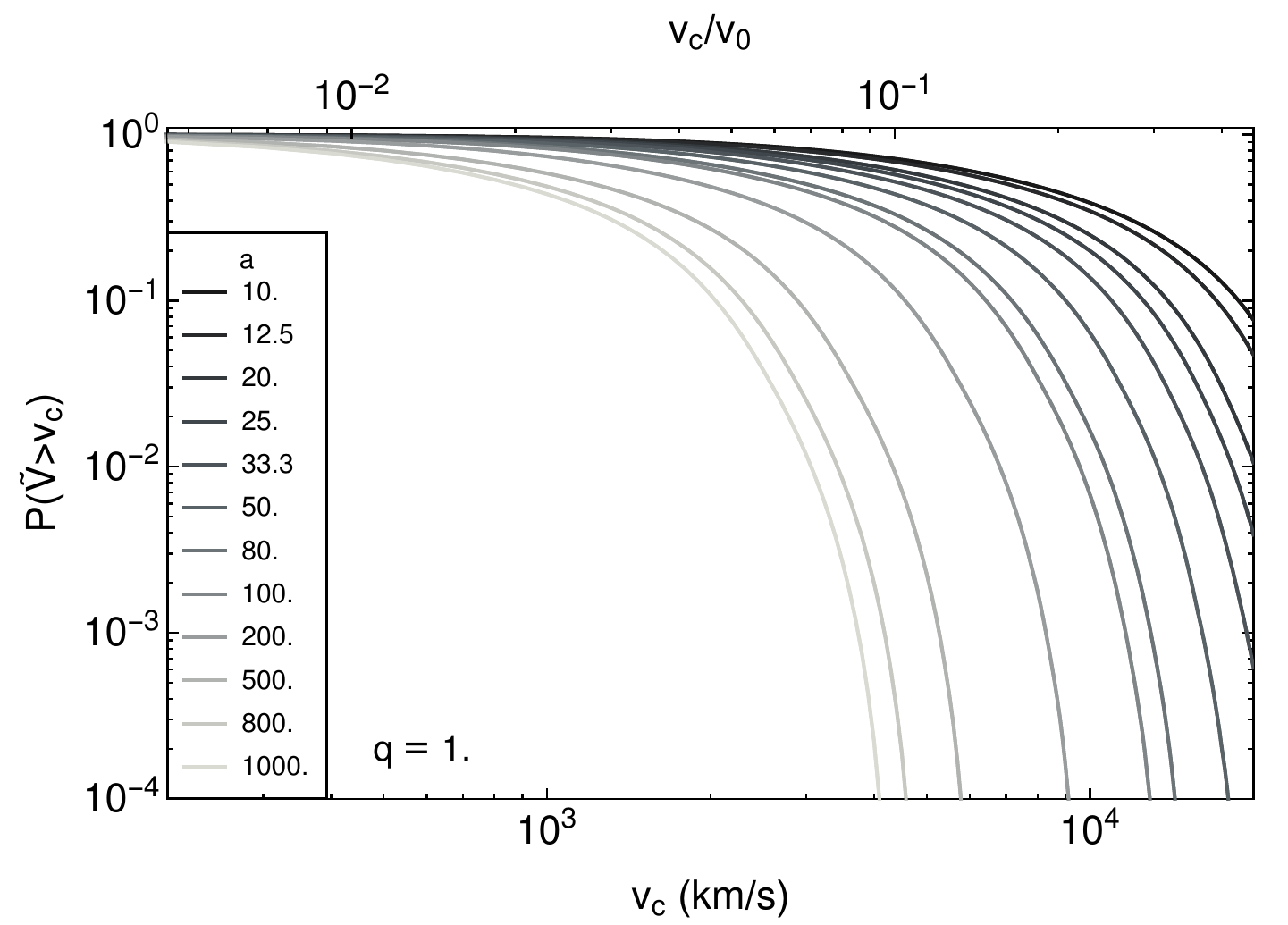}}\hfill
\subfloat{\includegraphics[scale=0.58]{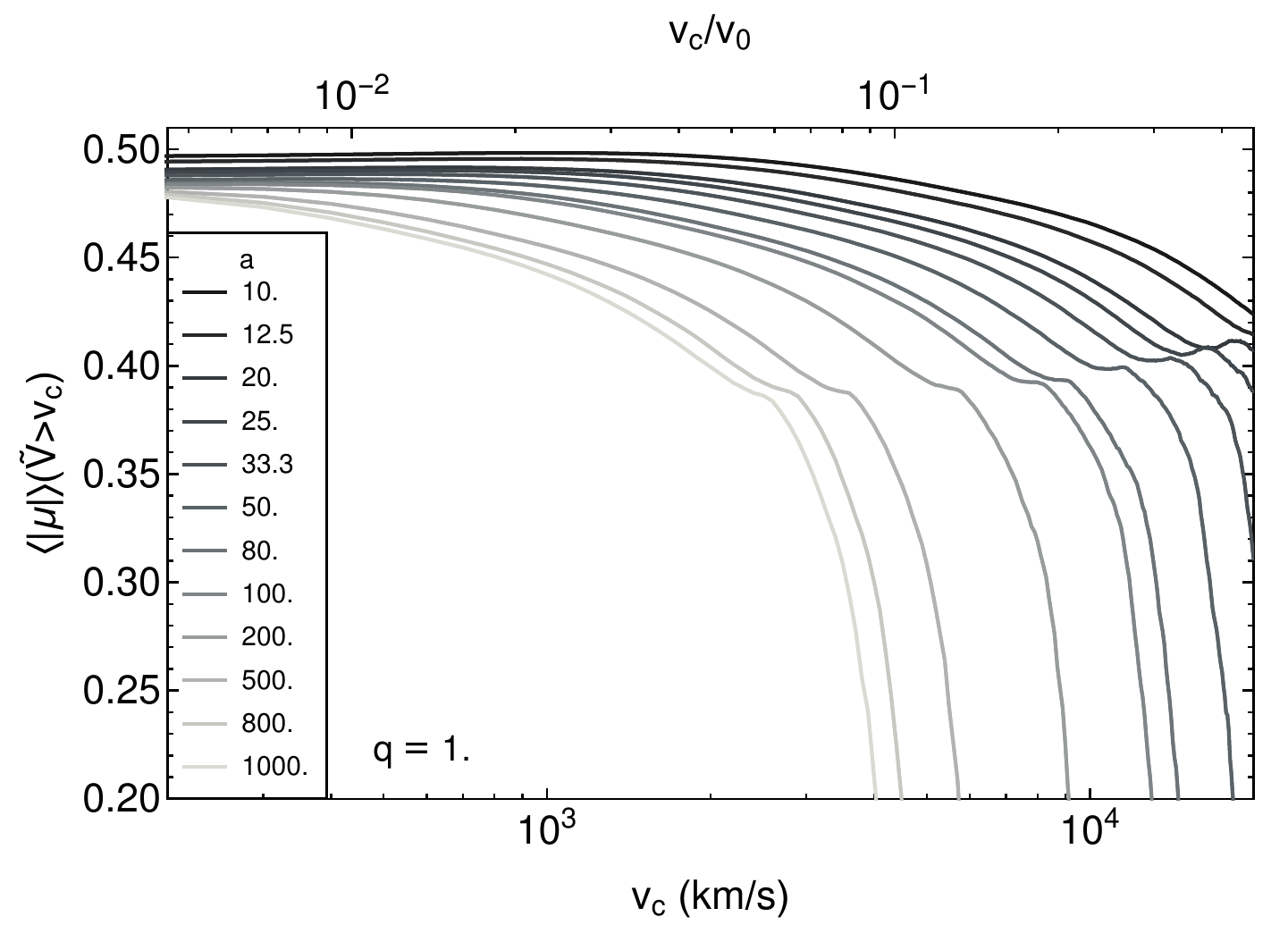}}
\caption{
Statistics of ejected stars with a range of velocity cutoffs $\tilde{V} > v_c$. The top panels show the results for $q = 0.01$, the center ones for $q = 0.1$, and the bottom ones for $q = 1$. The left panels show the probability $P(\tilde{V}>v_c)$ that an ejected star has a velocity $\tilde{v}_\infty > v_c$. The right panels show the orientation of ejected stars with velocities $v_\infty > v_c$; the orientation is parameterized by $\mu = \cos\theta = v_{z,\infty} / v_\infty$, where $\theta$ is the polar angle measured from the direction normal to the binary SMBH orbital plane. For an isotropic distribution, $\langle |\mu| \rangle = 0.5$.
}
\label{fig:velcutoffsappendix}
\end{figure*}

\bsp	
\label{lastpage}
\end{document}